\documentclass[12pt]{article}
\usepackage{fullpage,amsmath,amssymb,mathtools,natbib,hyperref}
\usepackage{titling,enumitem}
\usepackage[skip=0pt]{caption} 
\usepackage{datetime,bigints}
\usepackage[dvipsnames]{xcolor}
\usepackage[most]{tcolorbox}
\usdate
\usepackage{multirow,algorithm} 
\usepackage{float}
\usepackage{algorithm,bbm}
\usepackage[]{algpseudocode}
\usepackage{arydshln}
\usepackage[section]{placeins}
\usepackage[titletoc,toc,title]{appendix}
\usepackage{fancyvrb} 
\usepackage{listings}
\usepackage{bm}
\usepackage{color, colortbl} 
\usepackage{wrapfig}

\newcommand{\notes}[1]{%
    \linespread{0.2}\vspace{0.1em}%
    \captionsetup{justification=justified}%
    \caption*{\footnotesize #1}%
}

\definecolor{lightgray}{gray}{0.9}
\definecolor{gray}{gray}{0.85}
\xdefinecolor{blue}{RGB}{58,95,205}
\DeclareCaptionFormat{listing}{\rule{\dimexpr\textwidth+17pt\relax}{0.4pt}\vskip1pt#1#2#3}
\usepackage{xr,caption}

\renewenvironment{thebibliography}[1]{
  \begin{oldthebibliography}{#1}
    \setlength{\itemsep}{0em}
    \setlength{\parskip}{0em}
}
{
  \end{oldthebibliography}
}

\newcommand{\argmin}{\text{argmin}}

\usepackage{thmtools} 
\usepackage{amsthm}
{
      \theoremstyle{plain}
      \newtheorem{definition}{Definition}
      \newtheorem{theorem}{Theorem}
      \newtheorem{example}{Example}
      \newtheorem{proposition}{Proposition}
      \newtheorem{lemma}{Lemma}
      
      \newtheorem{assumption}{Assumption}
      
  }

\renewcommand{\arraystretch}{1.5}

\def\lQ{\scalebox{-1}[1]{''}}

\makeatletter
\renewenvironment{abstract}{%
    \if@twocolumn
      \section*{\abstractname}%
    \else 
      \begin{center}%
        {\bfseries \normalsize\abstractname\vspace{\z@}}
      \end{center} \vspace{-0.5cm}%
      \quotation
    \fi}
    {\if@twocolumn\else\endquotation\fi}
\makeatother

\begin{document}

  \title{Detecting Identification Failure   in\\  Moment Condition Models}
  \author{\Large Jean-Jacques Forneron\thanks{Department of Economics, Boston University, 270 Bay State Road, Boston, MA 02215 USA.\newline Email: \href{mailto:jjmf@bu.edu}{jjmf@bu.edu}, Website: \href{http://jjforneron.com}{http://jjforneron.com}.   \newline I would like to thank Serena Ng for discussions that initiated this project. I thank Francesca Molinari for suggestions  that  greatly  improved  this  paper. I also greatly benefited from comments and discussions with Tim Christensen, Pavel Cizek, Greg Cox, Iv\`an Fern\`andez-Val, Hiro Kaido, Nour Meddahi, Arthur Lewbel, Demian Pouzo, Zhongjun Qu, Eric Renault, Yichong Zhang and the participants of the BU-BC econometric workshop, the seminar participants at Brown, Chicago, CREST, NUS, NYU, UC Berkeley, Universit\'e de Montr\'eal, University of Rochester, SMU, Toulouse School of Economics and conferences. I would also like to thank Joachim Grammig for kindly sharing his replication files for the long-run risks model.}
  } \date{\today}
  \maketitle 

  \begin{abstract}  
    This paper develops an approach to detect identification failure in moment condition models. This is achieved by introducing a \textit{quasi-Jacobian} matrix computed as the slope of a linear approximation of the moments on an estimate of the identified set. It is asymptotically singular when local and/or global identification fails, and equivalent to the usual Jacobian matrix which has full rank when the model is globally and locally identified. Building on this property, a simple test with chi-squared critical values is introduced to conduct subvector inferences allowing for strong, semi-strong, and weak identification without \textit{a priori} knowledge about the underlying identification structure. Monte-Carlo simulations and an empirical application to the Long-Run Risks model illustrate the results.
  \end{abstract}
  

  \bibliographystyle{ecta}
  \baselineskip=18.0pt
  \thispagestyle{empty}
  \setcounter{page}{0}
  
\newpage

\section{Introduction}

The Generalized Method of Moments (GMM) of \citet{Hansen1982} is a powerful estimation framework which does not require the model to be fully specified parametrically. Under regularity conditions, the estimates are consistent and asymptotically Gaussian. In particular, the moments should uniquely identify the finite-dimensional parameters. This is very difficult to verify in practice and, as noted in \citet{Newey1994a}, is often assumed. Yet, when identification fails or nearly fails, the Central Limit Theorem provides a poor finite sample approximation for the distribution of the estimates. This has motivated a vast amount of research on tests which are robust to identification failure. An empirically relevant problem, which remains less explored, is of determining, for a given set of estimating moments, whether local and global identification actually hold. 

The contribution of this paper is two-fold: first, it introduces a \textit{quasi-Jacobian} matrix which is singular under both local (first-order) and global identification failure and is informative about the coefficients involved in the identification failure. This is the main contribution of the paper as it provides an approach similar to \citet{Cragg1993} and \citet{Stock2005} but in a non-linear setting. Second, the information is used to construct an identification robust subvector test which does not require \textit{a priori} knowledge of the identification structure. The test is asymptotically non-conservative under strong identification. It is asymptotically efficient for strongly just-identified models.

The quasi-Jacobian matrix is the best linear approximation of the sample moment function over a region of the parameters where these moments are close to zero. To find the best linear approximation, a sup-norm (or $\ell_\infty$-norm) loss is used to minimize the largest deviation from the linear approximation. This is known as a Chebyshev approximation problem which can be solved fairly quickly using convex optimization software. In the population, the quasi-Jacobian has full rank if, and only if, the parameters are both globally and locally identified. When either global or local identification fails, it is singular in all directions associated with the identification failure. (Non)-singularity of the quasi-Jacobian can be used to check whether identification holds numerically when it is not feasible analytically. 

The asymptotic behaviour of the quasi-Jacobian matrix is studied under three identification regimes: including strong, semi-strong, and weak (or set) identification. Under strong identification, the moment conditions are informative, have a unique solution, under semi-strong identification are less informative but sufficiently so that for estimates to be consistent and asymptotically Gaussian. \citet{Antoine2009}, \citet{Andrews2012} showed that: under (semi)-strong identification, standard inference methods such as the t-test with standard normal critical values are asymptotically valid.\footnote{The term (semi)-strong will refer to cases where identification can be either strong or semi-strong. \citet{Antoine2009} further distinguish between nearly-strong and nearly-weak identification. Under the latter, the limiting distribution may be non-Gaussian. Here, when this is the case, it will be referred to as higher-order local identification.} Under weak and set identification, the moments are insufficiently informative compared to sampling uncertainty and multiple distant solutions to the moment conditions appear plausible, even in large samples, so that the parameters cannot be consistently estimated and standard inference methods are not asymptotically valid. The Supplement also considers higher-order local identification, where the solution is unique but not locally identified; it can be consistently estimated but with non-Gaussian limiting distribution.    
Under (semi)-strong identification, the quasi-Jacobian is shown to be asymptotically equivalent to the usual Jacobian: after re-scaling, it is asymptotically non-singular. Under higher-order and weak identification the quasi-Jacobian is asymptotically singular with eigenvalues vanishing in directions where identification fails. It is thus informative about the presence of identification failures and which directions are not identified. 
  
Building on these results, this paper constructs a simple test procedure for subvector hypotheses on the parameters $\theta=(\theta_1^\prime,\theta_2^\prime)^\prime \in \mathbb{R}^{d_\theta}$ of the form:
\begin{align} \label{eq:hypothesis}
  H_0:\, \theta_1 = \theta_{10} \text{ vs. } H_1: \, \theta_1 \neq \theta_{10}.
\end{align}
Subvector inference as described in (\ref{eq:hypothesis}) is quite prevalent in empirical work where  only a few structural parameters $\theta_1$ are typically of interest. The remaining $\theta_2$ nuisance parameters  describe other features of the data generating process needed for estimation. For instance, in the empirical application only $2$ preference parameters are of interest while the remaining $10$ coefficients parameterize the law of motion for consumption and dividends which is not of immediate interest. 
The paper relies on the \citet[AR]{Anderson1949} test statistic for simplicity. The critical values take the form $\chi^2_{d_g-d}$ where $d_g$ is the number of moments and $d$ is determined using an Identification Category Selection (ICS) procedure based on the singular values of the quasi-Jacobian matrix. This is a projection inference procedure where the ICS step estimates the number of (semi)-strongly identified nuisance parameters to reduce the degrees of freedom.  

Monte-Carlo simulations illustrate the results for a simple consumption-based asset pricing model. In the empirical application, the procedure is used to conduct joint inference on risk-aversion and the inverse elasticity of substitution in the long-run risks model of \citet{bansal2004}. The results suggest that several nuisance parameters are weakly identified but not all; some are (semi)-strongly identified. This implies that standard inferences based on t or Wald statistics are not asymptotically valid and full projection inference is valid, but conservative. 
Given the number of parameters in the application, the standard approach of performing test inversion using a grid search is very computationally demanding. Instead, an adaptive sampling procedure based on the Population Monte Carlo (PMC) principle draws uniformly on level sets of the objective function. This makes it possible to conduct robust inference on more complex models like the empirical application: the quasi-Jacobian and 5,000 uniform draws on the confidence set are computed in about 4 hours on a desktop computer.

\subsection*{Structure of the Paper} 
After a review of the literature and an overview of the notation, Section \ref{sec:setting} introduces the setting, the procedure and provides more details about the quasi-Jacobian, the test, and the identification regimes. Section \ref{sec:asym_jac} derives the asymptotic behaviour of the quasi-Jacobian matrix, and Section \ref{sec:asym_test} results for the test. Section \ref{sec:MonteCarlo} gives Monte-Carlo evidence for the results, and Section \ref{sec:empirical} the empirical application. Appendices \ref{apx:prelim}, \ref{appx:proof_main} provide proofs for the main results. The Supplement includes sample R code to compute the quasi-Jacobian and for inference, a description of the PMC algorithm used to generate draws, and additional results for higher-order identification.

\subsection*{Related Literature}
The literature on the identification of economic models is quite vast, and an extensive review is given in \citet{Lewbel2018}. Within this literature, this paper mainly relates to three topics: local and global identification of finite-dimensional parameters in the population, detection of identification failure in finite samples, and identification robust inference. 

\citet{Koopmans1950} provide one of the earliest general formulations of the identification problem at the population level. To paraphrase the authors, the main problem is to determine whether the distribution of the data, assumed to be generated from a given class of models, is consistent with a unique set of structural parameters. In the likelihood setting, \citet{Fisher1967}, \citet{Rothenberg1971} introduced sufficient conditions for local and global identification. \citet{Komunjer2012} provides weaker global identification conditions for GMM.

In linear models, global identification amounts to a rank condition on the slope of the moments. This insight was used in pre-testing linear IV models for identification failure using a first-stage F-statistic or rank tests, \citet{Cragg1993}, \citet{Stock2005}, \citet{kleibergen2006}. Pre-tests based on the null of strong identification appear in \citet{Hahn2002} for linear IV and \citet{Inoue2011}, \citet{Bravo2012} for non-linear models. Pre-testing for strong identification could make size control difficult when the pre-test has low power. For non-linear models, \citet{Wright2003} uses a rank test and \citet{antoine2020} a distorted J-statistic to detect local identification failure. \citet{Arellano2012} develop a test for underidentification of a single coefficient.

Given the impact of (near) identification failure on standard inferences, a large body of literature has developed identification robust tests. Much of the literature is concerned with inference on the full parameter vector, e.g. \citet{Anderson1949}, \citet{Stock2000}, \citet{Kleibergen2005}, \citet{Andrews2016}. Projection inference can be used to conduct subvector inference from these tests \citep{dufour1997}. Alternatively, Bonferroni methods combined with a $C(\alpha)$ test can be used, \citet{Chaudhuri2011}, \citet{andrews2017}. For homoskedastic linear IV models, \citet{guggenberger2012a} propose critical values for a subset Anderson-Rubin test which improve power over full projection inference. In the same setting, \citet{guggenberger2019} propose a data-driven choice of critical values based on a measure of identification strength of the nuisance parameters, and \citet{kleibergen2021} considers subvector conditional Likelihood-Ratio inference. This paper relies on the Anderson-Rubin statistic for inference, which is the simplest to implement. More powerful test statistics exist such as the conditional quasi-Likelihood Ratio. The main challenge there is in computing the critical values by simulation, which requires to repeatedly minimize non-linear and potentially multi-modal objective functions.\footnote{This is difficult for non-convex problems, see e.g. \citet[Section 1.6.2]{Nemirovsky1983} for the complexity of the minimization problem and \citet[p14-16]{Nesterov2018} for the practical implications and software limitations.} 

Given knowledge about the source of a potential identification failure, and a specific structure in the underlying model \citet{Andrews2012,Andrews2013,Andrews2014}, \citet{Cheng2015}, \citet{han2019}, \citet{Cox2020} propose identification robust tests which are asymptotically non-conservative and powerful under strong identification. These papers rely on a data-driven choice of critical value; it is determined by an ICS statistic built from model-specific knowledge about the source and form of the identification failure. This paper proposes and studies an ICS statistic which does not rely on model-specific information to determine identification status. The choice of robust critical values can coincide with \citet{Andrews2012}'s least-favorable critical value, see Appendix \ref{apx:MC_additional_simu} for an example. \citet{andrews2017} proposes an ICS based on the singular values of sample Jacobian which measures local but not global identification strength. His test applies to GMM and likelihood problems. 

Under higher-order identification, estimates are consistent but the delta-method is not valid. The limiting distribution is non-standard \citep{Rotnitzky2000}, \citet{Dovonon2018}. This issue is known but much less studied than weak and set identifications. \citet{Dovonon2019} study identification robust tests under second-order identification, and \citet{Lee2018} conduct standard inference under known second-order identification structure. 

\subsection*{Notation}  
For any matrix (or vector) $A$, $\|A\| = \sqrt{\sum_{i,j} A_{i,j}^2} = \sqrt{\text{trace}(AA^\prime)}$ is the Frobenius (Euclidian) norm of $A$. For any square matrix $A$, $\lambda_j(A)$ refers to the j-th eigenvalues of $A$, in increasing order if $A$ is symmetric positive semi-definite; $\lambda_{\max}(A)$ and $\lambda_{\min}(A)$ refer to its largest and smallest eigenvalue, respectively, $\lambda_{1}(A),\dots,\lambda_{d}(A)$ are the first d eigenvalues of $A$ in increasing order. For a weighting matrix $W_n(\theta)$, the norm $\|\bar g_n(\theta)\|^2_{W_n}$ is computed as $\bar g_n(\theta)^\prime W_n(\theta) \bar g_n(\theta)$. The abbreviation wpa 1 will be used to abreviate ``with probability approaching 1.'' For $\varepsilon >0$, $B_{\varepsilon}(\theta)$ is a closed $\varepsilon$-ball around $\theta$.  

\section{Setting and Assumptions} \label{sec:setting}
Following \citet{Hansen1982}, the econometrician wants to estimate the solution vector $\theta_0$ to the system of unconditional moment equations:
\begin{align} g(\theta_0,\gamma_0) \overset{def}{=}  \mathbb{E}_{\gamma_0}(\bar g_n(\theta_0))=0, \label{eq:gmm} \end{align}
where $\theta_0 = (\theta_{10}^\prime,\theta_{20}^\prime)^\prime \in \overline{\Theta} = \overline{\Theta}_1 \times \overline{\Theta}_2$, a compact subset of $\mathbb{R}^{d_\theta}$, $\text{dim}(\bar{g}_n) = d_g \geq d_\theta$. $\bar g_n(\theta)= 1/n\sum_{i=1}^n g(z_i,\theta)$ is the sample vector of moment conditions, $(z_i)_{i=1,\dots,n}$ is a sample of iid or stationary random variables. The parameter $\gamma_0 \in \Gamma$ indexes the true distribution of the data $(z_i)$, including the true $\theta_0$. It has the form $\Gamma = \{ \gamma=(\theta,\omega), \theta \in \overline{\Theta}, \omega\in\Omega\}$. $\Omega$ indexes features of the data generating process beyond $\theta$ that are relevant to identification and weak convergence. $\Gamma = \overline{\Theta} \times \Omega$ is a compact subset of a metric space with a metric $\|\theta-\tilde \theta\| + d(\omega,\tilde\omega)$ between $\gamma=(\theta,\omega)$ and $\tilde\gamma=(\tilde\theta,\tilde\omega)$ that induces weak convergence for $(z_i,z_{i+m})$ for any $i,m \geq 1$.\footnote{For reduce the number of coefficients involved in the notation below, this distance will be written as $\|\theta-\tilde\theta\| + d(\gamma,\tilde \gamma)$.  See \citet[p2162]{Andrews2012} for a discussion of these conditions.} The operator $\mathbb{E}_{\gamma_0}$ denotes the expectation under $\gamma=\gamma_0$. $g(\theta,\gamma_0) = \mathbb{E}_{\gamma_0}(\bar g_n(\theta))$ is then the population vector of moment conditions evaluated at the true $\gamma_0 \in \Gamma$ and a coefficient $\theta$.  Throughout, it is assumed that $\theta_0$ is such that $g(\theta_0,\gamma_0)=0$. The function $g(\cdot,\gamma)$ is assumed to be continuously differentiable on $\Theta$ for all $\gamma$. 

Given the sample moments $\bar{g}_n$ and a sequence of positive definite weighting matrices $W_n(\theta)$ converging to $W(\theta)$, the GMM estimator $\hat \theta_n$ solves the sample minimization problem:
\begin{align}\hat \theta_n = \text{argmin}_{\theta \in \Theta} \,  \bar g_n(\theta)^\prime W_n(\theta)\bar g_n(\theta),\label{eq:estimator} \end{align}
where $\Theta = \Theta_1 \times \Theta_2$ is the optimization space. 
\begin{assumption}[Parameter Space, Sample Moments, Weighting Matrix] \label{ass:moments}\,
  i. $\Gamma$ and $\overline{\Theta} \subset \mathbb{R}^{d_\theta}$ are compact; $\Theta$ is a convex, compact subset of $\mathbb{R}^{d_\theta}$ such that $\overline{\Theta} \subset \Theta$ and $\cup_{\theta \in \overline{\Theta} } B_\eta(\theta) \subseteq \Theta$ for some $\eta >0$; for all $(\theta,\gamma) \in \overline{\Theta} \times \Gamma$ and all $\varepsilon >0$, $B_\varepsilon(\theta,\gamma) \cap (\overline{\Theta}\times\Gamma)$ is non-singleton and connected,
  ii.  for any sequence $(\theta_n,\gamma_n) \to (\theta_0,\gamma_0) \in \overline{\Theta} \times \Gamma$: $\sup_{\theta \in \Theta} \sqrt{n}\|\bar{g}_n(\theta) - g(\theta,\gamma_n)\| = O_p(1),$ and $\sqrt{n}\bar{g}_n(\theta_n) \overset{d}{\to} \mathcal{N}(0,V_0)$ where $V_0$ is finite and non-singular,
  iii. $\sup_{\theta \in \Theta}\|W_n(\theta)-W(\theta)\| = o_p(1)$. $W_n$ and $W$ are Lipschitz continuous in $\theta$; there exists $\underline{\lambda}_W,\overline{\lambda}_W$ such that $0<\underline{\lambda}_W \leq \lambda_{\min}(W_n(\theta)) \leq \lambda_{\max}(W_n(\theta)) \leq \overline{\lambda}_W <\infty$, for all $\theta$. 
\end{assumption}

Assumption \ref{ass:moments} i. implies that $\Theta$ strictly contains $\overline{\Theta}$ so that issues arising when a parameter is on the boundary are not considered here.\footnote{See \citet{Cox2020} for results on identification and boundary robust inference.} The connected neighborhood condition plays the role of Assumption ACP iv. in \citet[p2165]{Andrews2012}. It implies that we can find sequences $\gamma_n$ along a continuous path in $\Gamma$ leading to $\gamma_0$ such that $0<\|\gamma_n-\gamma_0\| \to 0$. Together with a continuity condition in Assumption \ref{ass:identification} below, it allows to interpolate converging subsequences into converging sequences of parameters in one of the desired identification categories. This is similar to Assumption B2 in \citet{andrews2020} and Assumption 14 in \citet{Cox2020}. Condition ii. is a uniform convergence condition, implied by a uniform CLT. Condition iii. ensures that $\|\cdot\|_{W_n}$ is equivalent to $\|\cdot\|$ so that the choice of $W_n$ does not alter the identifiability of the parameters.


\subsection{Outline of the Procedure} \label{sec:outline}
The following steps provide a general overview of the computation of the quasi-Jacobian matrix, the ICS, and test procedure used in the paper. In the following, the matrix $P_{\theta_1}^\perp$ is an orthogonal projection matrix, projecting on the space orthogonal to $\theta_1$. It can be written as $P_{\theta_1}^\perp = \text{diag}(0_{d_{\theta_1}},1_{d_{\theta_2}})$ so that it only selects elements associated with $\theta_2$. The matrix $\overline{V}_n$ is a weighted average of estimates of $\text{var}[\sqrt{n}\overline{g}_n(\theta_b)]$, with weights proportional to $\hat{K}_n(\theta_b)$, described in more details below.\footnote{For iid data, $\text{var}[\sqrt{n}\overline{g}_n(\theta_b)]$ is approximated using $\frac{1}{n} \sum_{i=1}^n g(z_i,\theta_b)g(z_i,\theta_b)^\prime - \overline{g}_n(\theta_b) \overline{g}_n(\theta_b)^\prime$; for dependent data a HAC estimator is used.}

\begin{tcolorbox}[
  colback=white!30,
  sharp corners,
]  \textbf{Computing the quasi-Jacobian and the test statistic:} \vspace{0.15cm}
 { \small 
 \begin{enumerate}[itemsep=0.5pt,parsep=0.5pt,topsep=0.5pt] 
  \item \textbf{Inputs} bandwidth $\kappa_n$, kernel $K$, cutoff $\underline{\lambda}_n$, number of draws $B$
  \item \textbf{quasi-Jacobian Matrix}
  \begin{enumerate}[itemsep=0.5pt,parsep=0.5pt,topsep=0.5pt] 
    \item[i.] Draw $(\theta_b)_{b=1,\dots,B}$ uniformly on the level set $\{ \theta \in \Theta, \|\bar{g}_n(\theta)\|_{W_n} \leq \kappa_n \}$
    \item[ii.] Compute the intercept $A_{n,\infty}$ and slope $B_{n,\infty}$ in the $\ell_\infty$-norm regression:
    \begin{align}
     (A_{n,\infty},B_{n,\infty}) =  \argmin_{A,B} \left( \sup_{b \in \{1,\dots,B\}} \|\overline{g}_n(\theta_b) -A - B \theta_b\| \hat{K}_n(\theta_b) \right),\label{eq:minimax}
    \end{align}
    where $\hat{K}_n(\theta_b) = K( \|\overline{g}_n(\theta_b)\|_{W_n}/\kappa_n )$.
    \item[iii.] Compute the variance $\Sigma_n$:
    \begin{align}
      (\mu_n,\Sigma_n) =  \argmin_{\Sigma,\mu} \left( \sup_{b \in \{1,\dots,B\}} (\log|\Sigma| + \|\theta_b-\mu\|^2_{\Sigma^{-1}})\hat{K}_n(\theta_b)\right),\label{eq:variance}
     \end{align}
  \end{enumerate}
  \item \textbf{Identification Category Selection} 
  \begin{enumerate}[itemsep=0.5pt,parsep=0.5pt,topsep=0.5pt] 
    \item[i.] Compute the singular values $(\lambda_{jn})_{j=1,\dots,d_{\theta}}$ of $\overline{V}_n^{-1/2} B_{n,\infty} P_{\theta_1}^\perp \Sigma_n^{-1/2} P_{\theta_1}^\perp$
    \item[ii.] Compute $\hat{d}_n$, the number of singular values $\lambda_{jn}$ greater than $\underline{\lambda}_n$
  \end{enumerate}
    \item  \textbf{Subvector Inference} 
    \begin{enumerate}[itemsep=0.5pt,parsep=0.5pt,topsep=0.5pt] 
      \item[i.] Compute the test statistic: $\text{AR}(\theta_{10}) = \inf_{\theta_2 \in \Theta_2} n\|\bar{g}_n(\theta_{10},\theta_2)\|^2_{\hat{V}_n^{-1}}$
      \item[ii.] Reject $H_0: \theta_1 = \theta_{10}$ at the $1-\alpha$ confidence level if $\text{AR}(\theta_{10}) > \chi^2_{d_g - \hat{d}_n}(1-\alpha)$
    \end{enumerate}
  \end{enumerate}
 }
\end{tcolorbox}

In the procedure, $\chi^2_{d_g - \hat{d}_n}(1-\alpha)$ is the $1-\alpha$ quantile of a $\chi^2$ distribution with $d_g - \hat{d}_n$ degrees of freedom, $d_g$ is the number of moment conditions. In the following, the number of draws $B$ is assumed to be sufficiently large for the finite-$B$ approximation error to be negligible. The $\ell_\infty$ regression (\ref{eq:minimax}) is known as a Chebyshev (or minimax) approximation problem and can be cast as a linear programming problem \citep[p293]{boyd2004}. It can be solved with a few lines of code using the \textsc{cvx} convex optimization toolkit.\footnote{See Supplemental Appendix \ref{sec:code} for sample R code which implements the method.} (\ref{eq:variance}) is also solved using \textsc{cvx}. Finally, note that, in the procedure, the intercept $A_{n,\infty}$ and the mean $\mu_n$ are nuisance parameters, only $B_{n,\infty}$ and $\Sigma_n$ are used in steps 3-4. On the computation side: Appendix \ref{apx:PMC} outlines a sequential Algorithm to sample on the level set (step 2i.), the quasi-Jacobian is only computed once; it is defined whether the sample moments are differentiable, or not. The standard Jacobian requires differentiability and needs to be evaluated at every grid point. Instead of the $\ell_\infty$ loss, one could use the $\ell_2$-norm which yields least-squares solutions $(A_{n,LS},B_{n,LS})$. Some technical difficulties arise because the identified set typically has measure zero, and stronger assumptions are required to derive the properties of $B_{n,LS}$ compared to $B_{n,\infty}$. The re-scaling in step 3 is discussed below. 
The following provides further details about the steps outlined above.  

\subsection{Linear Approximations and the quasi-Jacobian Matrix}

The \textit{quasi-Jacobian} matrix $B_{n,\infty}$ is defined as the slope of a local linear approximation for $\bar g_n (\cdot)$ over an estimate of the identified set. 

\begin{definition}{(Sup-Norm Approximation)} \label{def:Approx} Let $K$ be a kernel function and $\kappa_n$ a bandwidth. 
  The sup-norm approximation $(A_{n,\infty},B_{n,\infty})$ solves: 
  \begin{align}
      (A_{n,\infty},B_{n,\infty}) &= \argmin_{A,B} \left(\sup_{\theta \in \Theta} \left[\|A+B\theta-\bar g_n(\theta)\| \hat{K}_n(\theta)\right] \right), \label{eq:sup_norm}
  \end{align}
  where $\hat{K}_n(\theta)=K\left(\|\bar g_n(\theta)\|_{W_n}/\kappa_n\right)$. The quasi-Jacobian refers to the slope matrix $B_{n,\infty}$.
\end{definition}

In practice, the minimization problem (\ref{eq:sup_norm}) is solved over a finite grid as in (\ref{eq:minimax}). The grid can be generated using Monte-Carlo or quasi-Monte-Carlo methods \citep{Robert2004,Lemieux2009}. In the simulations, the Sobol sequence was used. In the empirical application, $d_\theta=12$ is relatively large, and the set of $\theta$ where $\hat{K}_n(\theta) >0$ is fairly narrow; the acceptance rate is very low. A very large number of draws would be needed to find sufficiently many $\theta_b$ with non-zero weight, i.e. $\hat{K}_n(\theta_b)>0$. The empirical application relies on a sequential sampling principle called Population Monte Carlo \citep{cappe2004}. It constructs a sequence of proposal distributions that approximate the target distribution with increasing accuracy, see Appendix \ref{apx:PMC} for details. These proposals can be re-purposed to compute confidence sets, reducing the additional time required for test inversion. It can also be used to compute $B_{n,\infty}$ for different values of $\kappa_n$ as a sensitivity analysis.

\begin{assumption}[Kernel, Bandwidth] \label{ass:kernel_bandwidth} \,
  i. $K(x)>0$ if $x \in [0,1)$, $K(x)=0$ if $x \geq 1$. $K$ is continuous on $[0,1)$,
  ii. $\sqrt{n}\kappa_n \to \infty$, $\sqrt{n}\kappa_n^2 \to 0$.
\end{assumption}

The kernel is assumed to have compact support. 
The uniform kernel, $K(x) = \mathbbm{1}_{x \in [-1,1]}$, was used in the simulations and empirical results.\footnote{The estimated $B_{n,\infty}$ is nearly numerically identical using the cosine or Epanechnikov kernels.} The first condition ensures that $\hat{K}_n(\cdot)$ selects the identified set with wpa 1 under weak identification. The second ensures that $B_{n,\infty}$ only captures the first-order Jacobian term in local expansions under (semi)-strong identification. Otherwise, it would also capture nonlinear terms from the remainder.


\subsection{Test Procedure} \label{sec:test}
To illustrate the usefulness of detecting identification failure, consider the following simple data-driven test procedure. It is based on the Anderson-Rubin statistic for non-linear GMM models as described in \citet{Stock2000}. To test null hypotheses of the form $H_0: \theta_1 = \theta_{10}$, compute the sample statistic:
\[ \text{AR}_n(\theta_{10}) = \inf_{\theta_2 \in \Theta_2} n \left( \bar{g}_n(\theta_{10},\theta_2)^\prime \hat{V}^{-1}_n(\theta_{10},\theta_2) \bar{g}_n(\theta_{10},\theta_2) \right), \]
where $\hat V_n(\theta)$ consistently estimates the asymptotic variance  $\lim_{n\to\infty} \text{var}(\sqrt{n}\bar{g}_n(\theta))$. The test rejects at a nominal level $\alpha \in (0,1)$ if $\text{AR}_n(\theta_{10}) > \chi^2_{d_g - \hat d_n}(1-\alpha)$ where $\chi^2_{d_g - \hat d_n}(1-\alpha)$ is the $1-\alpha$ quantile of a chi-square distribution with $d_g - \hat d_n$ degrees of freedom. $\hat d_n \in \{0,\dots,d_{\theta_2}\}$ is computed using an identification category selection (ICS) procedure based on the quasi-Jacobian and its singular values. The procedure, described below, evaluates the number of nuisance parameters in $\theta_2$ which are potentially weakly/set identified. Using $\hat d_n = 0$ yields the largest critical value and amounts to full projection inference \citep{Dufour2005}. Using $\hat d_n = d_{\theta_2}$ yields the smallest critical value which provides valid, non-conservative inferences when all of the nuisance parameters are strongly identified. Intermediate values of $\hat d_n$ improve power compared to full projection while ensuring robustness if a subset of the nuisance parameters is weakly identified. A confidence set for $\theta_1$ collects all values of $\theta_1$ for which $\text{AR}_n(\theta_{1}) \leq \chi^2_{1-\alpha}(d_g - \hat d_n)$ using the same $\hat d_n$.  

The choice of $\hat d_n$ should be invariant to rescaling the sample moments $\bar{g}_n$ and/or the parameters $\theta$. To this end, the procedure relies on two normalization matrices: $\bar{V}_n = \int_\Theta \hat{V}_n(\theta) \hat \pi_n(\theta)d\theta$ and $\Sigma_n$, where $\hat \pi_n(\theta) = \hat{K}_n(\theta)/\int_{\Theta} \hat{K}_n(\theta) d\theta$. $\bar{V}_n$ an average of asymptotic variance estimators for $\lim_{n\to\infty} n \text{var}(\bar{g}_n(\theta))$. It is used to ensure the procedure is invariant to re-scaling and rotating the sample moments. $\Sigma_n$ is the $\ell_\infty$-covariance matrix minimizing $\sup_{\theta \in \Theta} \left( \log|\Sigma| + \|\theta-\mu\|_{\Sigma^{-1}}^2 \right) \hat{K}_n(\theta)$ over $\mu$ and $\Sigma$. These quantities are readily available from the steps required to compute $B_{n,\infty}$. It is important to use an estimate of the variance $\Sigma_n$ of $\theta$ on $\Theta_n$ rather than the variance of $B_{n,\infty}$ or of the sample Jacobian. When the model is set or weakly identified, the variance $\Sigma_n$ - which measures the size of the set $\Theta_n$ - does not go to zero in directions where identification fails.\footnote{Lemma \ref{lem:Sigma_weak} shows that $\Sigma_n^{-1/2}$ is bounded above under weak identification in directions where identification fails.} The variance of $B_{n,\infty}$ or the Jacobian could be arbitrarily small, however.\footnote{Take $g(z_i;\theta) = 0$, for all $\theta$, $z_i$ a.s.. The variances of both $B_{n,\infty}$ and the Jacobian are zero; yet, $\Sigma_n \neq 0$.} Hence, $B_{n,\infty} \Sigma_n^{-1/2}$ is vanishing in directions where identification fails and is invariant to rescaling the coefficients $\theta$.  

Let $P_{\theta_1}^\perp$ be the projection matrix on the orthogonal of the span of $\theta_1$, compute the singular values of the normalized $\bar{V}_n^{-1/2} B_{n,\infty} P_{\theta_1}^\perp \Sigma_n^{-1/2}$:
\[ \lambda_{jn} = \lambda_j( P_{\theta_1}^\perp \Sigma_n^{-1/2} P_{\theta_1}^\perp B_{n,\infty}^\prime \bar{V}_n^{-1} B_{n,\infty} P_{\theta_1}^\perp \Sigma_n^{-1/2} P_{\theta_1}^\perp )^{1/2}, \]
where $\lambda_{j}$ denotes the j-th eigenvalue in increasing order so that $0 \leq \lambda_{1n} \leq \dots \leq \lambda_{d_{\theta}n}$. By projection, the smallest $d_{\theta_1}$ singular values are equal to zero. Take $\underline{\lambda}_n \to 0$, a decreasing sequence such that $\kappa_n = o(\underline{\lambda}_n)$, and compute:
\[ \hat d_n = \#\{ j \in \{ d_{\theta_1}+1,\dots,d_\theta\}, \lambda_{jn} > \underline{\lambda}_n \}, \]
where $\#$ counts the number of singular values $\lambda_{jn}$ which are greater than the threshold $\underline{\lambda}_n$. 

\paragraph{Choice of Tuning Parameters:} A default choice is the uniform kernel $K(x) = \mathbbm{1}_{x \in [-1,1]}$. Then, the role of the pair $(\kappa_n, W_n)$ is to estimate the solution set of parameter(s) $\theta_0$ such that $g(\theta_0,\gamma_0)=0$. For this choice of kernel, if a law of the iterated logarithm applies, then, pointwise, $\text{liminf}_{n\to\infty} \hat{K}_n(\theta_0) =1$ almost surely using $W_n(\theta_0) = \text{var}[\sqrt{n}\overline{g}_n(\theta_0)]^{-1}$ and $\kappa_n = \sqrt{2 \log[\log(n)]/n}$.\footnote{A law of the iterated logarithm implies $\text{limsup}_{n\to\infty} \sqrt{n/(2\log[\log(n)])}\|\overline{g}_n(\theta_0)\|_{W_n} = 1$ almost surely, also $K(x) = 1$ for all $x \in [-1,1]$, see e.g. \citet[Ch7]{petrov1995}; and \citet[p31]{kosorok2008}, \citet[p379, footnote b]{VanderVaart1996} for references applying to empirical processes, which are not pointwise.} In that sense, efficient weighting and $\kappa_n = \sqrt{2\log\log(n)/n}$ are asymptotically optimal and makes $\hat{K}_n(\theta)$ invariant to linear transformations of the moments. 

The role of the normalized $B_{n,\infty}$ and the threshold $\underline{\lambda}_n$ is analogous to the ICS procedure in \citet[Section 5.2]{Andrews2012}, and the subsequent literature. Here, it is shown that if $d$ nuisance parameters are weakly identified then at least $d$ singular values are $O_p(\kappa_n)$. Hence, wpa 1, they are smaller than $\underline{\lambda}_n$, if $\underline{\lambda}_n = o(\kappa_n)$. As a result, $\hat d_n$ is no greater than the number (semi)-strongly identified nuisance parameters wpa 1, which leads to valid inferences under weak identification. Typically, using larger values of $\underline{\lambda}_n$ in an ICS procedure is desirable for robust inference since it correctly detects identification failures with greater probability in finite samples. However, it also makes the test more conservative under semi-strong identification since it incorrectly detects identification failure with greater probability. This implies a trade-off between power for semi-strongly identified models with robustness for weakly identified models. The normalization $\Sigma_n^{-1/2}$ in the procedure improves on this by making the behaviour of the ICS statistic more distinct between these two regimes. The normalization $B_{n,\infty} P_{\theta_1}^\perp\Sigma_n^{-1/2} P_{\theta_1}^\perp$ preserves the asymptotic singularity under weak identification but the normalized matrix diverges at a $\kappa_n^{-1}-$rate when identification is strong. 

\subsection{The quasi-Jacobian}
The main component of the procedure is the quasi-Jacobian. To better understand the main differences with the Jacobian, the following derives its properties for $n = \infty$, using a positive definite $W(\theta)$ and the uniform kernel $K(x) = \mathbbm{1}_{x \in [-1,1]}$. Take \[ (A_\infty,B_\infty) = \lim_{\kappa \to 0} \left( \text{argmin}_{A,B} (\sup_{\|g(\theta,\gamma_0)\|_W \leq \kappa} \|g(\theta,\gamma_0)-A-B \theta\|) \right),\] where the $\sup$ is taken over $\theta \in \Theta$ with $\kappa>0$. To compute the Jacobian, $\partial_\theta g(\theta_0,\gamma_0)$, one would use the set $\|\theta-\theta_0\| \leq \kappa$; the main difference is the choice of neighborhood.

This difference suggests that, unlike the Jacobian, the properties of the quasi-Jacobian depend on the set $\Theta_0 = \{ \theta \in \Theta, g(\theta,\gamma_0)=0  \}$, which collects all solutions to the moment condition. For any given value $\gamma_0 \in \Gamma$, there are three possibilities, either: i. $\Theta_0$ is non-singleton, ii. $\Theta_0 = \{\theta_0\}$ is singleton and $\partial_\theta g(\theta_0,\gamma_0)$ is singular, or iii. $\Theta_0 = \{\theta_0\}$ is singleton and $\partial_\theta g(\theta_0,\gamma_0)$ has full rank. Under i., $\theta_0$ is not globally identified. Under ii. and iii. $\theta_0$ is globally identified but only locally identified under iii. Consistency and asymptotic normality require iii., i.e. strong identification, and standard inference need not be asymptotically valid under i. or ii. The following Theorem relates the rank of $B_\infty$ to identifications i., ii., and iii.

\begin{theorem}[quasi-Jacobian, $n = \infty$] \label{th:qJ} Take $\gamma_0 \in \Gamma$. Suppose $0 < \underline{\lambda}_W \leq \lambda_{\min}(W(\theta)) \leq \lambda_{\max}(W(\theta)) \leq \overline{\lambda} < \infty$ and $g(\cdot,\gamma_0)$ is continuously differentiable for all $\theta \in \Theta$.  Suppose there are $\overline{\varepsilon},\overline{C} > 0$ and $\alpha >1$ such that when $\Theta_0 = \{\theta_0\}$ is singleton: $\|g(\theta)-g(\theta_0) - \partial_\theta g(\theta_0)(\theta-\theta_0)\| \leq \overline{C} \|\theta-\theta_0\|^{\alpha}$ for all $\|\theta-\theta_0\| \leq \overline{\varepsilon}$. Then the quasi-Jacobian $B_\infty$ is such that: 
\begin{itemize}  \itemsep0em 
  \item[(1)] $B_\infty$ singular if, and only if: $\Theta_0$ non-singleton or, $\Theta_0$ singleton and $\partial_\theta g(\theta_0,\gamma_0)$ singular, 
  \item[(2)] For $\Theta_0$ singleton and $\partial_\theta g(\theta_0,\gamma_0)$ full rank: $B_{\infty} = \partial_\theta g(\theta_0,\gamma_0)$,
  \item[(3)] For $\Theta_0$ non-singleton: $B_\infty (\theta^1_0 - \theta_0^2) = 0$ for all $\{\theta_0^1,\theta_0^2\}\subseteq \Theta_0$,
  \item[(4)] For $\Theta_0$ singleton and $\partial_\theta g(\theta_0,\gamma_0)$ singular: $B_\infty v = 0$ whenever $\partial_\theta g(\theta_0,\gamma_0) v = 0$.
\end{itemize}
\end{theorem}
The dependence of $\Theta_0$, $B_\infty$ on $\gamma_0$ is omitted to simplify notation. The condition for globally identified models holds with $\alpha=2$ if $g(\cdot,\gamma_0)$ is twice continuously differentiable with bounded second derivative around $\theta_0$. Theorem \ref{th:qJ} shows that $B_\infty$ is singular as soon as $\gamma_0$ is such that global or local identification fails (1). An immediate implication of Theorem \ref{th:qJ} is that for all $v \neq 0$ such that $B_\infty v \neq 0$, $P_v \Theta_0 = P_v \{\theta_0\}$; i.e. the parameter is point identified in direction $v$. This contrasts with the Jacobian which can have full rank without global identification. $B_\infty$ is singular in all directions in which global identification fails (3), or local identification fails (4); these directions may vary depending on $\gamma_0$. $B_\infty$ has full rank only if $\gamma_0$ is such that both global and local identification hold (2). Theorem \ref{th:qJ} holds for both just and over-identified models. The identified set $\Theta_0$ can be arbitrary, e.g. discrete. The results do require correct specification, $\Theta_0$ non-empty, for the quasi-Jacobian $B_\infty$ to be well defined. Even though Theorem \ref{th:qJ} is fairly general, the main results will be restricted to settings where either i. $\Theta_0$ is non-singleton, or iii. $\Theta_0$ is singleton and $\partial_\theta g(\theta_0,\gamma_0)$ has full rank. Additional results for ii. are given in Appendix \ref{apx:ho}.

The Jacobian generally does not have property (1) or (3). When using projection methods for subvector inference, one can concentrate out nuisance parameters that are both globally and locally identified. The Jacobian can only determine the latter which is not sufficient for consistency. The following illustrates (3) and gives a sketch of the proof using a simple non-linear model where $\Theta_0$ is non-singleton but the Jacobian has full rank for all $\theta \in \Theta_0$.

\paragraph{Intuition for linear models.} For linear models, the sup-norm approximation is exact with $B_{\infty} = \mathbb{E}(x_i x_i^\prime)$ and $\mathbb{E}(z_i x_i^\prime)$ for OLS and IV, respectively. The quasi-Jacobian coincides with the Jacobian and it is singular when the regressors are multicollinear or the instruments are not relevant. Both are singular in directions where the rank condition fails. 


\paragraph{Non-linear models: a pen and pencil example.} Consider a simple MA(1) process:
\[ y_t = \sigma( e_t + \vartheta e_{t-1}), \quad e_t \overset{iid}{\sim} (0,1), \]
where $\theta = (\vartheta,\sigma^2) \in \mathbb{R} \times \mathbb{R}_{+}$ are the parameters of interest. The model is estimated using the following set of moment conditions (the dependence on $\gamma$ is omitted in this example):
\[ g(\theta) := \mathbb{E}\left( \begin{array}{cc} y_t^2 - \sigma^2(1+\vartheta^2), \quad y_t y_{t-1} - \vartheta \sigma^2 \end{array} \right)^\prime = 0. \]
Whenever $\vartheta_0 \not\in \{-1,0,1\}$ and $\sigma_0^2 >0$, this system of equations has two distinct solutions: $\theta_0^1 = (\vartheta_0,\sigma_0^2)$ and $\theta_0^2 = (1/\vartheta_0,\vartheta_0^2\sigma_0^2)$. Imposing invertibility (i.e. $|\vartheta_0|<1$), or non-invertibility (i.e. $|\vartheta_0|>1$) restores identification so that, intuitively, only one dimension is unidentified. Both solutions are locally identified: the Jacobian $\partial_\theta g (\theta)$ has full rank at both values; it is uninformative about the global identification failure in this example. The goal of this example is to show that $B_{\infty}$ is informative about the lack of global identification and the direction in which identification fails. Without the quasi-Jacobian, one would need to check with pen and pencil whether $g(\theta)=0$ has multiple solutions, or not. 

The first step is to find a one-to-one linear reparameterization $\beta = (\beta_1^\prime,\beta_2^\prime)^\prime$ such that $\beta_1$ is uniquely identified but $\beta_2$ is not. Let $v_2 = (\theta_0^1-\theta_0^2)/\|\theta_0^1-\theta_0^2\|$ and pick any orthogonal $v_1 \perp v_2$ such that $\|v_1\|=1$. By construction: $v_1^\prime (\theta_0^1-\theta_0^2) = 0$ and $v_2^\prime (\theta_0^1-\theta_0^2) = \|\theta_0^1-\theta_0^2\|^2 >0$. This implies that $\theta_0^1$ and $\theta_0^2$ are equal in direction $v_1$ but distinct in direction $v_2$. Pick $\beta_1 = v_1^\prime \theta$, $\beta_2 = v_2^\prime \theta$. As desired: the mapping is one-to-one, with $\beta_1$ uniquely and $\beta_2$ set identified. Property (3) in Theorem \ref{th:qJ} implies that directions in which $B_\infty$ is non-singular must be associated with a unique value for $\theta_0$. This first step illustrates how these directions can be constructed from the set $\Theta_0$. Importantly, the linear reparametrization need not be computed explicitly in practice, as explained below.

The second step is to show that $B_\infty$ is informative about the identification failure and contains information about the reparametrization above.
In the MA(1) model, the set $\Theta_0 = \{\theta_0^1,\theta_0^2\}$ has two points. Take $\kappa >0$ and compute the intercept and slope $A_{\kappa,\infty},B_{\kappa,\infty}$:
\[ (A_{\kappa,\infty},B_{\kappa,\infty}) =  \argmin_{A,B} \left( \sup_{ \theta \in \Theta, \|g(\theta)\| \leq \kappa} \|g(\theta) - A - B\theta\| \right), \]
here using the uniform kernel $K(x) = \mathbbm{1}_{x \in [-1,1]}$, and $W = I$ for analytical simplicity. Notice that for $(A,B) = 0$, $\sup_{ \theta \in \Theta, \|g(\theta)\| \leq \kappa} \|g(\theta)\| \leq \kappa$. Also, because $\|g(\theta_0^1)\| = \|g(\theta_0^2)\| = 0 \leq \kappa$, the solution $A_{\kappa,\infty},B_{\kappa,\infty}$ is such that $\|g(\theta) - A_{\kappa,\infty} - B_{\kappa,\infty}\theta\| \leq \kappa$ for $\theta \in \{\theta_0^1,\theta_0^2\}$. Using the triangular inequality and its reverse, this implies $\|B_{\kappa,\infty}(\theta_0^1-\theta_0^2)\| \leq 2\kappa + \|g(\theta_0^1)\| + \|g(\theta_0^2)\| = 2\kappa$. Now, express this in terms of the direction vector $v_2$ constructed above:
\[ \|B_{\kappa,\infty} v_2\| \leq \frac{2\kappa}{\|\theta_0^1-\theta_0^2\|} \to 0, \text{ as } \kappa \searrow 0. \]

In the limit, the quasi-Jacobian $B_{\infty}$ is singular in the direction $v_2$ where identification fails. This implies that $v_2$ is a right-singular vector associated with the singular value $0$. The singular value decomposition of $B_{\infty}$ is informative about the directions of identification failure and the linear reparametrization from the first step. While the linear reparamerization requires knowledge of $\Theta_0$ and computing all possible $\theta_0^1 - \theta_0^2$ with $\{\theta_0^1,\theta_0^2\} \subseteq \Theta_0$, Theorem \ref{th:qJ} implies that the right-singular vectors of $B_\infty$ associated with the singular value $0$ span all directions of identification failure $\theta_0^1 - \theta_0^2$.

\begin{figure}[ht] \caption{MA(1): singular values of the Jacobian, unscaled and scaled quasi-Jacobian} \label{fig:MA1}
  \includegraphics[scale=0.55]{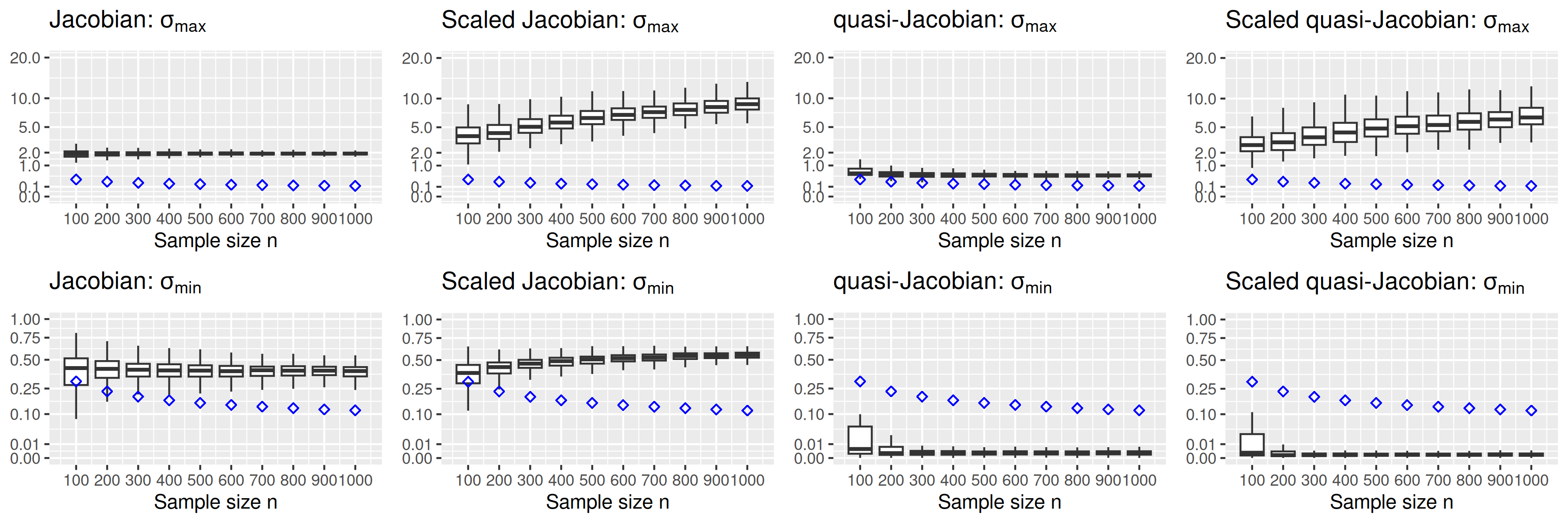}
  \notes{\textbf{Note:} bandwidth $\kappa_n = \sqrt{2\log(\log[n])/n}$, cutoff $\underline{\lambda}_n = \sqrt{2\log[n]/n}$ (blue diamonds). Model: MA(1) with true value $\theta_0^1 = (\vartheta_0,\sigma_0^2) =  (0.5,1)$. $\sigma_{\max},\sigma_{\min}$: largest and smallest singular values. $W_n = \hat{V}_n^{-1}$ where $\hat{V}_n = $ HAC estimate of $\text{var}[\sqrt{n}\overline{g}_n(\theta)]$.  }
\end{figure}

In large samples and under Assumptions \ref{ass:moments}-\ref{ass:kernel_bandwidth}, $\|B_{n,\infty} v_2\| \leq 2 \kappa_n / \|\theta_0^1-\theta_0^2\|$ wpa $1$ (using the same $K$, $W$). As a result, for any sequence $\underline{\lambda}_n$ such that $\kappa_n = o(\underline{\lambda}_n)$, $\sigma_{\min}(B_{n,\infty}) \leq \underline{\lambda}_n$ wpa $1$ which signals the identification failure, as desired. 
To illustrate, Figure \ref{fig:MA1} compares the distribution of the largest and smallest singular values of the Jacobian $\partial_\theta \overline{g}_n(\hat\theta_n)$, quasi-Jacobian $B_{n,\infty}$, and scaled quasi-Jacobian $\overline{V}_n^{-1/2} B_{n,\infty} \Sigma_n^{-1/2}$ with the same cutoff $\underline{\lambda}_n$. The scaling makes the singular values scale invariant. The Jacobian fails to detect the lack of identification, even for large $n$ (left panel) and also with the scaling $\overline{V}_n^{-1/2} \partial_\theta \overline{g}_n(\hat\theta_n) \Sigma_n^{-1/2}$. The quasi-Jacobian detects the identification failure since the smallest singular value is below the cutoff. However, the largest singular value is also close to the cutoff. With the scaling, the largest singular value diverges while the smallest one shrinks to zero (right panel). 

\subsection{Drifting Sequences of Parameters, Identification Regimes}

The test procedure described above is said to be robust to identification failure if it has asymptotic null rejection probability bounded above by the nominal size, i.e.:
\[ \limsup_{n\to\infty} \sup_{\gamma \in \Gamma, \theta = (\theta_{10}^\prime,\theta_2^\prime)^\prime \in\overline{\Theta}} \mathbb{P}_{\gamma}\left(\text{AR}_n(\theta_{10}) > \chi_{1-\alpha}^2(d_g - \hat d_n) \right) \leq \alpha. \]
In the limit, the worst-case rejection rate should be no greater than the nominal size $\alpha$. Following \citet{Andrews2012}, this can be determined from the asymptotic properties of the test for specific sequences of parameters $(\theta_n,\gamma_n) \in \overline{\Theta} \times \Gamma$. 
\begin{assumption}[Identification] \label{ass:identification} There exists a continuous function $\delta(\cdot) \geq 0$ and a strictly positive function $h(\cdot) >0$ such that for any $(\theta_n,\gamma_n)\in \overline{\Theta} \times \Gamma$ where $g(\theta_n,\gamma_n)=0$ and $\varepsilon >0$:
  \[ \inf_{ \theta \in \Theta, \|\theta-\theta_n\| \geq \varepsilon } \|g(\theta,\gamma_n)\| \geq \delta(\gamma_n)h(\varepsilon). \]
There exists a $\overline{\varepsilon} >0$ and a constant $C>0$ such that for $0<\varepsilon \leq \overline{\varepsilon}$:
\[ \inf_{ \theta \in \Theta, \|\theta-\theta_n\| \geq \varepsilon } \|g(\theta,\gamma_n)\| \leq C \delta(\gamma_n)h(\varepsilon). \]
\end{assumption}
The function $\delta$ indicates whether the solution $\theta_0$ to the moment condition $g(\theta,\gamma_0)=0$ is unique for a given $\gamma = \gamma_0$. The second part of the assumption implies that when $\delta(\gamma_0)=0$, there is at least one $\theta \neq \theta_0$ such that $g(\theta,\gamma_0)=0$. Sequences such that $\gamma_n \to \gamma_0$, $\delta(\gamma_0)=0$, satisfy $\delta(\gamma_n)\to 0$ since $\delta$ is continuous. The properties of $\hat\theta_n$ depend on the rate at which $\delta(\gamma_n)$ converges to zero. Under Assumptions \ref{ass:moments} and \ref{ass:identification}, Lemma \ref{lem:consistency} shows that $\|\hat\theta_n-\theta_n\|=o_p(1)$, the estimator is consistent, if $\sqrt{n}\delta(\gamma_n)\to \infty$. When $\sqrt{n}\delta(\gamma_n) = O(1)$, the estimator is generally not consistent, see e.g. \citet{Stock2000}. 

In the MA(1) example, pick $\overline{\Theta} = \{\vartheta_0,\sigma_0^2\} \in (\mathbb{R}/\{-1,0,1\}) \times (\mathbb{R}_+/\{0\})$, then $\delta(\gamma)=0$ regardless of $\gamma$ as long as the two distinct solutions $\theta_0^1,\theta_0^2 \in \Theta$. The second inequality only holds for $\varepsilon < \|\theta_0^1-\theta_0^2\|$, with $C=1$, which implies $\overline{\varepsilon} \in (0,\|\theta_0^1-\theta_0^2\|)$.

To give another example, consider a linear IV regression: $g(\theta,\gamma) = \mathbb{E}_{\gamma}[z_i(y_i - x_i^\prime\theta)] = \mathbb{E}_{\gamma}[z_ix_i^\prime](\theta_n - \theta)$ using $y_i = x_i^\prime \theta_n + u_i$ and $\mathbb{E}_{\gamma}(u_i z_i)=0$. Here $\|g(\theta,\gamma)\| \geq \sigma_{\min}(\mathbb{E}_{\gamma}[z_ix_i^\prime])\|\theta_n-\theta\|$ so that $\delta(\gamma) = \sigma_{\min}(\mathbb{E}_{\gamma}[z_ix_i^\prime])$ and $h(\varepsilon) = \varepsilon$. The inequality holds with equality, i.e. $C=1$, when $\theta_n-\theta$ is the right singular vector of $\mathbb{E}_{\gamma}[z_ix_i^\prime]$ associated with the smallest singular value. Here $\delta(\gamma_0)=0$ implies $\mathbb{E}_{\gamma_0}[z_ix_i^\prime]$ singular, and the model is underidentified.\footnote{Additional derivations for a non-linear regression model are given in Appendix \ref{apx:conds_NLS}.}

The dichotomy between $\delta$ and $h$ in Assumption \ref{ass:identification} allows to construct a measure of global identification strength used to categorize the sequences $\gamma_n$.\footnote{A similar decomposition can be found in \citet[p5589]{Chen2007} to isolate the effect of the sieve dimension $k$ on the shape of the objective in nonparametric estimation.}  Let $\Gamma_0 = \{ \gamma \in \Gamma, \delta(\gamma)=0\}$ and $\Gamma_1 = \Gamma / \Gamma_0$. $\Gamma_0$ collects all DGPs such that $\theta_0$ is not uniquely identified, and in $\Gamma_1$ those that are point identified. Let $\Gamma_0(\infty) = \{ \gamma_n \in \Gamma, \gamma_n \to \gamma_0 \in \Gamma_0, \sqrt{n}\delta(\gamma_n)\to\infty\}$, $\Gamma_0(b) = \{ \gamma_n \in \Gamma, \gamma_n \to \gamma_0 \in \Gamma_0, \lim_{n\to\infty}\sqrt{n}\delta(\gamma_n) = b <\infty\}$. In the following, any converging sequence $\gamma_n$ will be assumed to belong to one of $\Gamma_0(b)$ for some $b \geq 0$, $\Gamma_0(\infty)$, or converges in $\Gamma_1$. 
These will be referred to as weak, semi-strong, and strong sequences.

\begin{assumption}[Strong and Semi-Strong Sequences] \label{ass:ss} Let $(\theta_n,\gamma_n) \to (\theta_0,\gamma_0)$ where $\gamma_0 \in \Gamma_1$, or $\gamma_n \in \Gamma_0(\infty)$. Let $H_n = \left( \partial_\theta g(\theta_n,\gamma_n)^\prime \partial_\theta g(\theta_n,\gamma_n) \right)^{-1/2}$. For any $r_n = o(1)$, suppose the following holds: i. $\partial_\theta g(\theta,\gamma)$ is continuous in $\theta$ and $\gamma$; $\partial_\theta g(\theta_n,\gamma_n)$ has full rank for all $n\geq 1$,
    ii. $n \times \lambda_{\min}\left( \partial_\theta g(\theta_n,\gamma_n)^\prime \partial_\theta g(\theta_n,\gamma_n) \right) \to \infty$, $\lambda_{\max}\left( \partial_\theta g(\theta_n,\gamma_n)^\prime \partial_\theta g(\theta_n,\gamma_n) \right) \leq \overline{\lambda} < \infty$,
    iii. $\sup_{\|\partial_\theta g(\theta_n,\gamma_n)(\theta-\theta_n)\| \leq r_n} \sqrt{n} \| [\bar{g}_n(\theta)-\bar{g}_n(\theta_n)] - [g(\theta,\gamma_n) - g(\theta_n,\gamma_n)] \| \overset{p}{\to} 0$,
    iv. there exists $\varepsilon >0$, $\underline{C}>0$ such that for $\|\theta-\theta_n\| \leq \varepsilon$, $\|g(\theta,\gamma_n)\| \geq \underline{C}\|\partial_\theta g(\theta_n,\gamma_n)(\theta-\theta_n)\|$, and  $\sup_{\|\partial_\theta g(\theta_n,\gamma_n)(\theta-\theta_n)\| \leq r_n} \| g(\theta,\gamma_n) - g(\theta_n,\gamma_n) - \partial_\theta g(\theta_n,\gamma_n)(\theta-\theta_n) \|  = O(r_n^2)$,
    v,  $\partial_\theta g(\theta_n,\gamma_n)H_n \to R_0$, where $R_0$ is a full rank matrix.
\end{assumption}

Assumption \ref{ass:ss} provides sufficient conditions to establish asymptotic normality of $\hat\theta_n-\theta_n$ at a potentially slower than $\sqrt{n}$-rate. Condition i. is standard and ensures the model is locally identified. Condition ii. allows the Jacobian to be vanishing at a slower than $\sqrt{n}$-rate in some directions. Conditions iii. is a stochastic equicontinuity condition. Condition iv. implies that the Taylor remainder is quadratic under the weaker norm $\|\partial_\theta g(\theta_n,\gamma_n)(\cdot)\|$, which is the relevant norm for convergence when $\gamma_n\in \Gamma_0(\infty)$. Indeed, Lemma \ref{lem:asym_normal} establishes that $\sqrt{n}\|\partial_\theta g(\theta_n,\gamma_n)(\hat\theta_n-\theta_n)\|=O_p(1)$. Condition iv. excludes settings where the non-linear remainder dominates the first-order term.\footnote{These second or higher-order identification issues are not considered in the main text, additional results for the quasi-Jacobian under higher-order identification are given in the Supplement.} Condition v. is analogous to Assumption 3iv in \citet{Antoine2012}. It requires a rescaling for which the Jacobian is non-singular in the limit. For instance, under a singular value decomposition of the form $\partial_\theta g(\theta_n,\gamma_n) = U D_n V^\prime$, we have $\partial_\theta g(\theta_n,\gamma_n) H_n = UV^\prime = R_0$. The rescaling corrects for the possibly vanishing, but non-zero, terms in the diagonal $D_n$. \citet[Sec2.2]{antoine2021} discuss conditions relating to Assumption \ref{ass:ss} in more detail.

\begin{proposition}[Asymptotic Distribution for (Semi)-Strong Sequences] \label{prop:test_ss} Let $(\theta_n,\gamma_n) \to (\theta_0,\gamma_0)$. Let $\text{AR}_n(\theta_{1n}) = \inf_{\theta_2 \in \Theta_2} \|\bar{g}_n(\theta_{1n},\theta_2)\|_{V_n}^2$, if Assumptions \ref{ass:moments}, \ref{ass:identification} and \ref{ass:ss} hold then:
  \[ \text{AR}_n(\theta_{1n}) \overset{d}{\to} \chi^2_{d_g - d_{\theta_2}}.\]
\end{proposition}

Proposition \ref{prop:test_ss} implies that the test is asymptotically valid for any choice of $\hat{d}_n \in \{0,\dots,d_{\theta_2}\}$ and asymptotically non-conservative if $\hat{d}_{n}=d_{\theta_2}$ wpa 1. Furthermore, for just-identified models $\text{QLR}_n(\theta_1) = \text{AR}_n(\theta_1)$, and the test is asymptotically efficient if $\hat{d}_{n}=d_{\theta_2}$ wpa 1.

\paragraph{Linear reparameterization.} As in the MA(1) example, the derivations rely on a one-to-one linear reparameterization $\beta = M\theta = (\beta_1^\prime,\beta_2^\prime)^\prime$ with $\beta_1$ uniquely and $\beta_2$ set identified. The following steps construct the reparameterization, which is not implemented in practice: the span of right-singular vectors associated with singular values below $\underline{\lambda}_n$ consistently estimates the span of identification failure. The following applies to just and over-identified models.

First, take $\gamma_0 \in \Gamma$, collect all solutions to the moment conditions $\Theta_0 = \{ \theta \in \Theta, g(\theta,\gamma_0) = 0 \}$. Let $V_2 = \text{span}( \{ v_2 = \theta_0^1 - \theta_0^2, (\theta_0^1,\theta_0^2) \in \Theta_0 \times \Theta_0 \})$ and $V_1 = V_2^\perp$. If $V_2 = \{0\}$, then $V_1 = \mathbb{R}^{d_\theta}$ which implies that $\Theta_0$ is a singleton; i.e. the parameters are uniquely identified. This is the case when $\gamma_0 \in \Gamma_1$. If $\{0\} \subset V_2$ strictly, then $V_1 \subset \mathbb{R}^{d_\theta}$ strictly; i.e. the parameters are set identified. This is the case when $\gamma_0 \in \Gamma_0$. As in the MA(1) example, by projection $P_{V_1}(\theta_0^1 - \theta_0^2) = 0$ for any two $\theta_0^1,\theta_0^2 \in \Theta_0$; i.e. the solution is unique on $V_1$. In contrast, for any non-zero $v_2 \in V_2$, there exists two distinct $\theta_0^1,\theta_0^2 \in \Theta_0 \times \Theta_0$ s.t. $v_2^\prime(\theta_0^1-\theta_0^2) \neq 0$, by construction.  Define $\beta_1$ as the projection of $\theta$ on $V_1$ and $\beta_2$ the projection on $V_2$. The matrix $M$ combines the bases of $V_1$ and $V_2$. As illustrated by the MA(1) example, it may not be possible to improve on this linear reparameterization with a non-linear one without some further structure on the moments or the model. The reparameterization is defined up to a rotation on $V_1$ and $V_2$, respectively. 


For testing $H_0: \theta_1 = \theta_{10}$, the identification status of the nuisance parameters $\theta_2$ matters. Consider a further sub-decomposition $(\beta_1,\beta_{21},\beta_{22})$ where only $\beta_{22}$ is unidentified under the restriction $\theta_1 = \theta_{10}$. To find it, take $V_{22} = \text{span}( \{ \theta_0^1-\theta_0^2, (\theta_0^1,\theta_0^2) \in \Theta_0 \times \Theta_0, P_{\theta_1}\theta_0^1=P_{\theta_1}\theta_0^2 = \theta_{10} \} )$ and follow the same steps as above. By construction, $V_{22}$ is a subset of $V_2$, also $\theta_1$ is in $V_{22}^\perp$ and $\beta_{22}$ is the subset of $\theta_2$ which is unidentified under $H_0$.\footnote{Note that, by linearity and by construction, $\text{span}(P_{V_{22}}) = \text{span}(P_{V_{22}} P_{\theta_1}^\perp) \subseteq \text{span}(P_{V_{2}} P_{\theta_1}^\perp)$.}

Now consider sequences $(\theta_n,\gamma_n) \to (\theta_0,\gamma_0)$ with $\gamma \in \Gamma_0$. Combine the linear reparameterization with the continuity of $g$ with respect to $\theta$ and $\gamma$ to find, using the Maximum Theorem, that for all $(\theta_n,\gamma_n) \to (\theta_0,\gamma_0)$, any $\varepsilon >0$, and letting $\beta_n = M\theta_n$:\footnote{To apply the Maximum Theorem, note that by continuity of $g(\cdot,\gamma_0)$ and compactness of $\Theta$, both $\Theta_0$ and $\mathcal{B}_2^0$ are compact subsets of $\mathbb{R}^{d_\theta}$ and $\mathbb{R}^{d_{\beta_2}}$, respectively. Similar equations can be derived for $(\beta_1,\beta_{21},\beta_{22})$ with the added constraint $\theta_1 = \theta_{1n}$.}
\begin{align}
    \inf_{\|\beta_1-\beta_{1n}\| \geq \varepsilon, \beta_2} \|g(\beta_1,\beta_2,\gamma_n)\| &\overset{n\to\infty}{\longrightarrow} \inf_{\|\beta_1-\beta_{10}\| \geq \varepsilon, \beta_2} \|g(\beta_1,\beta_2,\gamma_0)\| >0, \label{eq:point_idt}\\
    \inf_{d(\beta_2,\mathcal{B}_2^0) \geq \varepsilon, \beta_1} \|g(\beta_1,\beta_2,\gamma_n)\| &\overset{n\to\infty}{\longrightarrow} \inf_{d(\beta_2,\mathcal{B}_2^0) \geq \varepsilon, \beta_1} \|g(\beta_1,\beta_2,\gamma_0)\| >0, \label{eq:idt_set}\\
    \sup_{\beta_2 \in \mathcal{B}_2^0} \|g(\beta_{1n},\beta_2,\gamma_n)\| &\overset{n\to\infty}{\longrightarrow} \sup_{\beta_2 \in \mathcal{B}_2^0} \|g(\beta_{10},\beta_2,\gamma_0)\| =0, \label{eq:flat}
\end{align}
where $\mathcal{B}_2^0 = P_{V_2}\Theta_0$ is the identified set for $\beta_2$ when $(\theta,\gamma) = (\theta_0,\gamma_0)$. The first limit implies $\beta_1$ is consistently estimable, while the second and third imply that the population objective function becomes flat (only) on $\mathcal{B}_2^0$. The decomposition so far separates $\beta_1$ point identified from $\beta_2$ set unidentified when $\gamma=\gamma_0$.\footnote{Note that for the class of models considered in \citet{Andrews2012}, their parameter $\beta$ which is point identified and determines identification strength is included in the vector $\beta_1$ constructed here.}

If there is a single source of identification failure, then $\sup_{\beta_2 \in \mathcal{B}_2^0} \sqrt{n}\|g(\beta_{1n},\beta_2,\gamma_n)\|$ is determined by a scalar subset of $\gamma_n$, and is bounded above for weak sequences. To illustrate, consider the linear IV example again with a single endogenous regressor $x_i$ and one instrument $z_i$. In this case $\sqrt{n}\|g(\beta_{1n},\beta_2,\gamma_n)\| = \sqrt{n}|\text{cov}(x_i,z_i)| \times |\beta_2-\beta_{2n}|$ depends on the scalar $\sqrt{n}|\text{cov}(x_i,z_i)|$ being bounded which characterizes weak sequences. In the case with multiple sources of identification failure, there may be mixed identification strength, and some components of $\beta_2$ may be (semi)-strongly identified, so the reparameterization needs to be further refined. This is deferred to Appendix \ref{apx:linpar}.

\begin{assumption}[Weak Sequences] \label{ass:weak} Let $(\theta_n,\gamma_n) \to (\theta_0,\gamma_0)$ with $\gamma_n \in \Gamma_0(b)$. Let $\mathcal{B}_n = \{ \beta \in \mathcal{B}, M^{-1} \beta = (\theta_{1n}^\prime, \theta_2^\prime)^\prime, \theta_2 \in \Theta_2 \}$ be the null-constrained space for $\beta$. There exists $\tilde \delta(\cdot) \geq 0$ continuous satisfying $\sqrt{n}\tilde \delta(\gamma_n) \to \infty$ and $\tilde h(\cdot)$ strictly positive, and two non-empty and non-singleton sets $\mathcal{B}_{2}^0 \subset \mathbb{R}^{d_{\beta_{2}}}$ and $\mathcal{B}_{22}^0 \subset \mathbb{R}^{d_{\beta_{22}}}$ such that for any $\varepsilon>0$:
  \begin{itemize}
    \item[i.] $\inf_{\beta_2, \|\beta_1 - \beta_{1n}\| \geq \varepsilon} \|g(\beta_1,\beta_2,\gamma_n)\| \geq \tilde \delta(\gamma_n)\tilde h(\varepsilon),$ $\limsup_{n\to\infty} \sup_{\beta_2 \in \mathcal{B}_{2}^0} \sqrt{n}\|g(\beta_{1n},\beta_2,\gamma_n)\| < \infty$ and $\inf_{\beta_1, d(\beta_2,\mathcal{B}_{2}^0) \geq \varepsilon} \|g(\beta_1,\beta_2,\gamma_n)\| \geq \tilde \delta(\gamma_n) \tilde h(\varepsilon)$,
    \item[ii.] $\inf_{\beta_{22}, \|\beta_1 - \beta_{1n}\| + \|\beta_{21} - \beta_{21n}\| \geq \varepsilon} \|g(\beta_1,\beta_{21},\beta_{22},\gamma_n)\| \geq \tilde \delta(\gamma_n)\tilde h(\varepsilon)$,\\ $\limsup_{n\to\infty} \sup_{\beta_{22} \in \mathcal{B}_{22}^0} \sqrt{n}\|g(\beta_{1n},\beta_{21n},\beta_{22})\| < \infty$, and $\inf_{\beta_{1},\beta_{21}, d(\beta_{22},\mathcal{B}_{22}^0) \geq \varepsilon} \|g(\beta_1,\beta_2,\gamma_n)\| \geq \tilde \delta(\gamma_n) \tilde h(\varepsilon)$, where the infs are taken over the constrained space $(\beta_1^\prime,\beta_{21}^\prime,\beta_{22}^\prime)^\prime \in \mathcal{B}_n$.
  \end{itemize}   
\end{assumption}

Assumption \ref{ass:weak} adds this additional structure to (\ref{eq:point_idt})-(\ref{eq:flat}), where $\beta_1$ are assumed semi-strongly and $\beta_2$ weakly identified.
The first part Assumption \ref{ass:weak}i. implies $\beta_1$ is consistently estimable, allowing for some components to be semi-strongly identified. The second and third part imply the objective function is flat with respect to $\beta_2$ but only on the identified set $\mathcal{B}_2^0$. For the quasi-Jacobian, the $\limsup_{n\to\infty} \sup_{\beta_2 \in \mathcal{B}_{2}^0} \sqrt{n}\|g(\beta_{1n},\beta_2)\| < \infty$ implies that $\|\overline{g}_n(\beta_{1n},\beta_2)\|_{W_n} \leq \kappa_n$ uniformly in $\beta_2 \in \mathcal{B}_2^0$ with increasing probability so that Step 2.i of the procedure consistently estimates the identified set and all directions of identification failure. Similarly, condition ii. repeats the conditions under the restriction that $\theta_1 = \theta_{1n}.$ The parameters $(\beta_1,\beta_{21})$ correspond to the directions that are consistently estimable. To simplify notation, the Proposition below denotes as $\phi$ these $d_\theta - d_{\theta_1} - d_{\beta_{22}} = d_\phi$ coefficients that are consistently estimable and semi-strongly identified under $H_0: \theta_1=\theta_{1n}$.

\begin{proposition}[Asymptotic Distribution for Weak Sequences] \label{prop:test_weak} Let $(\theta_n,\gamma_n) \to (\theta_0,\gamma_0)$. Suppose there is a linear reparameterization $M_\phi$ invertible, $M_\phi \theta = (\theta_1^\prime,\phi^\prime,\beta_{22}^\prime)^\prime$, such that the moment function $\phi \to \bar{g}_n(\theta_{1n},\phi,\beta_{22n})$ satisfies Assumptions \ref{ass:moments}, \ref{ass:identification} and \ref{ass:ss}, then:
  \[ \text{AR}_n(\theta_{1n}) \leq \inf_{\phi} \|\bar{g}_n(\theta_{1n},\phi,\beta_{22n})\|^2_{V_n^{-1}} \overset{d}{\to} \chi^2_{d_g - d_\phi}.\]
\end{proposition}

Proposition \ref{prop:test_weak} implies that the test procedure has limiting null rejection probability bounded by the nominal size for weak sequences as long as $\hat d_n \leq d_{\beta_1}+d_{\beta_{21}} - d_{\theta_1} = d_\phi$ wpa 1, since  $\mathbb{P}_{\gamma_n}\left(\text{AR}_n(\theta_{1n}) \geq \chi^2_{1-\alpha}(d_g-\hat d_n)\right) \leq \mathbb{P}_{\gamma_n}\left( \inf_{\phi \in \Phi}\|\bar{g}_n(\theta_{1n},\phi,\beta_{22n})\|_{V_n^{-1}} \geq \chi^2_{1-\alpha}(d_g-d_\phi)\right)+o(1) \to \alpha$. Note that Assumption \ref{ass:identification} with respect to $\phi$ is implied by Assumption \ref{ass:weak} ii.

\section{Asymptotic Behaviour of the quasi-Jacobian} \label{sec:asym_jac}

As discussed above, the properties of the ICS and test procedure are tied to those of the quasi-Jacobian under different identification regimes. The following derives the large sample behaviour of the sup-norm and least-squares quasi-Jacobian matrices $B_{n,\infty}$ under strong, semi-strong, and weak identification.

\subsection{Strong and Semi-Strong Sequences}
\begin{theorem}[quasi-Jacobian and Jacobian Equivalence] \label{th:qJ-ss} Let $B_{n,\infty}$ denote the quasi-Jacobian. Suppose $(\theta_n,\gamma_n) \to (\theta_0,\gamma_0)$ with $\gamma_0 \in \Gamma_1$ or $\gamma_n \in \Gamma_0(\infty)$. Suppose that Assumptions \ref{ass:moments}, \ref{ass:kernel_bandwidth}, and \ref{ass:ss} hold. If $\kappa_n^{-1}\delta(\gamma_n) \to \infty$ and $\kappa_n^2 = o \left( \lambda_{\min}(\partial_\theta g(\theta_n,\gamma_n)^\prime \partial_\theta g(\theta_n,\gamma_n)  \right)$, then:
  \begin{align*}  
    [B_{n,\infty}-\partial_\theta g(\theta_n,\gamma_n)]H_n &= o_p(n^{-1/2}\kappa_n^{-1}),\\
    A_{n,\infty} + B_{n,\infty}\theta_n - \bar{g}_n(\theta_n) &=  o_p(n^{-1/2}),
  \end{align*}
  where $n^{-1/2}\kappa_n^{-1} \to 0$ by assumption and $H_n = \left( \partial_\theta g(\theta_n,\gamma_n)^\prime \partial_\theta g(\theta_n,\gamma_n) \right)^{-1/2}$.
\end{theorem}

The proof is given in Appendix \ref{appx:proof_main}. Theorem \ref{th:qJ-ss} implies that, for (semi)-strong sequences, the quasi-Jacobian, and the Jacobian are asymptotically equivalent after re-scaling to a non-singular limit. For non-smooth moments, where the sample Jacobian is not defined as in quantile-IV regression or SMM estimation of discrete choice models, $B_{n,\infty}$ can be used in the sandwich formula to compute standard errors for $\hat\theta_n$. Assumption \ref{ass:ss} v. implies $\lambda_{\min}(H_n \partial_\theta g(\theta_n,\gamma_n)^\prime \partial_\theta g(\theta_n,\gamma_n) H_n) = \lambda_{\min}(R_0^\prime R_0) + o(1) \to 1$, hence:
\[ \lambda_{\min}( B_{n,\infty}^\prime B_{n,\infty} ) = \lambda_{\min}\Big(\partial_\theta g(\theta_n,\gamma_n)^\prime \partial_\theta g(\theta_n,\gamma_n)\Big) \left( 1 + o_p(1) \right). \]
For sufficiently strong sequences such that $\underline{\lambda}_n^2 = o\left(\lambda_{\min}(\partial_\theta g(\theta_n,\gamma_n)^\prime \partial_\theta g(\theta_n,\gamma_n))\right)$, where $\underline{\lambda}_n$ is the cutoff in Section \ref{sec:test}, this implies that $\hat d_n = d_{\theta_2}$ wpa 1.

\subsection{Weak Sequences} 

\begin{theorem}[Asymptotic Singularity of the quasi-Jacobian] \label{th:weak_sup} Suppose $(\theta_n,\gamma_n) \to (\theta_0,\gamma_0)$ with $\gamma_{n} \in \Gamma_0(b), b \in [0,\infty)$ and Assumptions \ref{ass:moments}, \ref{ass:kernel_bandwidth}, \ref{ass:identification}, \ref{ass:weak} hold. For any $v = (0_{d_{\beta_1}},\beta_{2}^{1\prime} - \beta_{2}^{2\prime})^\prime$, with $\beta_2^1,\beta_2^2 \in \mathcal{B}_2^0 \times \mathcal{B}_2^0$ in the identified set for $\beta_2$, $\|B_{n,\infty} M^{-1}v\| \leq O_p(\kappa_n)$. Let $\lambda_{j}(B_{n,\infty}^\prime B_{n,\infty} ) \geq 0$ denote the eigenvalues of $B_{n,\infty}^\prime B_{n,\infty}$ in increasing order, then:
  \[ \sum_{j=1}^{d_{\beta_2}}\lambda_{j}(B_{n,\infty}^\prime B_{n,\infty} ) \leq O_p(\kappa_n^2). \]
  In particular, $\lambda_{\min}(B_{n,\infty}^\prime B_{n,\infty} ) \leq O_p(\kappa_n^2)$.
\end{theorem}
Theorem \ref{th:weak_sup} shows that when $\theta$ is not uniquely identified, the quasi-Jacobian vanishes at a $\kappa_n$ rate in all directions associated with the identification failure. The span of these directions has dimension $d_{\beta_2}$ so that $B_{n,\infty}$ vanishes on a subspace of dimension $d_{\beta_2}$. Hence, small singular values are indicative of an identification failure, and the number of weakly identified coefficients. The constants involved in the $O_p$ terms are made explicit in the proof. The following Proposition extends these results to $B_{n,\infty}P_{\theta_1}^\perp$, which focuses on the identification status of the nuisance parameters only. For both results, the proof is similar to the derivations used for the MA(1) example.

\begin{proposition}[quasi-Jacobian after Projection] \label{prop:weak_sup} Suppose $(\theta_n,\gamma_n) \to (\theta_0,\gamma_0)$ with $\gamma_{n} \in \Gamma_0(b), b \in [0,\infty)$ and Assumptions \ref{ass:moments}, \ref{ass:kernel_bandwidth}, \ref{ass:identification}, \ref{ass:weak} hold. For any $v = (0_{d_{\beta_1}+d_{\beta_{21}}},\beta_{22}^{1\prime} - \beta_{22}^{2\prime})^\prime$, with $\beta_{22}^1,\beta_{22}^2 \in \mathcal{B}_{22}^0 \times \mathcal{B}_{22}^0$ the identified set for $\beta_{22}$ under the null, $\|B_{n,\infty} M^{-1}v\| \leq O_p(\kappa_n)$. Let $\lambda_{j}(P_{\theta_1}^\perp B_{n,\infty}^\prime B_{n,\infty} P_{\theta_1}^\perp ) \geq 0$ denote the eigenvalues of $ P_{\theta_1}^\perp B_{n,\infty }^\prime B_{n,\infty} P_{\theta_1}^\perp$ in increasing order:
  \[ \sum_{j=1}^{d_{\theta_1} + d_{\beta_{22}}}\lambda_{j}(P_{\theta_1}^\perp B_{n,\infty}^\prime B_{n,\infty} P_{\theta_1}^\perp ) \leq O_p(\kappa_n^2). \]
\end{proposition}

\section{Asymptotic Properties of the Test Procedure} \label{sec:asym_test}

As discussed above, the ICS procedure used to compute $\hat d_n$ relies on two normalizations that ensure invariance to rescaling of the sample moments and/or the parameters. The first normalizing matrix is $\Sigma_n$ computed in the procedure outlined above. $\Sigma_n^{-1/2}$ is shown to be bounded above in directions associated with the identification failure in Lemma \ref{lem:Sigma_weak}, so that Proposition \ref{prop:weak_sup} extends to the normalized $B_{n,\infty}P_{\theta_1}^\perp \Sigma_n^{-1/2} P_{\theta_1}^\perp$. Under strong identification, Lemma \ref{lem:Sigma_ss} implies that $\Sigma_n^{-1/2} = O(\kappa_n^{-1})$ so that $B_{n,\infty}P_{\theta_1}^\perp \Sigma_n^{-1/2} P_{\theta_1}^\perp$ diverges at a $\kappa_n^{-1}$ rate in $d_{\theta}-d_{\theta_1}$ directions. As a result, $B_{n,\infty}P_{\theta_1}^\perp \Sigma_n^{-1/2} P_{\theta_1}^\perp$ vanishes at a $\kappa_n$-rate in directions where identification fails, and diverges at a $\kappa_n^{-1}$-rate when all parameters are strongly identified.

The second normalizing matrix is $\overline{V}_n = \int_\Theta \hat V_n(\theta) \hat\pi_n(\theta)d\theta$, where $\hat V_n(\theta)$ is an estimator of the asymptotic variance $\lim_{n \to \infty} \text{var}_{\gamma_n}(\sqrt{n}\overline{g}_n(\theta))$. The Assumption below requires $\hat V_n(\theta)$ consistent and asymptotically non-singular so that the normalization does not alter the asymptotic properties of $B_{n,\infty}P_{\theta_1}^\perp \Sigma_n^{-1/2}$.

\begin{assumption} \label{ass:variance_mat} $V(\theta,\gamma) = \lim_{n \to \infty} \text{var}_{\gamma}(\sqrt{n}\overline{g}_n(\theta))$ is non-singular and $0<\underline{\lambda}_V \leq \lambda_{\min}(V(\theta,\gamma)) \leq \lambda_{\max}(V(\theta,\gamma)) \leq \overline{\lambda}_V < \infty$ for all $\theta\in \Theta$, $\gamma \in \Gamma$, $\sup_{\theta\in\Theta,\gamma \in \Gamma}\|\hat V_n(\theta) - V(\theta,\gamma) \| = o_p(1)$. 
\end{assumption}

\begin{theorem}[Asymptotic Size] \label{th:asym_size} Suppose Assumptions \ref{ass:moments}-\ref{ass:weak} hold. Let $\underline{\lambda}_n \to 0$ such that $\kappa_n = o(\underline{\lambda}_n)$. Let $\hat d_n = \#\big\{ j \in \{ d_{\theta_1}+1,\dots,d_\theta\},\, \lambda_j( P_{\theta_1}^\perp \Sigma_n^{-1/2} P_{\theta_1}^\perp B_{n,\infty}^\prime \overline{V}_n^{-1} B_{n,\infty} P_{\theta_1}^\perp \Sigma_n^{-1/2} P_{\theta_1}^\perp ) > \underline{\lambda}_n^{2} \big\},$ then for any $\alpha \in (0,1)$:
\[ \limsup_{n\to\infty} \sup_{\gamma \in \Gamma, \theta = (\theta_{10}^\prime,\theta_2^\prime)^\prime \in \overline{\Theta}} \mathbb{P}_{\gamma} \left(  \text{AR}_n(\theta_{10}) > \chi^2_{d_\theta - \hat d_n}(1-\alpha) \right) \leq \alpha. \]
For any sequence $\gamma_n \in \Gamma_0(\infty) \cup \Gamma_1$ such that $\underline{\lambda}^2_n = o(\lambda_{\min}(\partial_\theta g(\theta_n,\gamma_n)^\prime \partial_\theta g(\theta_n,\gamma_n))$:
\[  \lim_{n\to\infty} \mathbb{P}_{\gamma_n} \left(  \text{AR}_n(\theta_{1n}) > \chi^2_{d_\theta - \hat d_n}(1-\alpha) \right) = \alpha. \]
\end{theorem}

Theorem \ref{th:asym_size} establishes the uniform validity of the test procedure described in Section \ref{sec:test} under strong, semi-strong, and weak sequences. First, it is shown that the normalizations do not affect the predictions of Theorems \ref{th:qJ-ss}, \ref{th:weak_sup}, and Proposition \ref{prop:weak_sup}. Then, since $\overline{\Theta}$ and $\Gamma$ are compact, the worst-case rejection probability is attained by a converging subsequence which, using the stated assumptions, can be interpolated into a converging sequence in either $\Gamma_0(b)$ for some $b \in [0,\infty)$, $\Gamma_0(\infty)$, or converging in $\Gamma_1$. The result then relies on two properties. The first is that $\hat d_n \geq d_{\beta_{22}}$ under weak identification, and the second is that $\text{AR}_n(\theta_{1n}) = \inf_{\theta_2 \in \Theta_2} \text{AR}_n(\theta_{1n},\theta_2) \leq \text{AR}_n(\theta_{1n},\hat\phi_n,\beta_{22n})$ which has a standard chi-squared limiting distribution with degrees of freedom that only depend on the dimension of $\bar{g}_n$, and the number of identified nuisance parameters. For just-identified models, the resulting procedure is efficient under strong identification since it uses the smallest valid critical value, and is equivalent to a quasi-Likelihood ratio test. For over-identified model, the test uses the smallest valid critical value for the projected AR test so it is non-conservative within that class.  The results above can be extended to some other existing robust test statistics. For instance, the K-statistic of \citet{Kleibergen2005} is such that, under additional regularity conditions, $\text{K}_n(\theta_{1n}) = \inf_{\theta_2 \in \Theta_2} \text{K}_n(\theta_{1n},\theta_2) \leq \text{K}_n(\theta_{1n},\hat\phi_n,\beta_{22n})$ which also has a chi-squared limiting distribution with reduced degrees of freedom. 

\section{Monte-Carlo Simulations} \label{sec:MonteCarlo}

The finite-sample properties of the quasi-Jacobian matrix and the test procedure are illustrated using a consumption capital asset pricing model (CAPM) as in \citet[Sec3]{Wright2003}. 

Let $\delta$, $\gamma$ measure time preference and relative risk aversion. $C_t,D_t,R_t$ are real consumption, dividends, and the gross asset return at time $t$. The Euler equation is: $\mathbb{E}_t[ \delta R_{t+1}(C_{t+1}/C_t)^{-\gamma}-1 ] = 0$, 
where $C_{t+1}/C_t$ measures consumption growth. $R_{t}$ depends endogenously on $y_{t+1} = ( c_{t+1},d_{t+1} )^\prime$, where $c_{t+1} = \log( C_{t+1}/C_t )$ and $d_{t+1} = \log( D_{t+1}/D_t )$, which follows a first-order vector autoregressive (VAR) process: $y_{t+1} = \mu + \Phi y_t + u_{t+1}$, where $u_{t+1} \overset{iid}{\sim} \mathcal{N}(0,\Lambda)$. The sample moments are: \[ \overline{g}_n(\theta) = \frac{1}{n} \sum_{t=1}^n [\delta R_{t+1}(C_{t+1}/C_t)^{-\gamma}-1]Z_t,\] where $Z_{t} = (1,R_t,C_t/C_{t-1})^\prime$. 
 \citet{tauchen1986} illustrates how $(\mu,\Phi,\Lambda)$ affects the finite-sample properties of $\hat\theta_n = (\hat\delta_n,\hat\gamma_n)$. The following considers three DGPs: Rank Failure (RF), Near Rank Failure (NRF) and Full Rank (FR).\footnote{RF, NRF and FR correspond to RF1, NRF1 and FR in \citet[p326]{Wright2003}.} \citet{Wright2003} explains that they correspond to $\theta = (\delta,\gamma)$ being set, weakly, and strongly identified. NRF is calibrated to match annual U.S. data \citep[][Sec3]{kocherlakota1990}.

\begin{table}[h] \caption{CAPM - VAR parameters used in the simulations} \label{tab:DGPs}
  \begin{tabular}{c|c|c} \hline\hline
    Rank Failure & Near Rank Failure & Full Rank\\ \hline
    $\mu_{\text{RF}} = (0.018,0.013)^\prime$  & $\mu_{\text{NRF}} = (0.021,0.04)^\prime$ & $\mu_{\text{FR}} = (0.00,0.00)^\prime$\\
    $\Phi_{\text{RF}} = \left( \begin{array}{cc} 0 & 0 \\ 0 & 0 \end{array} \right)$ & $\Phi_{\text{NRF}} = \left( \begin{array}{cc} -0.161 & 0.017 \\ 0.414 & 0.117 \end{array} \right)$ & $\Phi_{\text{FR}} = \left( \begin{array}{cc} -0.5 & 0 \\ 0 &-0.5 \end{array} \right)$\\
    $\Lambda_{\text{RF}} = \left( \begin{array}{cc} 0.0012 & 0.0017 \\ 0.0017 & 0.0146 \end{array} \right)$ & $\Lambda_{\text{NRF}} = \left( \begin{array}{cc} 0.0012 & 0.00177 \\ 0.00177 & 0.014 \end{array} \right)$ & $\Lambda_{\text{FR}} = \left( \begin{array}{cc} 0.01 & 0 \\ 0 & 0.01 \end{array} \right)$ \\ \hline\hline
    \end{tabular}
    \notes{\textbf{Note:} the parameters $(\mu,\Phi,\Lambda)$ describe the dynamics of consumption and dividend growth $y_t = (c_t,d_t)^\prime$.}
\end{table} 

Table \ref{tab:CAPM} reports rejection rates for the method in Section \ref{sec:outline} (Proj$_1$), full projection inference using $\chi^2_{3}$ (Proj$_2$) and $\chi^2_{2}$ (Proj$_3$) critical values as well a t-test with standard normal critical value ($t_n$). The empirically relevant sample sizes are $n=100,250$. $n=500,1000$ illustrate large sample properties. The parameter space is $\Theta = [0.7,1.1] \times [0,10]$. The t-test does not control size in RF and NRF. It is closer to nominal size for FR. However, as Figure \ref{fig:CAPMest} in Appendix \ref{apx:MC_additional_simu} shows, another global solution $\hat\theta_n \simeq (0.7,10)$ is estimated in about 1\% and 0.05\% of the replications for $n=100,250$. Here, the parameters are locally strongly identified, but not globally. The sample Jacobian would not detect this issue which leads to some over-rejection for the t-test. In comparison, the proposed procedure (Proj$_1$) has null rejection rates below nominal size across sample sizes and DGPs. 

\begin{figure}[h] \caption{CAPM - distribution of largest and smallest singular values} \label{fig:CAPMsing} \centering
  \setlength\tabcolsep{2.5pt}
\renewcommand{\arraystretch}{0.9} 
  \includegraphics[scale=0.45]{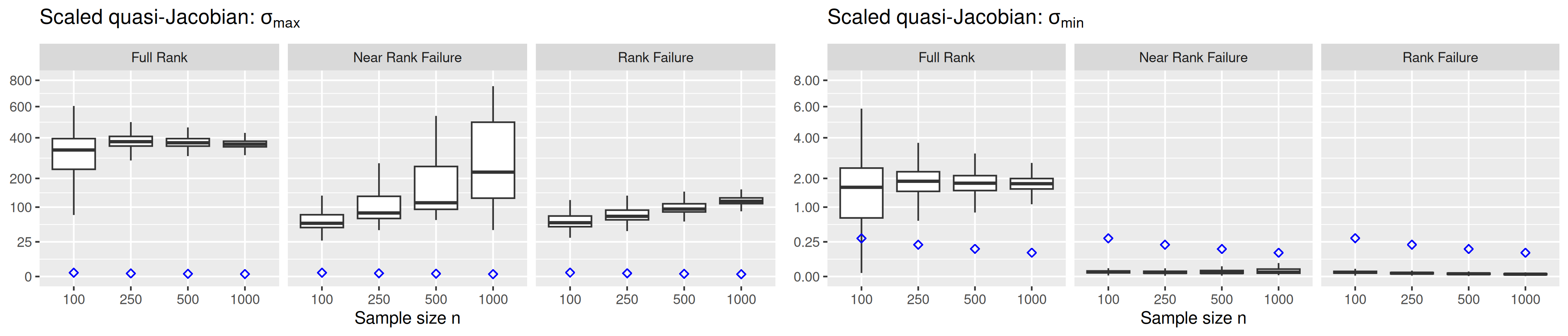}
  \notes{\textbf{Note:} True value $(\delta_0,\gamma_0) = (0.97,1.3)$. 2000 Monte Carlo replications. $W_n = \hat{V}_n(\theta)^{-1}$ where $\hat{V}_n$ is a HAC estimate of $\text{var}[\sqrt{n}\overline{g}_n(\theta)]$. bandwidth $\kappa_n = \sqrt{2\log(\log[n])/n}$, cutoff (blue diamonds) $\underline{\lambda}_n = \sqrt{2\log[n]/n} =0.30, 0.21, 0.16, 0.12$ for $n=100,250,500,1000$. $\sigma_{\max},\sigma_{\min}$: largest and smallest singular values. Median values of $\sigma_{\min}$ for $n=100,250,500,1000$: $4\cdot 10^{-3},2\cdot 10^{-3},2\cdot 10^{-3},1\cdot 10^{-3}$ (RF), $4\cdot 10^{-3}$, $3\cdot 10^{-3}$, $3\cdot 10^{-3}$, $4\cdot 10^{-3}$ (NRF), and $1.7$, $1.9$, $1.8$, $1.8$ (FR).}
\end{figure}

\citet[Sec3]{Wright2003}, \citet[Sec5]{Antoine2009} explain that one coefficient is always strongly identified. Table \ref{tab:CAPM} and Figure \ref{fig:CAPMsing} confirm this. The procedure finds $\gamma$ to be weakly identified in nearly all replications for RF, NRF, and $\delta$ strongly identified. For FR, the procedure finds $\gamma$ weakly identified in $14\%$ and $0.5\%$ of replications when $n=100,250$.

\newcolumntype{g}{>{\columncolor{gray}}c}
\begin{table}[ht] \caption{CAPM -- rejection rates, frequency for detecting identification failure}
  \label{tab:CAPM}
  \centering \setlength\tabcolsep{2.5pt}
  \renewcommand{\arraystretch}{0.9} 
  {\small
  \begin{tabular}{c|c|gccc|g||gccc|g||gccc|g}
    \hline \hline
   & & \multicolumn{5}{c||}{Rank Failure} & \multicolumn{5}{c||}{Near Rank Failure} & \multicolumn{5}{c}{Full Rank}\\ \hline \rowcolor{white}
   n & & AR$_1$ & AR$_2$ & AR$_3$ & $t_n$ & $<\underline{\lambda}_n$ & AR$_1$ & AR$_2$ & AR$_3$ & $t_n$ & $<\underline{\lambda}_n$ & AR$_1$ & AR$_2$ & AR$_3$ & $t_n$ & $<\underline{\lambda}_n$ \\ 
    \hline
    \multirow{2}{*}{100} & $\delta$ & 0.01 & 0.01 & 0.03 & 0.02 & 1.00 & 0.01 & 0.01 & 0.03 & 0.02 & 1.00 & 0.04 & 0.02 & 0.05 & 0.07 & 0.14  \\ 
    & $\gamma$ & 0.05 & 0.02 & 0.05 & 0.00 & 0.00 & 0.05 & 0.02 & 0.05 & 0.00 & 0.00 & 0.04 & 0.01 & 0.04 & 0.06 & 0.00 \\ \hline
    \multirow{2}{*}{250} & $\delta$ & 0.02 & 0.02 & 0.05 & 0.09 & 1.00 & 0.02 & 0.02 & 0.03 & 0.07 & 1.00 & 0.04 & 0.02 & 0.04 & 0.05 & 0.01 \\ 
    & $\gamma$ & 0.05 & 0.02 & 0.05 & 0.00 & 0.00 & 0.05 & 0.02 & 0.05 & 0.00 & 0.00 & 0.04 & 0.02 & 0.04 & 0.06 & 0.00 \\ \hline
    \multirow{2}{*}{500} & $\delta$ & 0.02 & 0.02 & 0.04 & 0.17 & 1.00 & 0.01 & 0.01 & 0.04 & 0.08 & 1.00 & 0.05 & 0.02 & 0.05 & 0.05 & 0.00 \\ 
    & $\gamma$ & 0.04 & 0.02 & 0.04 & 0.00 & 0.00 & 0.04 & 0.02 & 0.04 & 0.00 & 0.00 & 0.06 & 0.02 & 0.06 & 0.05 & 0.00 \\ \hline
    \multirow{2}{*}{1000} & $\delta$ & 0.02 & 0.02 & 0.05 & 0.22 & 1.00 & 0.02 & 0.02 & 0.04 & 0.05 & 0.98 & 0.05 & 0.02 & 0.05 & 0.05 & 0.00 \\ 
    & $\gamma$ & 0.05 & 0.02 & 0.05 & 0.00 & 0.00 & 0.04 & 0.02 & 0.04 & 0.00 & 0.00 & 0.05 & 0.02 & 0.05 & 0.05 & 0.00 \\ 
     \hline \hline
  \end{tabular}}
  \notes{ 
  \textbf{Note:} Nominal size = $5\%$. 2000 Monte Carlo replications. AR$_1$, AR$_2$, AR$_3$: projection inference using AR statistic and $\chi^2$ critical values with $3-\hat{d}_n$, $3$, and $2$ degrees of freedom; $\hat{d}_n \in \{0,1\}$. $t_n$: t-test with standard normal critical values. $<\underline{\lambda}_n$: frequency (in \%) of singular values below cutoff $\underline{\lambda}_n$ after projecting out the parameter of interest. Rows for $\delta$ show results for $H_0:\delta=\delta_0$. Rows for $\gamma$ show results for $H_0:\gamma=\gamma_0$. $\kappa_n = \sqrt{2\log(\log[n])/n}$, $\underline{\lambda}_n = \sqrt{2\log(n)/n}$, $W_n = \hat{V}_n(\theta)^{-1}$ where $\hat{V}_n$ is a HAC estimate of $\text{var}[\sqrt{n}\overline{g}_n(\theta)]$.  }
  \end{table}

  \begin{figure} \caption{CAPM - power comparison} \label{fig:CAPMpow} \centering
    \setlength\tabcolsep{2.5pt}
  \renewcommand{\arraystretch}{0.9} 
    {\small \begin{tabular}{cc} 
      $\bf{n=100}$ & $\bf{n=250}$ \\
      \includegraphics[scale=0.45]{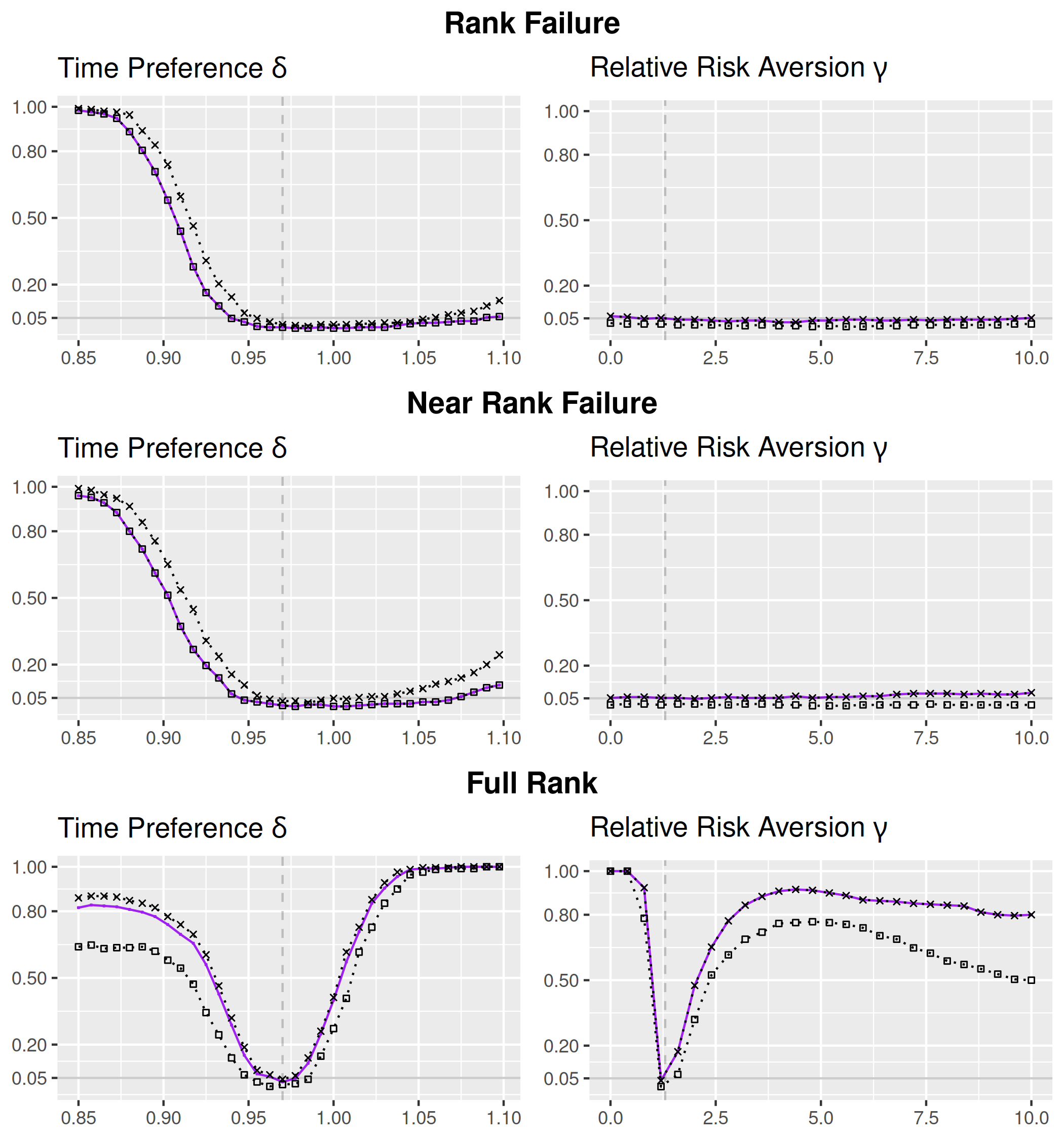} & \includegraphics[scale=0.45]{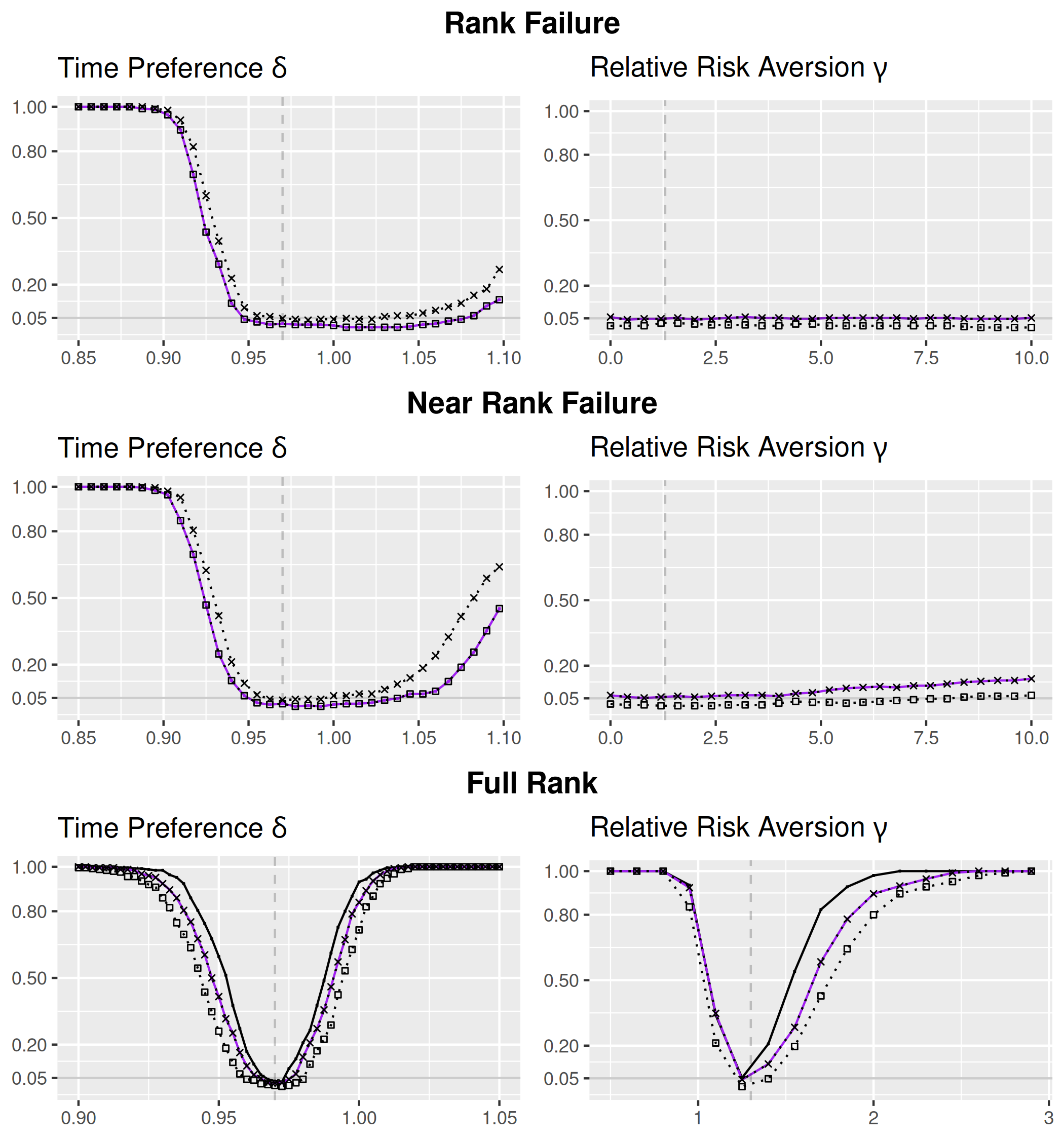}\\ 
      $\bf{n=500}$ & $\bf{n=1000}$ \\
      \includegraphics[scale=0.45]{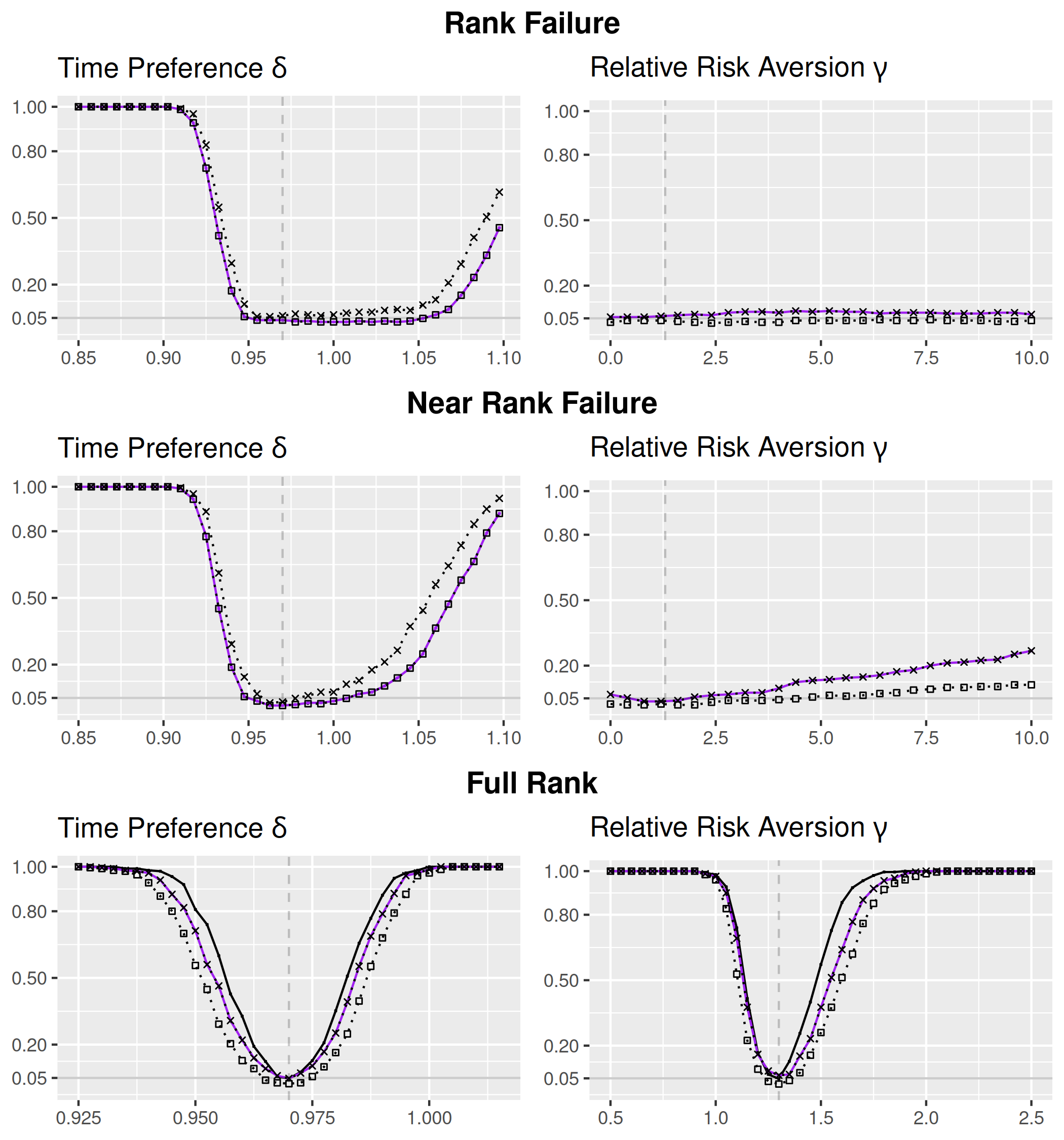} & \includegraphics[scale=0.45]{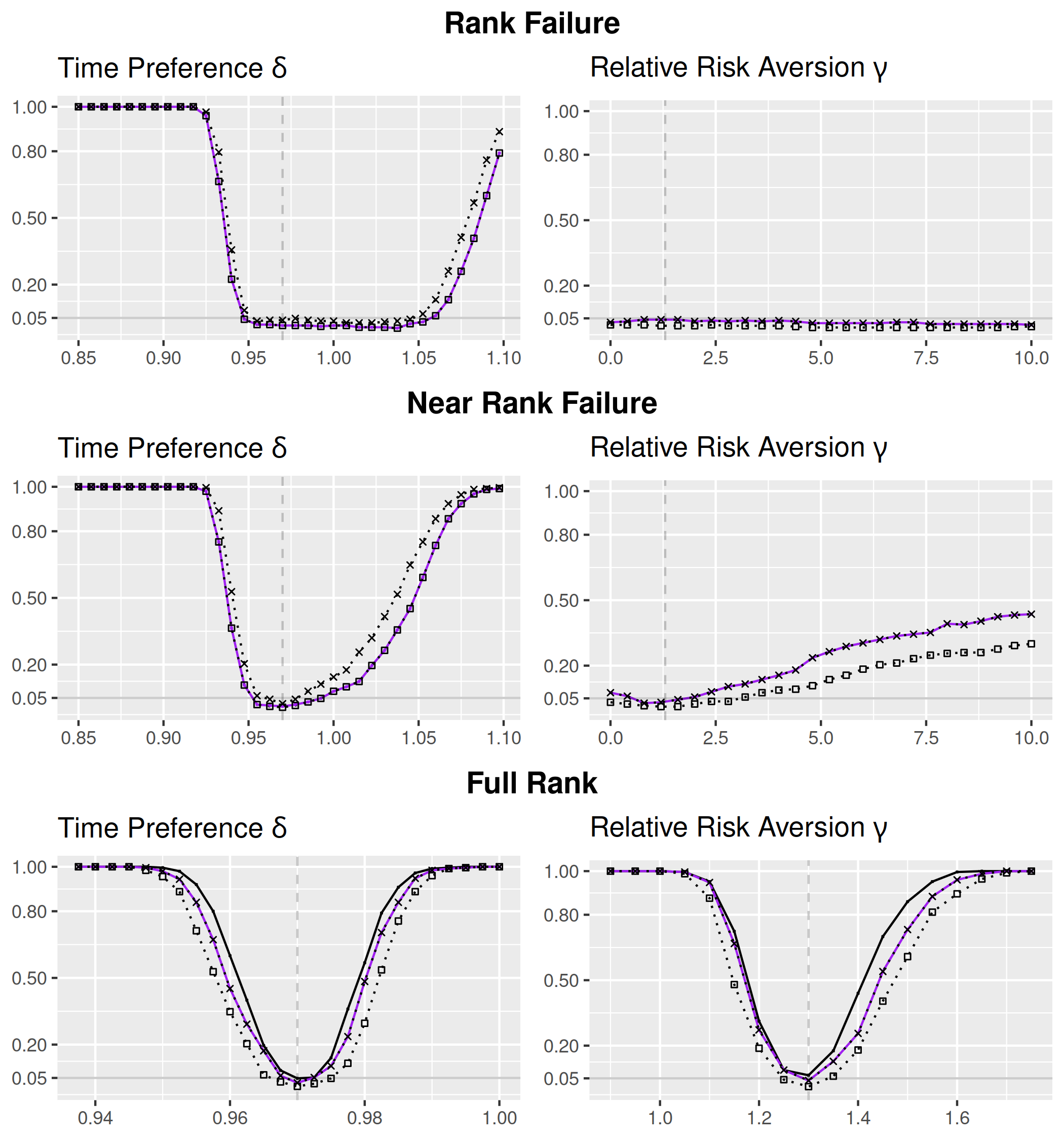}\\
    \includegraphics[scale=0.65]{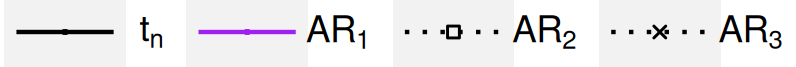}
    \end{tabular}}
    \notes{\textbf{Note:} Nominal size = $5\%$. Dashed vertical lines: true value $(\delta_0,\gamma_0) = (0.97,1.3)$. 250 Monte Carlo replications. Estimates computed for continuously-updated GMM with $W_n = \hat{V}_n(\theta)^{-1}$ where $\hat{V}_n$ is a HAC estimate of $\text{var}[\sqrt{n}\overline{g}_n(\theta)]$. AR$_1$, AR$_2$, AR$_3$: projection inference using AR statistic and $\chi^2$ critical values with $3-\hat{d}_n$, $3$, and $2$ degrees of freedom; $\hat{d}_n \in \{0,1\}$. $t_n$: t-test with standard normal critical values.}
  \end{figure}

Figure \ref{fig:CAPMpow} compares the power of the proposed procedure (AR$_1$) with full projection inference (AR$_3$), and projection inference with the nuisance parameter concentrated out (AR$_2$) as well as the t-test when appropriate (FR with $n = 250,500,1000$). The results show power improvement over full projection inference when the nuisance parameter is strongly identified, i.e. when testing hypotheses about $\gamma$. When the model is strongly identified (FR), the procedure is less powerful than the t-test because of over-identification. Result for a just-identified specification with $Z_t = (1,R_t)^\prime$ and a larger $\kappa_n$ are given in Appendix \ref{apx:MC_additional_simu}. Another example in that Appendix compares the procedure with \citet{Andrews2012} for a non-linear regression.


\section{Application to the Long-Run Risks Model} \label{sec:empirical}
To illustrate the empirical content which can be gained from the quasi-Jacobian for inference, consider a simulated method of moments estimation of the long-run risks (LRR) model \citep{bansal2004}. There are two latent variables representing a persistent component to the level of consumption growth $x_{1,t}$ and stochastic volatility $x_{2,t}$:
\begin{align*}
  x_{1,t} = \rho x_{1,t-1} + \phi_e f(x_{2,t-1})e_t, \quad x_{2,t} = \sigma^2 +  \nu (x_{2,t-1}-\sigma^2) + \sigma_w w_t,
\end{align*}
where $f(x) =\sqrt{x}$ if $x \geq \sigma^2$ and $f(x)=\sigma^2/\sqrt{2\sigma^2 - x}$ as in \citet[p346]{Calvet2015}. Consumption and dividend growth $g_t,d_{d,t}$ are then given by:
\begin{align*}
  g_t = \mu + x_{1,t-1} + f(x_{2,t-1})\eta_t, \quad g_{d,t} = \mu_d + \phi x_{1,t-1} + \phi_d f(x_{2,t-1})u_t,
\end{align*}
where $(e_t,w_t,\eta_t,u_t)\sim \mathcal{N}(0,I)$ iid. Given an Epstein-Zin utility function, equilibrium conditions imply that financial variables, log-price dividend ratio $z_{m,t}$, market return $r_{m,t}$ and the risk-free rate $r_{a,t}$ can be written as:
\begin{align*}
  z_{m,t} &= A_{0,m} + A_{1,m} x_{1t} + A_{2,m} x_{2,t},\\ r_{m,t} &= \kappa_{0,m} + \kappa_{1,m} z_{m,t+1} - z_{m,t} + g_{d,t+1},\\ r_{a,t} &= A_{0,r} + A_{1,r} x_{1,t} + A_{2,r} x_{2,t}
\end{align*}
where the coefficients $(A_{0,m},A_{1,m},A_{2,m},\kappa_{0,m},\kappa_{1,m},A_{0,r},A_{1,r},A_{2,r})$ are computed numerically as a solution of a non-linear system of equations involving the full vector of 12 parameters $\theta = (\rho,\phi_e,\sigma,\nu,\sigma_w,\mu,\mu_d,\phi,\phi_d,\delta,\gamma,\psi^{-1})$ where $\delta$ is the discount factor, $\gamma$ risk-aversion, and $\psi^{-1}$ the inverse intertemporal elasticity of sustitution (IES). See \citet{bansal2004} for details. The variables above need to be further time-aggregated from the monthly decision interval to match the quarterly frequency of the data. There are a number of estimations of this model using one of SMM and Indirect Inference,\footnote{See \citet{bansal2007,hasseltoft2012,Calvet2015,grammig2018}.} GMM,\footnote{See \citet{constantinides2011,bansal2012,bansal2016}.}, or Bayesian estimation\footnote{See \citet{schorfheide2018}.}  There are, however, several concerns for the identifiability of the parameters. \citet{Calvet2015} show that the latent variables $(x_{1,t},x_{2,t})$ cannot be recovered from the data for uncountably many values of $\theta$, resulting in highly irregular GMM and likelihood objective functions. \citet{grammig2018} find that the stochastic volatility component is poorly identified and calibrate $\nu = \sigma_w =0$. However, stochastic volatility in long-term consumption growth has important implications for asset prices \citep{schorfheide2018}. Several papers report estimates with very small standard errors \citep[see][Table 7, p24]{grammig2018}, but estimates can vary a lot across estimations. This suggests that some parameters are likely not globally identified but might be locally identified.

The following considers joint inference for the two preference parameters $\theta_1 = (\gamma,\psi^{-1})$. The remaining coefficients are $\theta_2 = (\rho,\phi_e,\sigma,\nu,\sigma_w,\mu,\mu_d,\phi,\phi_d,\delta)$. Amongst these nuisance parameters, it seems reasonable to think that several are (semi)-strongly identified. However, the asset pricing coefficients $(A_{0,m},\dots)$ are highly non-linear functions of $\theta$ so it is arguably more difficult to pin down exactly how many and which ones are well identified. Nevertheless, the results in this paper imply that $B_{n,\infty}P_{\theta_1}^\perp$ can determine how many nuisance parameters are weakly identified with high probability.

The moment conditions used for inference are based on matching the following sample with simulated moments: means of all variables, variances of $g_{t},g_{d,t},z_{m,t}$, AR$(2)$ coefficients of $g_{t}$, and autocorrelation of $g_{t}^2$.\footnote{A quasi-difference $z_{m,t} - 0.95 z_{m,t-1}$ is applied beforehand because $z_{m,t}$ is very persistent making $\hat V_n$ nearly singular, the quasi-differencing solves this issue and makes the estimation below more stable.} These just-identified moments match quantities of interest that are commonly reported in calibrations or post-estimation, see e.g. \citet{beeler2012}. The estimation is conducted using U.S. data shared by \citet{grammig2018} for $(g_t,g_{d,t},z_{m,t},r_{m,t},r_{a,t})$ over 1947Q2-2014Q4, totalling in $n=271$ observations. The simulated moments are computed over $S=2$ samples. The bounds for the optimization space $\Theta$ are $\rho \in [0.9,0.995], \phi_e \in [0,0.1], \sigma \in [10^{-4},0.1], \nu \in [0,0.995], 10^5 \times \sigma_w \in [0,2],\mu \in [-0.035,0.035],\mu_d \in [-0.035,0.035],\phi \in [0,10],\phi_d \in [0,10],\delta \in [0.93,1.2],\gamma \in [0.05,25],\psi^{-1} \in [0.01,3]$. Computations are conducted in R and C++ using Rcpp. 

\begin{table}[ht] \centering
  \caption{Long-Run Risks: singular values of Jacobian and quasi-Jacobian} \label{tab:LRR_eigen}
  \begingroup
  \setlength\tabcolsep{4.5pt}
    \renewcommand{\arraystretch}{0.935} 
  {\footnotesize
  \begin{tabular}{l|cccccccccccc} 
    \hline \hline
   & $\lambda_{1}$ & $\lambda_{2}$ & $\lambda_{3}$ & $\lambda_{4}$ & $\lambda_{5}$ & $\lambda_{6}$ & $\lambda_{7}$ & $\lambda_{8}$ & $\lambda_{9}$ & $\lambda_{10}$ & $\lambda_{11}$ & $\lambda_{12}$ \\ 
    \hline
  $\bar{V}_n^{-1/2} \partial_\theta \bar{g}_n(\hat\theta_n) \Sigma_n^{-1/2}$ & 
  $8 . 10^6$ & $4 . 10^6 $ & $7 . 10^5$ & $1. 10^5$ & $8 . 10^4$ & 255 & 22 & 1.68 & 0.30 & 0.04 & $<10^{-2}$ & $<10^{-2}$\\  
  $\bar{V}_n^{-1/2} B_{n,\infty} \Sigma_n^{-1/2}$ & 
  $2 . 10^8$ & $1 . 10^7$ & $1 . 10^6$ & $4 . 10^5$ & $2. 10^4$ & 208 & 0.94 & 0.42 & 0.06 & 0.01 & $<10^{-2}$ & $<10^{-2}$ \\ \rowcolor{gray}
  $\bar{V}_n^{-1/2} B_{n,\infty} P_{\theta_1}^\perp \Sigma_n^{-1/2} P_{\theta_1}^\perp$ & 
  $2 . 10^8$ & $1. 10^7$ & $1. 10^6$ & $4. 10^5$ & $2.10^4$ & 208 & 0.95 & 0.06 & 0.04 & 0.01 & 0.00 & 0.00 \\ 
     \hline \hline
  \end{tabular} }
  \endgroup
  \notes{\textbf{Note:} $B_{n,\infty}, \Sigma_n$ computed using $B=1000$ draws, $\kappa_n = \sqrt{2\log(\log(n_S))/n} = 0.12$, $n_s = n \times (1+1/S)$.}
\end{table}

Table \ref{tab:LRR_eigen} compares the spectrum of the normalized Jacobian and quasi-Jacobian.\footnote{The estimate $\hat\theta_n$ used for the Jacobian is computed by using the calibration in \citet{bansal2004} as starting value, and alternating between the Nelder-Mead and \textsc{bobyqa} optimizers until convergence. Note that different seeds for the simulated samples yield very different estimates but similar fitted moments. Also, the sample gradient is not available analytically; it is computed by finite differences which here is quite sensitive to the choice of step size.} 
Using the threshold $\underline{\lambda}_n = \sqrt{2\log(n_S)/n} = 0.21$ implies that $B_{n,\infty}$ detects $5$ directions of identification failure, with an additional singular value just above the threshold. In comparison, the gradient is small in $3$ directions.  After projecting out $\theta_1 = (\gamma,\psi^{-1})$, there are $7$ singular values above the threshold, indicating $7$ (semi)-strongly identified parameters. Hence, inference for $(\gamma,\psi^{-1})$ relies on a $\chi^2_{5}(0.95)=11.1$ critical value. In comparison, full projection relies on $\chi^2_{12}(0.95)=21$, and standard inference $\chi^2_{2}(0.95)=6$.

\begin{figure}[h] \centering \caption{Long-Run Risks: joint 95\% confidence set for $(\gamma,\psi^{-1})$} \label{fig:CS_gamma_psi}
  \includegraphics[scale=0.9]{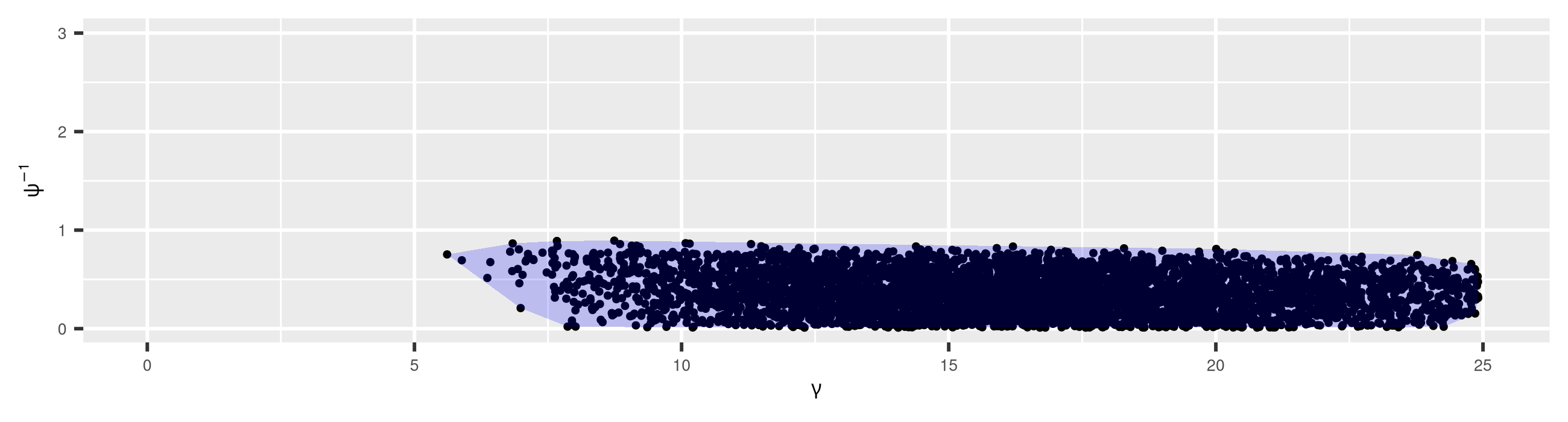}
\end{figure}

Figure \ref{fig:CS_gamma_psi} reports 5000 draws of $\theta_1$ such that $\text{AR}_n(\theta_1) \leq \chi^2_{5}(0.95)$ using the Population Monte Carlo algorithm in Appendix \ref{apx:PMC}, plus their convex hull in blue. Values for $\gamma$ are contained in $[5.41,25]$ and $\psi^{-1}\in [0.01,0.90]$. This excludes several regions of interest. First, we can reject $H_0: \psi \leq 1$ at the 95\% confidence level, i.e. the IES is strictly greater than unity. 
Second, we can reject $H_0:\gamma = \psi^{-1}$ and conclude that the utility function is not CRRA. Finally, the confidence set favours $H_1: \gamma > \psi^{-1}$ over $H_0: \gamma \leq \psi^{-1}$. Under $H_1$, households prefer an early resolution of uncertainty; their preference for consumption smoothing is less than their relative risk aversion. Although not reported here, note that full projection inference cannot reject some of these null hypotheses. As a robustness check with respect to tuning parameters, Appendix \ref{apx:extra_empirical} finds the same results using $\chi^2_6$ critical values (Figure \ref{fig:CS_gamma_psi_6}) and using a larger value for $\kappa_n$ (Table \ref{tab:extra_LRR_eigen}).

\section{Conclusion} \label{sec:conclusion}
This paper introduces a quasi-Jacobian matrix which is asymptotically equivalent to the usual Jacobian matrix under strong and semi-strong identification but is asymptotically singular when global identification fails. This can be useful because the Jacobian is not always informative about global identification failures. While the inference procedure relies on the AR statistic, extending the results to the robust score test is straightforward, as discussed earlier. For overidentified models, it could be interesting to extend the theory to more powerful test statistics such as the CQLR/AR test in \citet{andrews2017}. Another concern could be that a given choice of moments does not identify the parameters but another set of moments might. This is a moment selection problem. In that case, it could be interesting to extend the quasi-Jacobian to a continuum of moment conditions which can be used for conditional GMM estimation \citep{Carrasco2000}; allowing the use of all available information rather than selecting finite dimensional moments.
{ 
\bibliography{refs}  }
\begin{appendices}
  \renewcommand\thetable{\thesection\arabic{table}}
  \renewcommand\thefigure{\thesection\arabic{figure}}
  \renewcommand{\theequation}{\thesection.\arabic{equation}}
  \renewcommand\thelemma{\thesection\arabic{lemma}}
  \renewcommand\thetheorem{\thesection\arabic{theorem}}
  \renewcommand\thedefinition{\thesection\arabic{definition}}
    \renewcommand\theassumption{\thesection\arabic{assumption}}
  \renewcommand\theproposition{\thesection\arabic{proposition}}
    \renewcommand\theremark{\thesection\arabic{remark}}
    \renewcommand\thecorollary{\thesection\arabic{corollary}}


  \clearpage \baselineskip=18.0pt
  \appendix
\section{Preliminary Results} \label{apx:prelim}
\subsection{Preliminary results for Section \ref{sec:setting}}

  \begin{lemma}[Strong and Semi-Strong Sequences: Consistency] \label{lem:consistency}
    Let $(\theta_n,\gamma_n) \to (\theta_0,\gamma_0)$. If $\gamma_n \in \Gamma_0(\infty)$ or $\gamma_0 \in \Gamma_1$ and Assumptions \ref{ass:moments}, \ref{ass:identification} hold, then $\|\hat\theta_n-\theta_n\|=o_p(1)$.
  \end{lemma}

  \begin{lemma}[Strong and Semi-Strong Sequences: Asymptotic Normality] \label{lem:asym_normal}
    Let $(\theta_n,\gamma_n) \to (\theta_0,\gamma_0)$. If $\gamma_n \in \Gamma_0(\infty)$ or $\gamma_0 \in \Gamma_1$ and Assumptions \ref{ass:moments}, \ref{ass:identification}, \ref{ass:ss} hold, then \[\sqrt{n}H_n^{-1}(\hat\theta_n-\theta_n) \overset{d}{\to} \mathcal{N}(0,\Sigma_0),\] where $H_n = \left( \partial_\theta g(\theta_n,\gamma_n)^\prime \partial_\theta g(\theta_n,\gamma_n) \right)^{-1/2}$, $\Sigma_0 = (R_0^\prime W_0 R_0)^{-1} R_0^\prime W_0 R_0 (R_0^\prime W_0 R_0)^{-1}$, $W_0=W(\theta_0)$.
  \end{lemma}

\section{Proofs for the main results} \label{appx:proof_main}

\subsection{Proofs for Section \ref{sec:setting}}

\paragraph{Proof of Theorem \ref{th:qJ}:} For simplicity, the derivations for this Theorem rely on $K(x) = \mathbbm{1}_{x \in (-1,1)}$, see derivations for Section \ref{sec:asym_jac} for derivations with other kernels. For $\kappa>0$, let $(A_{\kappa,\infty},B_{\kappa,\infty}) =  \text{argmin}_{A,B} (\sup_{\|g(\theta)\|_W \leq \kappa} \|g(\theta)-A-B \theta\|)$. By construction, $B_\infty = \lim_{\kappa \to 0} B_{\kappa,\infty}$. To simplicify notation, denote $g(\theta) = g(\theta,\gamma_0)$ and $\partial_\theta g(\theta) = \partial_\theta g(\theta,\gamma_0)$. There are three cases to consider:\\
\underline{Case 1}) Take $\{\theta^1_0,\theta^2_0\} \subseteq \Theta_0$ non-singleton with $\theta^1_0 \neq \theta_0^2$. Take $\kappa > 0$, then $0=\|g(\theta^1_0)\|_W = \|g(\theta_0^2)\|_W \leq \kappa$, by construction. Also by construction, $\sup_{\|g(\theta)\|_W \leq \kappa} \|g(\theta)-A_{\kappa,\infty}-B_{\kappa,\infty}\theta\| \leq \sup_{\|g(\theta)\|_W \leq \kappa} \|g(\theta)\| \leq \underline{\lambda}^{-1/2}_W \kappa$. As a result, $\|g(\theta)-A_{\kappa,\infty}-B_{\kappa,\infty}\theta\| \leq \underline{\lambda}^{-1/2}_W \kappa$ for $\theta\in \{\theta^1_0,\theta_0^2\}$ and the triangular inequality implies $\|B_{\kappa,\infty}(\theta^1_0-\theta_0^2)\| \leq 2 \underline{\lambda}^{-1/2}_W \kappa.$ Take the limit as $\kappa \to 0$ to find $B_{\infty}(\theta^1_0-\theta_0^2)=0$ where $\theta^1_0-\theta_0^2 \neq 0$. Hence, $B_{\infty}$ is singular.\\
\underline{Case 2})  $\Theta_0=\{\theta_0\}$ is singleton and $\partial_\theta g(\theta_0)$ is singular. Take any vector $v \in \text{span}(\partial_\theta g(\theta_0))^\perp$ with $\|v\|=1$. For the following, consider $\theta = \theta_0 + \kappa^{1/\alpha} r v$ for some $r \in \mathbb{R}$ such that $\kappa^{1/\alpha}|r| \leq \overline{\varepsilon}$. Then $\|g(\theta)\|_W = \|g(\theta)-g(\theta_0) - \kappa^{1/\alpha} r \partial_\theta g(\theta_0) v\| \leq \overline{\lambda}_W \overline{C} \kappa |r|^\alpha \leq \kappa$ for all $|r| \leq (\overline{\lambda}_W \overline{C})^{-1/\alpha}$. As in Case 1),  $\|g(\theta) -A_{\kappa,\infty} - B_{\kappa,\infty}\theta\| \leq \underline{\lambda}^{-1/2}_W \kappa$ for all $\theta = \theta_0 + \kappa^{1/\alpha} r v$ with $|r| \leq (\overline{\lambda}_W \overline{C})^{-1/\alpha}$. Then $\|B_{\kappa,\infty}(\theta-\theta_0)\| \leq \|g(\theta) -A_{\kappa,\infty} - B_{\kappa,\infty}\theta\| + \|g(\theta_0) -A_{\kappa,\infty} - B_{\kappa,\infty}\theta_0\| \leq 2 \underline{\lambda}^{-1/2}_W \kappa$. Take $r \neq 0$, fixed, then $\theta-\theta_0 = r \kappa^{1/\alpha} v$ and $\|B_{\kappa,\infty} v\| \leq  2 r^{-1} \underline{\lambda}^{-1/2}_W \kappa^{1-1/\alpha} \to 0$ as $\kappa \to 0$ since $\alpha >1$. This implies $B_{\infty}v = 0$; $B_{\infty}$ is singular.\\
\underline{Case 3})  $\Theta_0=\{\theta_0\}$ is singleton and $\partial_\theta g(\theta_0)$ has full rank. Continuity and global identification imply $\|g(\theta)\|_W \geq \overline{\kappa}$ for some $\overline{\kappa}>0$ and all $\|\theta-\theta_0\| \geq \overline{\varepsilon}$. Consider $0 < \kappa < \overline{\kappa}$ so that $\|g(\theta)\|_W \leq \kappa$ implies $\|\theta-\theta_0\| \leq \overline{\varepsilon}$. Let $0 < \underline{\sigma} = \sigma_{\min}(\partial_\theta g(\theta_0)) \leq \sigma_{\max}(\partial_\theta g(\theta_0)) = \overline{\sigma} < \infty$. For these values of $\theta$, $\underline{\lambda}_W^{1/2}\underline{\sigma}\|\theta-\theta_0\| - \overline{\lambda}_W^{1/2}\overline{C}\|\theta-\theta_0\|^\alpha \leq \|g(\theta)\|_W \leq \overline{\lambda}_W^{1/2}\overline{\sigma}\|\theta-\theta_0\| + \overline{\lambda}_W^{1/2}\overline{C}\|\theta-\theta_0\|^\alpha$. We can further assume, without loss of generality, that $\overline{\kappa}$ and thus $\overline{\varepsilon}$ are sufficiently small that $1/2 \underline{\lambda}_W^{1/2}\underline{\sigma}\|\theta-\theta_0\| \leq \underline{\lambda}_W^{1/2}\underline{\sigma}\|\theta-\theta_0\| - \overline{\lambda}_W^{1/2}\overline{C}\|\theta-\theta_0\|^\alpha$ and $\overline{\lambda}_W^{1/2}\overline{\sigma}\|\theta-\theta_0\| + \overline{\lambda}_W^{1/2}\overline{C}\|\theta-\theta_0\|^\alpha \leq 2 \overline{\lambda}_W^{1/2}\overline{\sigma}\|\theta-\theta_0\|$. Re-write $\theta = \theta_0 + \kappa v$ for some vector $v$, then $\|v\| \leq (2 \overline{\lambda}_W^{1/2}\overline{\sigma})^{-1}$ implies $\|g(\theta)\|_W \leq \kappa$. Likewise, $\|v\| > (1/2 \underline{\lambda}_W^{1/2}\underline{\sigma})^{-1}$ implies $\|g(\theta)\|_W > \kappa$. Pick $(A,B) = (-\partial_\theta g(\theta_0) \theta_0, \partial_\theta g(\theta_0))$, then by construction: $\sup_{\|g(\theta)\|_W \leq \kappa}\|g(\theta) - A_{\kappa,\infty} - B_{\kappa,\infty}\theta\| \leq \sup_{\|g(\theta)\|_W \leq \kappa}\|g(\theta) - g(\theta_0) - \partial_\theta g(\theta_0)(\theta-\theta_0)\| \leq \sup_{\|g(\theta)\|_W \leq \kappa} \overline{C} \|\theta-\theta_0\|^\alpha \leq \overline{C} (1/2 \underline{\lambda}_W^{1/2}\underline{\sigma})^{-\alpha} \kappa^\alpha$. Pick any $\|v\| \leq (2 \overline{\lambda}_W^{1/2}\overline{\sigma} )^{-1}$, then $\|g(\theta) - A_{\kappa,\infty} - B_{\kappa,\infty} \theta \| \leq \overline{C} (1/2 \underline{\lambda}_W^{1/2}\underline{\sigma})^{-\alpha} \kappa^\alpha$ and $\|g(\theta_0) - A_{\kappa,\infty} - B_{\kappa,\infty} \theta_0 \| \leq \overline{C} (1/2 \underline{\lambda}_W^{1/2}\underline{\sigma})^{-\alpha} \kappa^\alpha$. Then $\| g(\theta) - g(\theta_0) -  B_{\kappa,\infty} [\theta- \theta_0] \| \leq 2 \overline{C} (1/2 \underline{\lambda}_W^{1/2}\underline{\sigma})^{-\alpha} \kappa^\alpha$. This implies $\| [\partial_\theta g(\theta_0) - B_{\kappa,\infty}] \kappa v \| \leq [2 \overline{C} (1/2 \underline{\lambda}_W^{1/2}\underline{\sigma})^{-\alpha} + \overline{C}  \|v\|^\alpha]\kappa^\alpha$ since $\theta-\theta_0 = \kappa v$. Then $\| [\partial_\theta g(\theta_0) - B_{\kappa,\infty}] v \| \leq [2 \overline{C} (1/2 \underline{\lambda}_W^{1/2}\underline{\sigma})^{-\alpha} + \overline{C}  \|v\|^\alpha]\kappa^{\alpha-1} \to 0$ using $\alpha>1$. Since this holds for any vector $v \neq 0$ with $\|v\| \leq (2 \overline{\lambda}_W^{1/2}\overline{\sigma} )^{-1}$, this implies that $B_{\infty} = \partial_\theta g(\theta_0)$ and, in addition, $B_{\kappa,\infty} = \partial_\theta g(\theta_0) + O(\kappa^{\alpha-1})$.\\
For the statements in the Theorem: Case 1) implies results (3), Case 2) implies result (4) and, case 3) implies result (2). For results (1), $\Theta_0$ non-singleton or, $\Theta_0$ singleton and $\partial_\theta g(\theta_0)$ singular implies $B_\infty$ singular. $\Theta_0$ singleton and $\partial_\theta g(\theta_0)$ full rank imply $B_\infty = \partial_\theta g(\theta_0)$ full rank. 
\qed

\paragraph{Proof of Proposition \ref{prop:test_ss}:} Note that Assumptions \ref{ass:moments}, \ref{ass:identification} and \ref{ass:ss} hold for the moment function $\theta_2 \to \bar{g}_n(\theta_{1n},\theta_2)$. Applying Lemma \ref{lem:asym_normal}, we have: \[\sqrt{n}H_{2n}^{-1}(\hat\theta_{2n} - \theta_{2n}) = -  \Big( \partial_{\theta_2} g(\theta_{n},\gamma_{n})^\prime V_0^{-1}\partial_{\theta_2} g(\theta_{n},\gamma_{n}) \Big)^{-1} \partial_{\theta_2} g(\theta_{n},\gamma_{n})^\prime V_0^{-1} \bar{g}_n(\theta_{n}) + o_p(1),\]
  where $H_{2n} = [\partial_{\theta_2} g(\theta_{n},\gamma_{n})^\prime \partial_{\theta_2} g(\theta_{n},\gamma_{n})]^{-1/2}$. 
By construction of the test statistic, we have: $\text{AR}_n(\theta_{1n}) = \|\bar{g}_n(\theta_{1n}\hat\theta_{2n})\|_{V_n^{-1}}^2$. We also have:
\begin{align*}
  &\bar{g}_n(\theta_{1n},\hat\theta_{2n})\\ &= \left( I -  \partial_{\theta_2} g(\theta_{n},\gamma_{n})H_{2n}\Big( H_{2n} \partial_{\theta_2} g(\theta_{n},\gamma_{n})^\prime V_0^{-1}\partial_{\theta_2} g(\theta_{n},\gamma_{n})H_{2n} \Big)^{-1} H_{2n} \partial_{\theta_2} g(\theta_{n},\gamma_{n})^\prime V_0^{-1} \right) \bar{g}_n(\theta_{n})\\ &\quad+o_p(n^{-1/2}).
\end{align*}
The leading term converges to $(I - R_{20}(R_{20}^\prime V_0^{-1}R_{20})^{-1} R_{20}^\prime V_0^{-1})$, where $\partial_{\theta_2} g(\theta_{n},\gamma_{n})H_{2n} \to R_{20}$ which has rank $d_{\theta_2}$. This limit is an orthogonal projection matrix with rank $d_g-d_{\theta_2}$. Hence, by the continuous mapping theorem: $\|\bar{g}_n(\theta_{1n}\hat\theta_{2n})\|_{V_n^{-1}}^2 \overset{d}{\to} \chi^2_{d_g- d_{\theta_2}}$.
\qed

\subsection{Proofs for Section \ref{sec:asym_jac}}

\subsubsection{Strong and semi-strong sequences.}

\paragraph{Proof of Theorem \ref{th:qJ-ss} for $B_{n,\infty}$:}
  Pick a $\varepsilon > 0$ such that Assumption \ref{ass:ss} iv. holds, then using $\kappa_n^{-1}\delta(\gamma_n) \to \infty$:
\begin{align*}
    \inf_{\|\theta-\theta_{n}\| \geq \varepsilon}\|\bar{g}_n(\theta)/\kappa_n\|_{W_n} &\geq \underline{\lambda}_W \left[ \kappa_n^{-1}\delta(\gamma_{n})h(\varepsilon) - \frac{1}{\sqrt{n}\kappa_n}\sup_{\theta \in \Theta} \sqrt{n}\|\bar{g}_n(\theta)-g(\theta,\gamma_{n})\|   \right] \\ &= \underline{\lambda}_W \kappa_n^{-1}\delta(\gamma_{n})h(\varepsilon) + o_p(1) \to +\infty,
\end{align*}
which implies $\sup_{\|\theta-\theta_{n}\| \geq \varepsilon}\hat{K}_n(\theta)=0$ wpa 1. Take $\|\theta - \theta_{n}\| \leq \varepsilon$, using Assumption \ref{ass:ss} iv. and using the change of variable $\theta = \theta_{n} + \kappa_n H_n h$ with $\| \kappa_n H_n h\| \leq \varepsilon$ we have:
\[ \|\bar{g}_n(\theta)/\kappa_n\|_{W_n} \geq \underline{\lambda}\left( \underline{C}\|\partial_\theta g(\theta_{n},\gamma_{n})H_n h\| - \frac{1}{\sqrt{n}\kappa_n}\sup_{\theta \in \Theta} \sqrt{n}\|\bar{g}_n(\theta)-g(\theta,\gamma_{n})\| \right). \]
The term on the right-hand-side is a $o_p(1)$ by assumption. The squared norm $\|\partial_\theta g(\theta_{n},\gamma_{n})H_n h\|^2 = \text{trace}\left( h^\prime H_n \partial_\theta g(\theta_{n},\gamma_{n})^\prime \partial_\theta g(\theta_{n},\gamma_{n}) H_n h\right) = \|h\|^2$ by construction of $H_n$. Hence, $\|\bar{g}_n(\theta)/\kappa_n\|_{W_n} > 1$ wpa 1 uniformly in $\|h\| \geq 2$ so that $\hat{K}_n(\theta)=0$ wpa 1.

For any $\theta$ such that $\|h\| \leq 2$, $\|\partial_\theta g(\theta_{n},\gamma_{n})(\theta-\theta_{n})\| = \kappa_n\|\partial_\theta g(\theta_{n},\gamma_{n})H_n h \| \leq 2 \kappa_n$ so that Assumption \ref{ass:ss} iv. applies with $r_n = 2\kappa_n$. For any two candidates $A,B$ we have wpa 1:
\begin{align*}
    &\sup_{\theta \in \Theta}\|\bar{g}_n(\theta) - A - B\theta\|\hat{K}_n(\theta)\\ &= \sup_{\|h\|\leq 2}\|\bar{g}_n(\theta_{n} + \kappa_n H_nh) - [A-B\theta_{n}] - \kappa_nBH_nh \|\hat{K}_n(\theta_{n} + \kappa_n H_nh)\\
    &= \sup_{\|h\|\leq 2}\|[\bar{g}_n(\theta_{n}) - A + B\theta_{n}] + \kappa_n[\partial_\theta g(\theta_{n},\gamma_{n})-B]H_nh + o_p(n^{-1/2}) + o(\kappa_n)\|\hat{K}_n(\theta_{n} + \kappa_n H_nh)\\
    &\geq \sup_{\|h\|\leq 1/4}\|[\bar{g}_n(\theta_{n}) - A + B\theta_{n}] + \kappa_n[\partial_\theta g(\theta_{n},\gamma_{n})-B]H_nh + o_p(n^{-1/2}) + o(\kappa_n)\|\underline{K},
\end{align*}
for $\inf_{x \in [0,1/2]}K(x) = \underline{K}>0$ by Assumption \ref{ass:moments} ii., using $\hat{K}_n(\theta) \geq \inf_{x \in [0,1/2]}K(x)$ wpa 1 for $\|h\|\leq 1/4$ by similar derivations as above. 

Pick $B_n = \partial_\theta g(\theta_{n},\gamma_{n})$ and $A_n = \bar{g}_n(\theta_{n}) - B_n \theta_{n}$ then $\sup_{\theta \in \Theta}\|\bar{g}_n(\theta) - A_n - B_n\theta\|\hat{K}_n(\theta) = o_p(n^{-1/2})$. By contradiction, suppose $\sqrt{n}\|A_{n,\infty} + B_{n,\infty}\| \not \to 0$ and/or $\sqrt{n}\kappa_n\|[B_n - B_{n,\infty}]H_n \| \not\to 0$, in probability. Then for any $\theta = \theta_n + \kappa_n H_n h$ with $\|h\| < 1/4$, we have wpa 1:
\begin{align*}
  &n^{1/2}\|\bar{g}_n(\theta) - A_{n,\infty} - B_{n\infty}\theta \|\hat{K}_n(\theta)\\
  &= n^{1/2}\|\bar{g}_n(\theta_n) + \kappa_n\partial_\theta g(\theta_{n},\gamma_{n})H_n h - A_{n,\infty} - B_{n,\infty}[\theta_n + \kappa_n H_n h]\|\hat{K}_n(\theta) + O_p(\sqrt{n}\kappa_n^2)\\
  &\geq n^{1/2}\|\bar{g}_n(\theta_n) + \kappa_n\partial_\theta g(\theta_{n},\gamma_{n})H_n h - A_{n,\infty} - B_{n,\infty}[\theta_n + \kappa_n H_n h]\|\underline{K} + o_p(1)\\
  &= n^{1/2}\|[\bar{g}_n(\theta_n) - A_{n,\infty} - B_{n,\infty}\theta_n] + \kappa_n[\partial_\theta g(\theta_{n},\gamma_{n})-B_{n,\infty}]H_nh \|\underline{K} + o_p(1) \not\to 0,
\end{align*}
in probability for at least one $\|h\|<1/4$ while the same quantity converges in probability to zero when evaluated at $A_n,B_n$. For instance if $\sqrt{n}\|A_{n,\infty} + B_{n,\infty}\| \not \to 0$, pick $h=0$. This contradicts the approximate minimizer property of $A_{n,\infty},B_{n,\infty}$. We conclude that $[\partial_\theta g(\theta_{n},\gamma_{n})-B_{n,\infty}]H_n = o_p(n^{-1/2}\kappa_n^{-1})$ and $A_{n,\infty} + B_{n,\infty} \theta_{n} = \bar{g}_n(\theta_{n}) + o_p(n^{-1/2})$.


\qed

\subsubsection{Weak sequences.}
\begin{definition} \label{def:Vstar} Define the span of the identification failure in the full space $\mathcal{B}$ and the constrained space $\mathcal{B}_n$ respectively as:
  \begin{align*}
    V_\star &= \text{span}\left(  (0_{d_{\beta_1}},\beta_{2}^{1\prime}-\beta_{2}^{2\prime})^\prime, \beta_{2}^{1},\beta_{2}^{2} \in \mathcal{B}_2^0 \times \mathcal{B}_2^0  \right),\\
    V_\star^0 &= \text{span}\left(  (0_{d_{\beta_1}+d_{\beta_{21}}},\beta_{22}^{1\prime}-\beta_{22}^{2\prime})^\prime, \beta_{22}^{1},\beta_{22}^{2} \in \mathcal{B}_{22}^0 \times \mathcal{B}_{22}^0  \right).
  \end{align*}

\end{definition}

\paragraph{Proof of Theorem \ref{th:weak_sup}:}
  Let $\tilde B_{n,\infty} = B_{n,\infty} M^{-1}$. After applying the reparameterization, we have:
  \begin{align*}
      \sup_{\beta \in \mathcal{B}}\| \overline{g}_n(\beta) - A_{n,\infty}-\tilde B_{n,\infty} \beta \|\hat{K}_n(\beta) &\leq \inf_{A,B}\sup_{\beta \in \mathcal{B}}\| \overline{g}_n(\beta) - A-B \beta \| \hat{K}_n(\beta) + o(\kappa_n)\\
      &\leq \sup_{\beta \in \mathcal{B}}\| \overline{g}_n(\beta) \| \hat{K}_n(\beta) + o(\kappa_n).
  \end{align*}
  For any $\beta$ such that $\hat{K}_n(\beta) >0$, we have:
  $\|\overline{g}_n(\beta)\|_{W_n} \leq \kappa_n$ and then $\|\overline{g}_n(\beta)\| \leq \underline{\lambda}_W^{-1}\kappa_n.$
  By continuity of $K$ on $[0,1]$ we have $\hat{K}_n(\beta) \leq \overline{K}$ for some constant $\overline{K} >0$ so that:
  \begin{align*}
      \| \overline{g}_n(\beta) - A_{n,\infty}-\tilde B_{n,\infty} \beta \| \hat{K}_n(\beta) &\leq \overline{K}\underline{\lambda}_W^{-1}\kappa_n + o(\kappa_n),
  \end{align*}
  for any $\beta \in \mathcal{B}$. Then, using the reverse triangular inequality:
  \begin{align*}
      \| A_{n,\infty}-\tilde B_{n,\infty} \beta \| \hat{K}_n(\beta) &\leq \| \overline{g}_n(\beta) \| \hat{K}_n(\beta) + \overline{K}\underline{\lambda}_W^{-1}\kappa_n + o(\kappa_n) \leq 2 \overline{K}\underline{\lambda}_W^{-1}\kappa_n + o(\kappa_n),
  \end{align*}
  
  By definition of $V_\star$, we can find pairs $(\beta_n^j,\tilde \beta_n^j)$ $j=1,\dots,d_{\beta_2}$ with $\beta_n^j=(\beta_{1n},\beta_2^j)$, $\tilde \beta_n^j=(\beta_{1n},\tilde \beta_2^j)$ for two $(\beta_2^j,\tilde \beta_2^j) \in \mathcal{B}_{2}^0 \times \mathcal{B}_{2}^0$ such that the vectors $v^j = \beta_n^j-\tilde \beta_n^j$, $j=1,\dots,d_{\beta_2}$ are linearly independent. By assumption, we have:
  \[ \sup_{ \beta = (\beta_{1n},\beta_2),\beta_2 \in \mathcal{B}_{2} } \|\overline{g}_n(\beta)\|_{W_n} \leq \overline{\lambda}_W\left( \sup_{\beta \in \mathcal{B}} \|\overline{g}_n(\beta)-g(\beta,\gamma_{n})\| + \sup_{\beta_2 \in \mathcal{B}_{2}^0}\|g(\beta_{1n},\beta_2,\gamma_{n})\|\right), \]
  which is a $O_p(n^{-1/2}) = o_p(\kappa_n)$. This implies that $\|\overline{g}_n(\beta)/\kappa_n\|_{W_n} \leq 1/2$ with wpa 1 uniformly in $\beta = (\beta_{1n},\beta_2),\beta_2 \in \mathcal{B}_{2}^0$ so that $\hat{K}_n(\beta) \geq \inf_{x \in [0,1/2]} K(x) = \underline{K}$ with wpa 1 uniformly on the same set. In turn, we have wpa 1 for all $j$:
  \begin{align*}
      \| A_{n,\infty}-\tilde B_{n,\infty} \beta_n^j \|  &\leq 2 \overline{K}\underline{K}^{-1}\underline{\lambda}_W^{-1}\kappa_n + o(\kappa_n),\\
      \| A_{n,\infty}-\tilde B_{n,\infty} \tilde\beta_n^j \|  &\leq 2 \overline{K}\underline{K}^{-1}\underline{\lambda}_W^{-1}\kappa_n + o(\kappa_n).
  \end{align*}
  Using the triangular inequality, we have wpa 1 and uniformly in $j$:
  \begin{align*}
      \| B_{n,\infty} M ^{-1}v^j \| = \| \tilde B_{n,\infty} v^j \| \leq \| A_{n,\infty}-\tilde B_{n,\infty} \beta^j \| + \| A_{n,\infty}-\tilde B_{n,\infty} \tilde \beta^j \|  &\leq 4 \overline{K}\underline{K}^{-1}\underline{\lambda}_W^{-1}\kappa_n + o(\kappa_n).
  \end{align*}
  Let $V = M^{-1}(v^1,\dots,v^{d_{\beta_2}})$. By linear independence, $P_V = V(V^\prime V)^{-1}V^\prime$ is well defined and:
  \[ \| B_{n,\infty} P_V \|^2 = \|B_{n,\infty}V(V^\prime V)^{-1}V^\prime\|^2\leq [\lambda_{\min}(V^\prime V)]^{-1} d_{\beta_2} [4\overline{K}\underline{K}^{-1}\underline{\lambda}_W^{-1}\kappa_n + o(\kappa_n)]^2, \]
  wpa 1. For any $v \in V_\star$, $P_Vv = v$ hence $\| B_{n,\infty} v \| \leq \| B_{n,\infty} P_v \|\, \|v\| \leq O_p(\kappa_n)$ wpa 1. To find the other two results note that $B_{n,\infty}^\prime B_{n,\infty}$ is Hermitian, and $P_V$ is an orthogonal projection matrix by construction. Hence $P_V$ admits an eigen decomposition of the form $O \text{bckdiag}(I_{d_{\beta_2}},0_{d_{\theta}-d_{\beta_2}})O^*$ with $OO^* = I_d$; $O^*$ is the conjugate transpose of $O$ and $\text{bckdiag}$ builds a block-diagonal matrix. Using this decomposition we have:
  \[ \text{trace}\left( P_V B_{n,\infty}^\prime B_{n,\infty} P_V \right) = \text{trace}\left( O_{d_{\beta_2}}^* B_{n,\infty}^\prime B_{n,\infty} O_{d_{\beta_2}} \right), \]
  where $O_{d_{\beta_2}},O_{d_{\beta_2}}^*$ are the first $d_{\beta_2}$ columns/rows of $O$ and $O^*$, respectively, which satisfy $O_{d_{\beta_2}}^* O_{d_{\beta_2}} = I_{d_{\beta_2}}$. 
  As an implication of the minimax principle \citep[][Problem III.6.11, p77]{Bhatia1997} and the equality above, we have the following inequality:
  \begin{align*}
      \sum_{j=1}^{d_{\beta_2}} \lambda_{j}(B_{n,\infty}^\prime B_{n,\infty}) &= \min_{UU^* = I_{d_{\beta_2}}} \text{trace}\left( U B_{n,\infty}^\prime B_{n,\infty} U^* \right)\\ &\leq \text{trace}\left( P_V B_{n,\infty}^\prime B_{n,\infty} P_V \right) \leq [\lambda_{\min}(V^\prime V)]^{-1} d_{\beta_2}\left[ 4 \overline{K}\underline{K}^{-1}\underline{\lambda}_W^{-1}\kappa_n + o(\kappa_n) \right]^2,
  \end{align*}
  wpa 1. This concludes the proof.
\qed

\paragraph{Proof of Proposition \ref{prop:weak_sup}:} Following the steps in the proof of Theorem \ref{th:weak_sup}, we can construct a basis for $V^0_{\star} \subseteq V_\star$ using  $v^j =(0,\theta_{22}^j-\tilde \theta_{22}^j)$ with pairs $(\theta_{22}^j,\tilde \theta_{22}^j) \in \mathcal{B}_{22}^0 \times \mathcal{B}_{22}^0$. Since $\theta_{1}=\theta_{1n}$ is fixed, we have $P_{\theta_1} P_{V_\star} = 0$ and $P_{\theta_1}^\perp P_{V_\star} = P_{V_\star}$ for the basis $V_\star = M^{-1}(v^1,\dots,v^{d_{\beta_{22}}})$. Hence, $\|B_{n,\infty}P_{\theta_1}^\perp P_{V_\star}\| \leq O_p(\kappa_n)$ and $\|B_{n,\infty}P_{\theta_1}^\perp P_{\theta_1}\| = 0$. By the minimax principle, these imply the desired inequality :
$\sum_{j=1}^{d_{\theta_1} + d_{\beta_{22}}}\lambda_{j}(P_{\theta_1}^\perp B_{n,\infty}^\prime B_{n,\infty} P_{\theta_1}^\perp ) \leq O_p(\kappa_n^2).$
\qed

\subsection{Proofs for Section \ref{sec:asym_test}}

\paragraph{Proof of Theorem \ref{th:asym_size}:} First, we show that normalizations do not affect the results of Proposition \ref{prop:weak_sup} for weak sequences. This amounts to showing that $B_{n,\infty}P_{\theta_1}^\perp \Sigma_n^{-1/2}$ has $d_{\theta_1} + d_{\beta_{22}}$ singular values that are $O_p(\kappa_n)$. 
From Proposition \ref{prop:weak_sup}, there exists a linearly independent family $V_\star= M^{-1}(v^1,\dots,v^{d_{\beta_{22}}})$ such that $MV_\star \in V_\star^0 \subseteq V^0$ (from Definition \ref{def:Vstar}) and $\|B_{n,\infty}P_{V_\star}\| = O_p(\kappa_n)$ where $P_{V_\star} = V_\star(V_\star^\prime V_\star)^{-1} V_\star^\prime$. Similarly $\|\Sigma_n^{-1/2} M^{-1} P_{V_\star}\| \leq \tilde{C}$ from Lemma \ref{lem:Sigma_weak}. Also because $I_d = P_{\theta_1} + P_{\theta_1}^\perp$ where  $P_{\theta_1} M^{-1} P_{V_\star} = 0$ by design, we have: $\|\Sigma_n^{-1/2} P_{\theta_1}^\perp M^{-1} P_{V_\star}\| \leq \tilde{C}$. 
Then using the minimax characterization of singular values \citep[][Problem III.6.5]{Bhatia1997}, we have:
\[ \sigma_{i+j-1}(B_{n,\infty}P_{\theta_1}^\perp \Sigma_n^{-1/2} P_{\theta_1}^\perp M^{-1} P_{V_\star}) \leq \sigma_{i}(B_{n,\infty}P_{\theta_1}^\perp)\sigma_{j}(\Sigma_n^{-1/2} P_{\theta_1}^\perp M^{-1}P_{V_\star}). \]
Then $\sigma_{j}(\Sigma_n^{-1/2} P_{\theta_1}^\perp M^{-1}P_{V_\star}) \leq \sigma_{\max}(\Sigma_n^{-1/2} P_{\theta_1}^\perp M^{-1} P_{V_\star}) \leq \tilde{C}$, and $\sigma_{i}(B_{n,\infty}P_{\theta_1}^\perp) \leq O_p(\kappa_n)$ for $1 \leq i \leq d_{\theta_1}+d_{\beta_{22}}$ from Proposition \ref{prop:weak_sup}. Since $P_{V_\star}$ has rank $d_{\beta_{22}}$ and is orthogonal to the rank $d_{\theta_1}$ matrix $P_{\theta_1}$ for which $B_{n,\infty}P_{\theta_1}^\perp \Sigma_n^{-1/2} P_{\theta_1}^\perp P_{\theta_1}=0$, we have that $\sigma_{\ell}(B_{n,\infty}P_{\theta_1}^\perp \Sigma_n^{-1/2} P_{\theta_1}^\perp M^{-1}) \leq O_p(\kappa_n)$ for $1 \leq \ell \leq d_{\theta_1} + d_{\beta_{22}}$, in increasing order.\footnote{Pick $i=\ell$ and $j=1$ and notice that the matrix is bounded above on a subspace of dimension $d_{\theta_1}+ d_{\beta_{22}}$.} Also $\sigma_{\min}(M^{-1})$ is strictly positive and bounded below, because $M$ is invertible, so that: $\sigma_{\ell}(B_{n,\infty}P_{\theta_1}^\perp \Sigma_n^{-1/2} P_{\theta_1}^\perp) \leq O_p(\kappa_n)$ as well. Likewise, Assumption \ref{ass:variance_mat} implies that $\sigma_{\min}(\overline{V}_n^{-1/2}) \geq \overline{\lambda}_V^{-1/2} + o(1)$ which then also implies that $\sigma_{\ell}(\overline{V}_n^{-1/2}B_{n,\infty}P_{\theta_1}^\perp \Sigma_n^{-1/2} P_{\theta_1}^\perp) \leq O_p(\kappa_n)$ for $1 \leq \ell \leq d_{\theta_1}+d_{\beta_{22}}$ as desired.

Now we are interested in establishing the asymptotic size of the test. Let $(\theta_n,\gamma_n)$ be a sequence in $\overline{\Theta} \times \Gamma$ such that 
\begin{align*}
  \limsup_{n\to\infty}\, &\mathbb{P}_{\gamma_n}\left( \text{AR}_n(\theta_{1n}) > \chi^2_{d_g - \hat d_n}(1-\alpha) \right) \\ &= \limsup_{n\to\infty} \sup_{\gamma\in\Gamma,\theta=(\theta_{10}^\prime,\theta_2^\prime)^\prime \in \overline{\Theta}} \mathbb{P}_{\gamma}\left( \text{AR}_n(\theta_{10}) > \chi^2_{d_g - \hat d_n}(1-\alpha) \right),
\end{align*}
as noted in \citet[p501]{andrews2020}, such a sequence always exists. There always exists at least one subsequence of $(\theta_n,\gamma_n)$ which achieves the $\limsup$ above, i.e. for some $\varphi_1: \mathbb{N} \to \mathbb{N}$ strictly increasing: $\lim_{n\to\infty}\mathbb{P}_{\gamma_{\varphi_1(n)}}\left( \text{AR}_{\varphi_1(n)}(\theta_{1 \varphi_1(n)}) > \chi^2_{d_g - \hat d_{\varphi_1(n)}}(1-\alpha) \right) = \limsup_{n\to\infty}\mathbb{P}_{\gamma_n}\left( \text{AR}_n(\theta_{1n}) > \chi^2_{d_g - \hat d_n}(1-\alpha) \right)$. Assumption \ref{ass:moments} i. implies that $\overline{\Theta}\times \Gamma$ is sequentially compact so that this subsequence admits a convergence sub-subsequence in $\overline{\Theta} \times \Gamma$, i.e. for some $\varphi_2: \mathbb{N} \to \mathbb{N}$ strictly increasing: $(\theta_{\varphi_2 \circ \varphi_1 (n)},\gamma_{\varphi_2 \circ \varphi_1 (n)}) \to (\theta_0,\gamma_0) \in \overline{\Theta} \times \Gamma$ and $\lim_{n\to\infty}\mathbb{P}_{\gamma_{\varphi_2 \circ \varphi_1(n)}}\left( \text{AR}_{\varphi_2 \circ \varphi_1(n)}(\theta_{1 \varphi_2 \circ \varphi_1(n)}) > \chi^2_{d_g - \hat d_{\varphi_2 \circ \varphi_1(n)}}(1-\alpha) \right)$ has the same limit. 

Now, if we can find a converging sequence  $(\theta_m,\gamma_m)$, $m\geq 1$, in one of $\Gamma_0(b)$, for some $b \geq 0$, $\Gamma_0(\infty)$, or converging in $\Gamma_1$ such that $(\theta_m,\gamma_m) = (\theta_{\varphi_2 \circ \varphi_1(n)},\gamma_{\varphi_2 \circ \varphi_1(n)})$ when $m = \varphi_2 \circ \varphi_1(n)$ then the limiting rejection probability for the subsequence can be derived from the limiting rejection probability of the full sequence $(\theta_m,\gamma_m)$. Suppose $(\theta_{\varphi_2 \circ \varphi_1(n)},\gamma_{\varphi_2 \circ \varphi_1(n)}) \to (\theta_0,\gamma_0) \in \overline{\Theta} \times \Gamma_1$. Pick $(\theta_m,\gamma_m) = (\theta_{\varphi_2 \circ \varphi_1(n)},\gamma_{\varphi_2 \circ \varphi_1(n)})$ when $m = \varphi_2 \circ \varphi_1(n)$ and $(\theta_m,\gamma_m) = (\theta_0,\gamma_0)$ otherwise. $(\theta_m,\gamma_m)$ is a converging sequence with $\gamma_0 \in \Gamma_1$. If $(\theta_0,\gamma_0) \in \overline{\Theta} \times \Gamma_0$, then $\sqrt{\varphi_2 \circ \varphi_1(n)}\delta(\gamma_{\varphi_2 \circ \varphi_1(n)})$ is a sequence taking values in $[0,+\infty) \cup \{+\infty\}$, the positive part of the extended real line which is a compact space. This implies that $\sqrt{\varphi_2 \circ \varphi_1(n)}\delta(\gamma_{\varphi_2 \circ \varphi_1(n)})$ admits at least one subsequence $\sqrt{\varphi_3 \circ \varphi_2 \circ \varphi_1(n)}\delta(\gamma_{\varphi_3 \circ \varphi_2 \circ \varphi_1(n)})$ which converges in $[0,+\infty) \cup \{+\infty\}$. Let $\varphi = \varphi_3 \circ \varphi_2 \circ \varphi_1$ index the resulting subsequence of $(\theta_n,\gamma_n)$. There are now two possibilities: either $\sqrt{\varphi(n)}\delta(\gamma_{\varphi(n)}) \to \infty$ or $\sqrt{\varphi(n)}\delta(\gamma_{\varphi(n)}) \to b \in [0,\infty)$.

Suppose $\sqrt{\varphi(n)}\delta(\gamma_{\varphi(n)}) \to \infty$. Pick $(\theta_m,\gamma_m) = (\theta_{\varphi(n)},\gamma_{\varphi(n)})$ when $m = \varphi(n)$. For $\varphi(n) < m < \varphi(n+1)$, pick $(\theta_m,\gamma_m) = (\theta_{\varphi(n)},\gamma_{\varphi(n)})$ as well. By construction $\sqrt{m}\delta(\gamma_m) = \sqrt{m}\delta(\gamma_{\varphi(n)}) > \sqrt{\varphi(n)}\delta(\gamma_{\varphi(n)}) \to \infty$ so that $\gamma_m \in \Gamma_0(\infty)$. By construction $\hat d_{n} \in \{0,\dots,d_{\theta_2}\}$, hence Proposition \ref{prop:test_ss} implies that \[\mathbb{P}_{\gamma_m}\left(  \text{AR}_m(\theta_{1m}) > \chi^2_{d_g - \hat d_m}(1-\alpha) \right) \leq \mathbb{P}_{\gamma_m}\left(  \text{AR}_m(\theta_{1m}) > \chi^2_{d_g - d_{\theta_2}}(1-\alpha) \right) \to \alpha,\] for any $\alpha\in (0,1)$. This in turn implies that:
\begin{align*}
  \lim_{n\to \infty} &\mathbb{P}_{\gamma_{\varphi(n)}}\left( \text{AR}_{\varphi(n)}(\theta_{1 \varphi(n)}) > \chi^2_{d_g - \hat d_{\varphi(n)}}(1-\alpha) \right)\\ &= \lim_{m \to \infty} \mathbb{P}_{\gamma_m}\left(  \text{AR}_m(\theta_{1m}) > \chi^2_{d_g - \hat d_m}(1-\alpha) \right) \leq \alpha.
\end{align*}
Suppose $\sqrt{\varphi(n)}\delta(\gamma_{\varphi(n)}) \to b \in [0,\infty)$. Pick  $(\theta_m,\gamma_m) = (\theta_{\varphi(n)},\gamma_{\varphi(n)})$ for any $m = \varphi(n)$. For $\varphi(n) < m < \varphi(n+1)$, define $b_{m} = \min[\sqrt{\varphi(n)}\delta(\gamma_{\varphi(n)}), \sqrt{\varphi(n+1)}\delta(\gamma_{\varphi(n+1)})]$; note that $\lim_{m\to\infty} b_m = b$. Suppose, without loss of generality, that $b_m = \sqrt{\varphi(n)}\delta(\gamma_{\varphi(n)})$. Take $\varepsilon = d(\gamma_0,\gamma_{\varphi(n)})$. If $\varepsilon =0$, then $\gamma_{\varphi(n)} = \gamma_0 \in \Gamma_0$ and $b_m=0$. If $b_m=0$, pick $(\theta_m,\gamma_m) = (\theta_{\varphi(n)},\gamma_{\varphi(n)})$. If $\varepsilon>0$ and $b_m >0$, Assumption \ref{ass:moments} i. implies that the closure of $B_{\varepsilon}(\gamma_0)\cap \Gamma$ is connected. Hence, there exists a continuous map: $(\theta,\gamma): [0,1] \to \overline{\Theta}\times \Gamma$ such that $(\theta(0),\gamma(0)) = (\theta_0,\gamma_0)$ and $(\theta(1),\gamma(1))=(\theta_{\varphi(n)},\gamma_{\varphi(n)})$ and $\|\theta(u)-\theta_0\| + d(\gamma(u)-\gamma(0)) \leq \varepsilon$ for any $u \in [0,1]$. By continuity of $\delta: \Gamma \to \mathbb{R}_+$, the image of $u \to \delta \circ \gamma(u)$ is a closed interval which contains $0 = \delta(\gamma_0)$ and $\delta(\gamma_{\varphi(n)})>0$. For each $m$, the values $0$ and $b_m$ are both contained in the image $\sqrt{m}[\delta \circ \gamma([0,1])]$, so that there exists a $u_m$ such that $\sqrt{m}\delta \circ \gamma(u_m) = b_m$. Pick $(\theta_m,\gamma_m) = (\theta(u_m),\gamma(u_m))$. If $b_m$ is attained at $\varphi(n+1)$, repeat the above with $\varphi(n+1)$ instead of $\varphi(n)$. By construction $\|\theta_m - \theta_0\| + d(\gamma_m,\gamma_0) \leq \max[\|\theta_{\varphi(n)} - \theta_0\| + d(\gamma_{\varphi(n)},\gamma_0),\|\theta_{\varphi(n+1)} - \theta_0\| + d(\gamma_{\varphi(n+1)},\gamma_0)] \to 0$ and $\lim_{m \to \infty} \sqrt{m}\delta \circ \gamma(u_m) = \lim_{m\to\infty} b_m = b \in [0,\infty)$. This implies that $\gamma_m \in \Gamma_0(b)$ with $b \in [0,\infty)$. As shown above, for this converging sequence we have $\hat d_m \leq d_{\phi}$ wpa 1. Using Proposition \ref{prop:test_weak}: 
\[\mathbb{P}_{\gamma_m}\left(  \text{AR}_m(\theta_{1m}) > \chi^2_{d_g - \hat d_m}(1-\alpha) \right) \leq \mathbb{P}_{\gamma_m}\left(  \text{AR}_m(\theta_{1m}) > \chi^2_{d_g - d_{\phi}}(1-\alpha) \right) + o(1) \to \alpha.\]
Then, we have: $\lim_{n\to \infty} \mathbb{P}_{\gamma_{\varphi(n)}}\left( \text{AR}_{\varphi(n)}(\theta_{1 \varphi(n)}) > \chi^2_{d_g - \hat d_{\varphi(n)}}(1-\alpha) \right) \leq \alpha$. Putting everything together, we have: $\limsup_{n\to\infty}\mathbb{P}_{\gamma_n}\left( \text{AR}_n(\theta_{1n}) > \chi^2_{d_g - \hat d_n}(1-\alpha) \right) \leq \alpha$ for the original sequence $(\theta_n,\gamma_n)$.

For the second part of the Theorem, note that $\kappa^2_n = o(\underline{\lambda}_n^2) = o(\lambda_{\min}(\partial_\theta g(\theta_n,\gamma_n)^\prime \partial_\theta g(\theta_n,\gamma_n))$ so Theorem \ref{th:qJ-ss} applies. Now, from the proof of Lemma \ref{lem:Sigma_ss}: $\kappa_n^{-2} H_n^{1/2}\Sigma_n H_n^{1/2} \overset{p}{\to} \tilde \Sigma,$
which is the arg-minimizer of the limiting sup-norm minimization and is non-singular because of the log-determinant. Hence, $\lambda_{\min}(\Sigma_n^{-1/2}) \geq \kappa_n^{-1} \lambda_{\min}(H_n^{-1}) \lambda_{\min}(\tilde \Sigma) + o_p(\kappa_n^{-1})$ . Now, this implies:
\begin{align*}
  \lambda_{d_{\theta_1}+1}&( P_{\theta_1}^\perp \Sigma_n^{-1/2} P_{\theta_1}^\perp B_{n,\infty}^\prime \overline{V}_n^{-1} B_{n,\infty} P_{\theta_1}^\perp \Sigma_n^{-1/2} P_{\theta_1}^\perp )\\ &\geq \lambda_{\min}(\Sigma_n^{-1/2})^2 \lambda_{d_{\theta_1}+1}(P_{\theta_1}^\perp B_{n,\infty}^\prime \overline{V}_n^{-1} B_{n,\infty} P_{\theta_1}^\perp)\\ &\geq \underline{\lambda}_V^{-1}[\lambda_{\min}(\Sigma_n^{-1/2})]^2 \lambda_{\min}(B_{n,\infty}^\prime  B_{n,\infty})(1+o_p(1))\\ &\geq \underline{\lambda}_V^{-1}[\lambda_{\min}(\Sigma_n^{-1/2})]^2 \lambda_{\min}(\partial_\theta g(\theta_n,\gamma_n)^\prime \partial_\theta g(\theta_n,\gamma_n))(1+o_p(1)),
\end{align*}
where the last inequality follows from Assumption \ref{ass:variance_mat} and the discussion after Theorem \ref{th:qJ-ss}. Since $\lambda_{\min}(\Sigma_n^{-1/2})$ is bounded below, we have $\lambda_{d_{\theta_1}+1}( P_{\theta_1}^\perp \Sigma_n^{-1/2} P_{\theta_1}^\perp B_{n,\infty}^\prime \overline{V}_n^{-1} B_{n,\infty} P_{\theta_1}^\perp \Sigma_n^{-1/2} P_{\theta_1}^\perp ) > \underline{\lambda}_n^2$ wpa 1. This implies $\hat d_n = d_{\theta_2}$ wpa 1 and:
\[ \mathbb{P}_{\gamma_n}( \text{AR}_n(\theta_{1n})>\chi^2_{d_g-\hat d_n}(1-\alpha)) = \mathbb{P}_{\gamma_n}( \text{AR}_n(\theta_{1n})>\chi^2_{d_g-d_{\theta_2}}(1-\alpha)) + o(1) \to \alpha, \]
which concludes the proof.
\qed

\newpage
\begin{titlingpage} 
  \emptythanks
  \title{ {Supplement to\\ \lQ Detecting Identification Failure   in\\  Moment Condition Models''}}
  \author{Jean-Jacques Forneron\thanks{Department of Economics, Boston University, 270 Bay State Rd, MA 02215 Email: jjmf@bu.edu}  }
  \setcounter{footnote}{0}
  \setcounter{page}{0}

  \clearpage 
  \maketitle 
  \thispagestyle{empty} 
  \begin{center}
  This Supplemental Material consists of Appendices \ref{apx:proof_prelim}, \ref{apx:PMC}, \ref{sec:code}, \ref{apx:conds_NLS}, and \ref{apx:ho} to the main text.
  \end{center}
\end{titlingpage}

\setcounter{page}{1}

\section{Proofs for the preliminary results} \label{apx:proof_prelim}
\subsection{Preliminary results for Section \ref{sec:setting}}
\paragraph{Proof of Lemma \ref{lem:consistency}:}
  First, using $(a-b)^2 \geq a^2/2 - b^2$ for any $(a,b)\in \mathbb{R}^2$ we have:
  \begin{align*}
     &\bar{g}_n(\theta)^\prime W_n(\theta) \bar{g}_n(\theta)\\ &\geq \frac{1}{2} \|g(\theta,\gamma_{n})\|^2(\underline{\lambda}_W + \|W_n(\theta)-W(\theta)\|) -  \|\bar{g}_n(\theta)-g(\theta,\gamma_{n})\|^2(\bar{\lambda}_W + \|W_n(\theta)-W(\theta)\|) \\
     &\geq \frac{1}{2} \|g(\theta,\gamma_{n})\|^2(\underline{\lambda}_W + o_p(1)) -  O_p(n^{-1}),
  \end{align*}
  uniformly in $\theta \in \Theta$. The second inequality is:
  \[ \bar{g}_n(\theta_{n})^\prime W_n(\theta_{n}) \bar{g}_n(\theta_{n}) = \|\bar{g}_n(\theta_{n}) -g(\theta_{n},\gamma_n)\|^2_{W_n} \leq O_p(n^{-1}), \]
  since $g(\theta_{n},\gamma_n)=0$. Pick any $\varepsilon >0$.  For any approximate minimizer $\hat\theta_n$ such that $\|\bar{g}_n(\hat\theta_n)\|^2_{W_n} \leq \inf_{\theta \in \Theta} \|\bar{g}_n(\theta)\|^2_{W_n} + o(n^{-1})$, using the two inequalities above:
  \begin{align*}
     \mathbb{P}\left( \|\hat\theta_n-\theta_{n}\| \geq \varepsilon \right) &\leq \mathbb{P}\left( \inf_{\|\theta-\theta_{n}\| \geq \varepsilon } \|\bar{g}_n(\theta)\|^2_{W_n} \leq \|\bar{g}_n(\theta_{n})\|^2_{W_n} + o(n^{-1})\right)\\
     &\leq  \mathbb{P}\left( \frac{1}{2} \inf_{\|\theta-\theta_{n}\| \geq \varepsilon } \|g(\theta,\gamma_{n})\|^2(\underline{\lambda}_W + o_p(1)) \leq O_p(n^{-1})\right)\\
     &\leq  \mathbb{P}\left(  [\sqrt{n}\delta(\gamma_{n})]^2\leq O_p(1)\right) \to 0,
  \end{align*}
 since $\sqrt{n}\delta(\gamma_{n}) \to \infty$ for sequences converging in $\Gamma_0(\infty)$ or $\Gamma_1$. 
\qed

\paragraph{Proof of Lemma \ref{lem:asym_normal}:} For any approximate minimizer $\hat\theta_n$ such that $\|\bar{g}_n(\hat\theta_n)\|^2_{W_n} \leq \inf_{\theta \in \Theta} \|\bar{g}_n(\theta)\|^2_{W_n} + o(n^{-1})$, we have $\|\hat\theta_n - \theta_n\|=o_p(1)$ by Lemma \ref{lem:consistency} and:
  \begin{align*}
    o(1) &\geq n\left[\bar{g}_n(\hat\theta_n)^\prime W_n(\hat\theta_n)\bar{g}_n(\hat\theta_n) - \bar{g}_n(\theta_{n})^\prime W_n(\theta_{n})\bar{g}_n(\theta_{n})\right]\\
      &= n\left[\bar{g}_n(\hat\theta_n)^\prime W(\theta_{n})\bar{g}_n(\hat\theta_n) - \bar{g}_n(\theta_{n})^\prime W(\theta_{n})\bar{g}_n(\theta_{n})\right]\left(1+o_p(1)\right)\\
      &= n\Big[(\bar{g}_n(\theta_{n})+g(\hat\theta_n,\gamma_{n})-g(\theta_{n},\gamma_{n}))^\prime W(\theta_{n})(\bar{g}_n(\theta_{n})+g(\hat\theta_n,\gamma_{n})-g(\theta_{n},\gamma_{n})) \\ &\quad- \bar{g}_n(\theta_{n})^\prime W(\theta_{n})\bar{g}_n(\theta_{n})\Big]\left(1+o_p(1)\right)\\
      &= n\Big[2\bar{g}_n(\theta_{n})^\prime W(\theta_{n})(g(\hat\theta_n,\gamma_{n})-g(\theta_{n},\gamma_{n})) \\ &\quad+ (g(\hat\theta_n,\gamma_{n})-g(\theta_{n},\gamma_{n}))^\prime W(\theta_{n})(g(\hat\theta_n,\gamma_{n})-g(\theta_{n},\gamma_{n}))\Big]\left(1+o_p(1)\right)\\
      &= n\Big[2\bar{g}_n(\theta_{n})^\prime W(\theta_{n})\partial_\theta g(\theta_{n},\gamma_{n})(\hat\theta_n-\theta_{n}) \\ &\quad+ [\partial_\theta g(\theta_{n},\gamma_{n})(\hat\theta_n-\theta_{n})]^\prime W(\theta_{n})[\partial_\theta g(\theta_{n},\gamma_{n})(\hat\theta_n-\theta_{n})]\Big]\left(1+o_p(1)\right).
  \end{align*}
  By assumption, $W(\theta_{n}) \to W(\theta_0)$ positive definite so the above implies:
  \[ n\|\partial_\theta g(\theta_{n},\gamma_{n})(\hat\theta_n-\theta_{n})\|^2 \leq o(1) + \sqrt{n}\|\partial_\theta g(\theta_{n},\gamma_{n})(\hat\theta_n-\theta_{n})\|O_p(1). \]
  As in \citet{Newey1994a} completing the square above implies $[\sqrt{n}\|\partial_\theta g(\theta_{n},\gamma_{n})(\hat\theta_n-\theta_{n})\| + O_p(1) ]^2 \leq O(1)$. Taking the square root on both sides yields:
  \[\sqrt{n}\|\partial_\theta g(\theta_{n},\gamma_{n})(\hat\theta_n-\theta_{n})\| = O_p(1).\] 
  Define $\tilde \theta_n = \theta_{n} - \Big( \partial_\theta g(\theta_{n},\gamma_{n})^\prime W(\theta_{n})\partial_\theta g(\theta_{n},\gamma_{n}) \Big)^{-1} \partial_\theta g(\theta_{n},\gamma_{n})^\prime W(\theta_{n}) \bar{g}_n(\theta_{n}).$   By continuity of $\partial_\theta g(\theta,\gamma)$ and $W$, we have: $\sqrt{n}H_n^{-1}(\tilde{\theta}_n - \theta_n) = (R_0^\prime W_0 R_0)^{-1} R_0^\prime \sqrt{n}\bar{g}_n(\theta_n) + o_p(1) \overset{d}{\to} \mathcal{N}(0,\Sigma_0)$. To conclude the proof we need to prove that $\sqrt{n}H_n^{-1}(\tilde{\theta}_n - \hat\theta_n) =o_p(1)$. Using similar calculations as above, we have:
  \begin{align*}
      &n\left[\bar{g}_n(\tilde\theta_n)^\prime W_n(\tilde\theta_n)\bar{g}_n(\tilde\theta_n) - \bar{g}_n(\theta_{n})^\prime W_n(\theta_{n})\bar{g}_n(\theta_{n})\right]\\
      &= n\Big[2\bar{g}_n(\theta_{n})^\prime W(\theta_{n})\partial_\theta g(\theta_{n},\gamma_{n})(\tilde\theta_n-\theta_{n}) \\ &\quad+ [\partial_\theta g(\theta_{n},\gamma_{n})(\tilde\theta_n-\theta_{n})]^\prime W(\theta_{n})[\partial_\theta g(\theta_{n},\gamma_{n})(\tilde\theta_n-\theta_{n})]\Big]\left(1+o_p(1)\right).
  \end{align*}
  By construction of $\tilde\theta_n$,
  $-\partial_\theta g(\theta_{n},\gamma_{n})^\prime W(\theta_{n})\bar{g}_n(\theta_{n}) = \left( \partial_\theta g(\theta_{n},\gamma_{n})^\prime W(\theta_{n}) \partial_\theta g(\theta_{n},\gamma_{n}) \right)\left( \tilde \theta_n - \theta_{n} \right).$
  This implies the following equalities:
  \begin{align*}
      \bar{g}_n(\theta_{n})^\prime W(\theta_{n})\partial_\theta g(\theta_{n},\gamma_{n})(\tilde\theta_n-\theta_{n}) &= \left( \tilde \theta_n - \theta_{n} \right)^\prime \left( \partial_\theta g(\theta_{n},\gamma_{n})^\prime W(\theta_{n}) \partial_\theta g(\theta_{n},\gamma_{n}) \right)\left( \tilde \theta_n - \theta_{n} \right),\\
      \bar{g}_n(\theta_{n})^\prime W(\theta_{n})\partial_\theta g(\theta_{n},\gamma_{n})(\hat\theta_n-\theta_{n}) &= \left( \tilde \theta_n - \theta_{n} \right)^\prime \left( \partial_\theta g(\theta_{n},\gamma_{n})^\prime W(\theta_{n}) \partial_\theta g(\theta_{n},\gamma_{n}) \right)\left( \hat \theta_n - \theta_{n} \right).
  \end{align*}
  Since $\hat\theta_n$ is an approximate minimizer, we have:
  \begin{align*}
      o(1) &\geq n\left[\bar{g}_n(\hat\theta_n)^\prime W_n(\hat\theta_n)\bar{g}_n(\hat\theta_n) - \bar{g}_n(\tilde\theta_n)^\prime W_n(\tilde\theta_n)\bar{g}_n(\tilde\theta_n)\right]\\
      &= n\left[\bar{g}_n(\hat\theta_n)^\prime W_n(\hat\theta_n)\bar{g}_n(\hat\theta_n) - \bar{g}_n(\theta_{n})^\prime W_n(\theta_{n})\bar{g}_n(\theta_{n})\right]\\
      &\quad-n\left[\bar{g}_n(\tilde\theta_n)^\prime W_n(\tilde\theta_n)\bar{g}_n(\tilde\theta_n) - \bar{g}_n(\theta_{n})^\prime W_n(\theta_{n})\bar{g}_n(\theta_{n})\right]\\
      &= n\Big[[\partial_\theta g(\theta_{n},\gamma_{n})(\hat\theta_n-\theta_{n})]^\prime W(\theta_{n})[\partial_\theta g(\theta_{n},\gamma_{n})(\hat\theta_n-\theta_{n})] \\&\quad+ [\partial_\theta g(\theta_{n},\gamma_{n})(\tilde\theta_n-\theta_{n})]^\prime W(\theta_{n})[\partial_\theta g(\theta_{n},\gamma_{n})(\tilde\theta_n-\theta_{n})] \\&\quad-2\left( \tilde \theta_n - \theta_{n} \right)^\prime \left( \partial_\theta g(\theta_{n},\gamma_{n})^\prime W(\theta_{n}) \partial_\theta g(\theta_{n},\gamma_{n}) \right)\left( \hat \theta_n - \theta_{n} \right)\Big]\left(1+o_p(1)\right)\\
      &\geq n \underline{\lambda}_W \|\partial_\theta g(\theta_{n},\gamma_{n})(\hat\theta_n-\tilde\theta_{n})\|^2\left(1+o_p(1)\right),
  \end{align*}
  which implies $\sqrt{n}\partial_\theta g(\theta_{n},\gamma_{n})(\hat\theta_n-\tilde\theta_{n}) = o_p(1)$ and concludes the proof.
 \qed

\newpage

\section{Supplemental Results} \label{apx:suppl_results}
The following results concern the matrix $\Sigma_n$ used for re-scaling in the procedure. The derivations follow very closely those in Theorems \ref{th:qJ-ss} and \ref{th:weak_sup}.
\begin{lemma} \label{lem:Sigma_ss}
  Suppose $K$ is the uniform kernel and the Assumptions for Theorem \ref{th:qJ-ss} hold, then $H_n^{1/2} \Sigma_n H_n^{1/2} \leq O_p(\kappa_n^2)$.
\end{lemma}

\paragraph{Proof of Lemma \ref{lem:Sigma_ss}.}
As in the proof of Theorem \ref{th:qJ-ss}, let $\theta = \theta_n + \kappa_n H_n h$. Take $\Sigma_n = \kappa_n^2 H_n^{-1/2}\tilde \Sigma_n H_n^{-1/2}$ for $\tilde \Sigma_n \geq 0$, $\mu_n =  \theta_n + \kappa_n H_n \tilde h_n$. Then wpa 1:
\begin{align*}
  (\mu_n,\tilde\Sigma_n) &= \argmin_{\tilde h,\tilde \Sigma} [\sup_{ \|h\|\leq 2 } \|  h - \tilde h \|^2_{\tilde \Sigma^{-1}} - \log|\tilde\Sigma^{-1}| - \log|\kappa_n H_n^{-1}|]\hat{K}_n(\theta_n + \kappa_n H_n h )\\
  &\to  \argmin_{\tilde h,\tilde \Sigma} [\sup_{ \|h\|\leq 2 } \|  h - \tilde h \|^2_{\tilde \Sigma^{-1}} - \log|\tilde\Sigma^{-1}| ]K( \|R_0 h\| ),
\end{align*}
using the argmax Theorem and taking the p-limit on the right-hand-side. The $- \log|\kappa_n H_n^{-1}|$ term can be removed because $\hat{K}_n \in \{0,1\}$ for the uniform kernel so it does not alter the supremum and the infimum. This implies that $\kappa_n^{-2} H_n^{1/2} \Sigma_n H_n^{1/2} = \tilde \Sigma_n = O_p(1)$ as desired. 
\qed

\begin{lemma} \label{lem:Sigma_weak}
Suppose $K$ is the uniform kernel and the Assumptions for Theorem \ref{th:weak_sup} hold, then there exists $C >0$ such that $\| \Sigma_n^{-1/2} M^{-1} v \|_2 \leq C,$ wpa 1 for any $v = (0,\beta_2^1-\beta_2^2)$ with $\beta_2^1,\beta_2^2 \in \mathcal{B}_2^0$. This implies that $\|\Sigma_n^{-1/2} M^{-1} P_{V_2}\| \leq \tilde C$, wpa 1 for some finite constant $\tilde C$.
\end{lemma}
\paragraph{Proof of Lemma \ref{lem:Sigma_weak}.} Pick $v \neq 0$ as stated in the Lemma. Let $\beta^1 = (\beta_{1n},\beta_2^1)$,$\beta^2 = (\beta_{1n},\beta_2^2)$ so that $v = \beta^1-\beta^2$. Because $K$ is the uniform Kernel, after a change of variable,  $\tilde\mu_n,\tilde\Sigma_n = M^{-1}\mu_n,M^{'-1}\Sigma_n M^{-1}$ are the minimizers of:
\begin{align*}
 &\inf_{\tilde\mu,\tilde\Sigma} \sup_{\hat{K}_n(\beta)=1} \|\beta-\tilde\mu\|^2_{\tilde\Sigma^{-1}} - \|\beta^1-\tilde\mu\|^2_{\tilde\Sigma^{-1}} \leq 2\sup_{\beta} \|\beta\|^2_2,\\
 &\inf_{\tilde\mu,\tilde\Sigma} \sup_{\hat{K}_n(\beta)=1} \|\beta-\tilde\mu\|^2_{\tilde\Sigma^{-1}} - \|\beta^2-\tilde\mu\|^2_{\tilde\Sigma^{-1}} \leq 2\sup_{\beta} \|\beta\|^2_2,
\end{align*}
wpa 1, because $\hat{K}_n(\beta^1) = \hat{K}_n(\beta^2) = 1$ wpa 1 under the Assumptions, and the infimum is less than for $(\tilde \mu,\tilde \Sigma)=(0,I)$. 
This implies that $| \|\beta^1-\tilde\mu_n\|^2_{\Sigma_n^{-1}} - \|\beta^2-\tilde\mu_n\|^2_{\tilde \Sigma_n^{-1}} | \leq 2\sup_{\beta} \|\beta\|$. Using $\|\beta^1-\tilde\mu_n\|^2_{\tilde \Sigma_n^{-1}} = \|v\|^2_{\tilde \Sigma_n^{-1}} + \|\beta^2-\tilde\mu_n\|^2_{\tilde \Sigma_n^{-1}} + 2 \langle \tilde\Sigma_n^{-1/2} v, \tilde\Sigma_n^{-1/2}(\beta_2-\tilde\mu_n)\rangle$, we have $| \|\beta^1-\tilde\mu_n\|^2_{\tilde \Sigma_n^{-1}} - \|\beta^2-\tilde\mu_n\|^2_{\tilde \Sigma_n^{-1}} | = | \|v\|^2_{\tilde \Sigma_n^{-1}} + 2 \langle \tilde\Sigma_n^{-1/2} v, \tilde\Sigma_n^{-1/2}(\beta_2-\tilde\mu_n)\rangle | = | \|v\|^2_{\tilde \Sigma_n^{-1}} - 2 \langle \tilde\Sigma_n^{-1/2} v, \tilde\Sigma_n^{-1/2}(\beta_1-\tilde\mu_n)\rangle |$. Apply the triangular inequality to find wpa 1: $\|v\|^2_{\tilde \Sigma_n^{-1}} \leq 4 \sup_{\|\beta\|^2_2} :=  C.$
To get the first inequality, note that $\|v\|^2_{\tilde \Sigma_n^{-1}} = \| v^\prime M^{'-1} \Sigma_n^{-1/2} \Sigma_n^{-1/2} M^{-1} v\|_2 = \|\Sigma_n^{-1/2} M^{-1} v\|_2^2$. The second inequality, can be derived using the same steps used in the proof of Theorem \ref{th:weak_sup} and the minimax principle.
\qed
\newpage

\section{Linear Reparameterization, Continued} \label{apx:linpar}

The following gives additional details about the linear reparameterization in Section \ref{sec:setting}, and describes the additional steps to use when there are multiple sources of identification. To simplify the discussion, it will focus on two specific examples. 

The main idea is that if there are multiple but finitely many sources of identification failure, we can construct a finite partition of $\mathcal{B}_2$ where each subset is associated with a common rate (semi-strong, weak). Then, refine the reparameterization by using only the subset(s) corresponding to weak identification. 
When there is a single (scalar) source of identification failure, the partition $\beta_1,\beta_2$ presented in the main text systematically has $\beta_2$ weakly identified for weak sequences because the objective function becomes flat at the same rate on the entire set $\mathcal{B}_2$. The partition only has one element which is $\mathcal{B}_2$ itself.

\paragraph*{Example 1: Linear IV regression} First, consider the linear IV regression, now with multiple instruments. Let $y_i = x_i^\prime \theta_0 + u_i$ with moment condition $\mathbb{E}_\gamma( z_i [y_i - x_i^\prime \theta] )=0$. It can be re-written as $\mathbb{E}_\gamma( z_i x_i^\prime [\theta_0 -  \theta] )=0$, if $\mathbb{E}_\gamma(z_iu_i)=0$. As seen from the discussion of Assumption \ref{ass:identification}, identification fails for any $\gamma_0$ such that $\sigma_{\min}[\mathbb{E}_{\gamma_0}( z_i x_i^\prime)]=0$. Because the moment condition is linear in $\theta$ for this example, the linear reparameterization described in Section \ref{sec:setting} is such that $V_2 = \text{kern}[\mathbb{E}_{\gamma_0}( z_i x_i^\prime)]$, where $\text{kern}$ is the kernel, or null space, of the matrix. When the matrix has full rank $V_2 = \{0\}$ and the solution $\theta_0$ is unique.

Consider sequences $\gamma_n \to \gamma_0$ such that $\mathbb{E}_{\gamma_n}( z_i x_i^\prime) = U \Lambda_n V^\prime$ where $\Lambda_n = \text{diag}(\lambda_{1n},\dots,\lambda_{kn})$ is diagonal, and $U,V$ are semi-unitary: $U^\prime U = V^\prime V = I_k$. The span $V_2$ covers directions associated with the singularity, i.e. all columns $V_j$, $j \in \{1,\dots,k\}$, of $V$ where $\lim_{n\to\infty} \lambda_{jn} = \lambda_{j0}=0$. Consider only sequences such that the limit $b_j = \lim_{n\to\infty} \sqrt{n}\lambda_{jn}$ exists in $\mathbb{R}_+ \cup \{+\infty\}$.\footnote{Note that $\sqrt{n}\lambda_{jn}$ takes values in the extended real line $\mathbb{R}_+ \cup \{+\infty\}$ which is compact so we can always find a converging subsequence in the extended real line. This step appears in the proof of Theorem \ref{th:asym_size}.} Split the indices in two sets: $J_1 = \{ 1 \leq j \leq k, b_j = +\infty\}$ and $J_2 = \{ 1 \leq j \leq k, b_j < +\infty\}$. Clearly $J_1 \cap J_2 = \emptyset$ and $J_1 \cup J_2 = \{1,\dots,k\}$. Take $V_2$ to be the span associated with the columns $V_j$, $j \in J_2$. Then complete the reparameterization by taking $\beta_1$ in the orthogonal of $V_2$. Since the reparameterization is defined up to a rotation, suppose for simplicity that $V$ is ordered such that $\beta = V^\prime \theta$ and, note that: $\|g(\beta,\gamma_n)\|^2 = (\beta_n-\beta)^\prime \Lambda_{n} (\beta_n-\beta)$, where $\beta_n = V^\prime \theta_n$. Assumption \ref{ass:weak} can now be verified from this representation.  Here the sources of identification failure are indexed by the singular values $\lambda_j$, $j \in \{1,\dots,k\}$, and the parameter space is partitioned into $k$ different directions: $V_j^\prime \theta$, $j\in\{1,\dots,k\}$, associated with the $\lambda_j$.

\paragraph*{Example 2: Non-Linear regression} Consider the regression setup in \citet{Cheng2015}: $y_i  = \sum_{j=1}^k g_j(x_i,\pi_j) \delta_j + w_i^\prime \xi + u_i$, $\theta = (\pi,\delta,\xi)$, each $\delta_j$ here is scalar. The coefficient $\pi_j$ is unidentified if the corresponding $\delta_j =0$. This is related to the example used in Section \ref{sec:MonteCarlo}. For a vector of instruments $z_i$, take the moment condition $\mathbb{E}_\gamma( z_i[y_i - \sum_{j=1}^k f_j(x_i,\pi_j) \delta_j - w_i^\prime \xi] ) =0$ which can be re-written as:
\[ \mathbb{E}_\gamma \left( z_i (\sum_{j=1}^k \left[f_j(x_i,\pi_{j0}) - f_j(x_i,\pi_{j})\right]  \delta_{j0} + \sum_{j=1}^k f_j(x_i,\pi_j) (\delta_{j0} - \delta_{j}) + w_i^\prime (\xi_0-\xi)]) \right). \]
Take $\gamma = \gamma_0$ such that $\delta_{j0}=0$ for at least one $j$. Then $\Theta_0$ is non-singleton and includes all possible values of $\pi_j$ for which $\delta_{j0}=0$. Suppose $\gamma$, $z_i,x_i,z_i$, and the functions $f_j$ are such that only the coefficients $\pi$ are potentially unidentified. The linear reparameterization based on $\gamma_0$ is such that $\beta_2$ include all coefficients $\pi_j$ for which $\delta_{j0}=0$, while $\beta_1$ includes $\delta,\xi$ and the remaining $\pi_j$, for which $\delta_{j0} \neq 0$.

Take a converging sequence $\gamma_n \to \gamma_0$. Following the same steps as in the previous example, let $b_j = \lim_{n\to\infty} \sqrt{n}|\delta_{jn}| \in \mathbb{R}_+ \cup \{+\infty\}$, define $J_1$ and $J_2$ in the same way as above. As before, apply the reparameterization but now $\beta_2$ includes the $\pi_j$ with $j \in J_2$ and $\beta_1$ collects all remaining coefficients. Here the sources of identification failure are indexed by $|\delta_j|$, $j \in \{1,\dots,k\}$. This time, the partition separates the directions $\pi_j$, associated with the different $\delta_j$.

\paragraph*{Linear Reparameterization with Mixed Identification Strength} The goal of the following is to refine the linear reparameterization give in the main text when there is mixed identification strength, so as to have $\beta_1$ semi-strongly and $\beta_2$ weakly identified. The procedure relies on having finitely many sources of identification failure as in the above examples.

In the previous two examples, there were $1 \leq k < \infty$ sources of identification failure. There, for a sequence $\gamma_n$ associated with weak identification, there are $K = 2^k-1$ possibilities for identification strength. For instance, in Example 2 we have $(b_1,\dots,b_k) \in ( \mathbb{R}_+ \cup \{+\infty\} )^k$ with at least one $b_j < +\infty$. For each $b_j$ there are two possibilities ($b_j < \infty$, $b_j = \infty$) leading to $2^k$ outcomes, minus $1$ where $b_j <\infty$ for all $j$, which precludes weak identification, in which case all parameters are (semi)-strongly identified. 

With these $K \geq 1$ possible combinations, there are $K$ possible subsets $S_1,\dots,S_K \subseteq \mathcal{B}_2$ on which the parameters can be weakly identified. In Example 2, one possible subset is associated with $b_1 <\infty$ and $b_j = +\infty$ for $j >1$; here $S_1 = \{ \pi_1 \in \mathbb{R}, \pi_j = \pi_{j0}, j>1 \}$. Then there are $\delta_j(\cdot),\overline{\delta}_j(\cdot)$ continuous and $h_j(\cdot)>0$ such that:
\begin{align*}
  \sup_{\beta_2 \in S_j \cup \{\beta_{2n}\}} &\|g(\beta_{1n},\beta_2,\gamma_n)\|_W \leq \overline{\delta}_j(\gamma_n)\\ 
  \inf_{d(\beta, \{\beta_{1n}\} \times (S_j \cup \{\beta_{2n}\})) \geq \varepsilon} &\|g(\beta_{1},\beta_2,\gamma_n)\|_W \geq \delta_j(\gamma_n)h_j(\varepsilon).
\end{align*}
Now take $j^\star \in \{1,\dots,K\}$ such that $\sqrt{n}\delta_{j^\star}(\gamma_n) \to \infty$ and $\limsup_{n\to\infty} \overline{\delta}_{j^\star}(\gamma_n) < \infty$. 

If $S_{j^\star} = \mathcal{B}_2$, all parameters in $\beta_2$ are weakly identified and Assumption \ref{ass:weak} i. follows from the properties of linear reparameterization and the Maximum Theorem, as explained in the main text. Otherwise, $S_{j^\star} \subset \mathcal{B}_2$, strictly; only some parameters in $\beta_2$ are weakly identified. 

Let $\tilde{V}_2 = \text{span} \left( \{ v_2 = (\beta_{1n},\beta_2^1) - (\beta_{1n},\beta_2^2), (\beta_2^1,\beta_2^2) \in S_{j^\star} \times S_{j^\star} \} \right)$ and $\tilde{V}_1 = \tilde{V}_2^\perp$.  Let $\tilde{\beta}_1 = P_{\tilde{V}_1}\beta$ and $\tilde{\beta}_2 = P_{\tilde{V}_2}\beta$. By construction, $S_{j^\star}$ is at most a singleton on $\tilde{V}_1$ and a set of dimension $\text{rank}(P_{\tilde{V}_2})$ on $\tilde{V}_2$, denoted $\tilde{S}_{j^\star}$. 

If $\text{rank}(P_{\tilde{V}_2}) = d_{\beta_2}$ then all directions of $\beta_2$ are weakly identified and $\tilde{\beta}_2 = \beta_2$ is unchanged. 
By construction, $\limsup_{n\to\infty} \sqrt{n} \sup_{\tilde{\beta}_2 \in \tilde{S}_{j^\star}} \|g(\tilde{\beta}_{1n},\tilde{\beta}_2,\gamma_n)\|_W = \limsup_{n\to\infty} \sqrt{n} \overline{\delta}_{j^\star}(\gamma_n) <  \infty$. 
To reduce notation, suppose $\beta_{2n} \in S_{j^\star}$, then $\|\beta_1 -\tilde{\beta}_{1n}\| \geq \varepsilon \Rightarrow d(\beta,\{\beta_{1n}\} \times (S_j \cup \{\beta_{2n}\})) \geq \varepsilon$, using $\|\beta^1- \beta^2\| = \|P_{\tilde{V}_1}(\beta^1- \beta^2)\| + \|P_{\tilde{V}_2}(\beta^1- \beta^2)\|$. This implies that $\inf_{\|\tilde{\beta}_1-\tilde{\beta}_{1n}\| \geq \varepsilon,\tilde{\beta}_2} \|g(\tilde{\beta}_1,\tilde{\beta}_2,\gamma_n)\|_W \geq \delta_{j^\star}(\gamma_n) h_{j^\star}(\varepsilon)$ with $\sqrt{n}\delta_{j^\star}(\gamma_n) \to \infty$ which yields Assumption \ref{ass:weak} i., i.e. $\tilde{\beta}_1$ is semi-strongly identified and $\tilde{\beta}_2$ is weakly identified on the set $\tilde{S}_{j^\star}$.





\section{Uniform Sampling on Level Sets} \label{apx:PMC}
As shown in Section \ref{sec:outline}, the computation of the quasi-Jacobian requires uniform draws over the level set $\Theta_n = \{ \theta \in \Theta, \|\bar{g}_n(\theta)\|_{W_n}  \leq \kappa_n \}$ and similarly test inversion amounts to finding the level set  $\{ \theta \in \Theta, \|\bar{g}_n(\theta)\|^2_{\hat{V}_n^{-1}}  \leq \chi^2_{d_g - \hat{d}_n}(1-\alpha)\}$ and projecting it onto $\theta_1$.

\paragraph{Direct approach:} the approach used in Section \ref{sec:MonteCarlo} amounts to importance sampling. Draw $\theta_1,\dots,\theta_B$ uniformly distributed on $\theta$ and assign weights proportional to $\mathbbm{1}_{\|\bar{g}_n(\theta_b)\|_{W_n}  \leq \kappa_n}$. The weighted sample is uniformly distributed on the level set. The draws $(\theta_b)_{b=1,\dots,B}$ can be random or pseudo-random using quasi-Monte Carlo sequences such as the Sobol or Halton sequence \citep[see][Section 5]{Lemieux2009}. The main drawback of this approach is that the effective sample size can be very small, i.e. few draws have non-zero weight, when the level set is small relative to the parameter space. In particular, the effective sample size is approximately $B \times\text{volume}(\Theta_n)/\text{volume}(\Theta)$ which tends to be small when the dimension of $\theta$ is moderately large. 

\paragraph{Adaptive Sampling by Population Monte Carlo:} the main idea here to is preserve the simplicity of importance sampling while constructing a sequence of proposal distributions with a higher acceptance rate. Algorithm \ref{algo:pmc} below is adapted from the Population Monte Carlo principle laid out in \citet{cappe2004}. Consider a sequence of level sets: $\Theta_{jn} = \{ \theta \in \Theta, \|\bar{g}_n(\theta)\|_{W_n}  \leq \kappa_{jn} \}$ with $\kappa_{1n} > \kappa_{2n} > \dots > \kappa_{Jn} = \kappa_n$ for some $J \geq 1$. By construction $\Theta_n =  \Theta_{Jn} \subseteq \dots \subseteq \Theta_{2n} \subseteq \Theta_{1n}$ and $\text{volume}(\Theta_{1n}) \geq \dots \geq \text{volume}(\Theta_{Jn}) = \text{volume}(\Theta_{n})$. This implies that it is easier to generate uniform draws on $\Theta_{1n}$ than on $\Theta_{n}$.

\begin{algorithm}
  \caption{Population Monte Carlo Sampling on Level Sets} \label{algo:pmc}
  \begin{algorithmic}
    \State \textbf{Inputs:} $\kappa_{1n} > \dots > \kappa_{Jn} = \kappa_n$, number of draws $B$, generating distributions $q_{jb}$
    \State \textbf{Initialization:} 
    \For{$b=1,\dots,B$}
        \State draw $\theta^1_{b} \sim \mathcal{U}_{\Theta}$ until $\|\bar{g}_n(\theta^1_{b})\|_{W_n} \leq \kappa_{1n}$
        \State set $w^1_b = 1/B$
    \EndFor 
    \State \textbf{Sequential Sampling:} 
    \For{ $j=2,\dots,J$} 
        \For{$b=1,\dots,B$}
        \State draw $\theta^{j\star}_b \sim (\theta^{j-1}_b,w^{j-1}_b)_{b=1,\dots,B}$ and $\theta^j_{b} \sim q_{jb}(\cdot|\theta_b^{j\star})$ until $\|\bar{g}_n(\theta^j_{b})\|_{W_n} \leq \kappa_{jn}$
        \State set $w^j_b \propto w(\theta^{j\star}_b)/q_{jb}(\theta^j_{b}|\theta_b^{j\star})$
        \EndFor 
  \EndFor
  \State \textbf{Output:} weighted sample  $(\theta^{J}_b,w^{J}_b)_{b=1,\dots,B}$
  \end{algorithmic}
\end{algorithm}

The following summarizes the algorithm in plain terms. The initialization step is a simple accept-reject algorithm to generate iid draws on $\Theta_{1n}$. Then given a set of draws $j-1 \geq 1$, draw uniformly $\theta^{j\star}_b$ from the weighted sample $(\theta^{j-1}_b,w^{j-1}_b)_{b=1,\dots,B}$ and generate $\theta^j_{b}$ using a transition kernel $q_{jb}$, for instance a random-walk step $\theta^j_{b} \sim \mathcal{N}(\theta^{j\star}_b,\Sigma^j_b)$. Re-draw both $\theta^{j\star}_b$ and $\theta^j_{b}$ until the criterion $\|\bar{g}_n(\theta^j_{b})\|_{W_n} \leq \kappa_{jn}$ is met and then set the weight according to the sampling probability $w^j_b \propto w(\theta^{j\star}_b)/q_{jb}(\theta^j_{b}|\theta_b^{j\star})$. Repeat this process for each $b=1,\dots,B$ and each $j=2,\dots,J$. The final weighted sample $(\theta^{J}_b,w^{J}_b)_{b=1,\dots,B}$ targets the desired distribution.

There are several choices of tuning parameters in the steps above. First, $\kappa_{jn}$ can be chosen adaptively to avoid decreasing it too fast or too slow which would result in poor computational performance. In the empirical application, $\kappa_{1n}$ is set according to median value of $\|\bar{g}_n(\theta^1_{b})\|_{W_n}^2$ from uniform draws $\theta_b$ on $\Theta$; this yields $\kappa_{1n}^2 = 5500$. Then $\kappa_{jn}$ is set according to $\kappa_{jn}^2 = \min( 0.9 \kappa_{j-1n}^2, q_{j-1}(0.6) )$ where $q_{j-1}(0.6)$ is the $60\%$ quantile of $\|\bar{g}_n(\theta_b^{j-1})\|_{W_n}^2$. This guarantees that $\kappa_{jn}$ is strictly decreasing but declines slowly enough to maintain a reasonable acceptance rate. To adapt to the shape of each $\Theta_{jn}$, the proposal $q_{jn}$ is also constructed adaptively. For each $j \geq 2$, a clustering algorithm is applied to the draws $(\theta^j_b)_{b=1,\dots,B}$ to split the draws into $K=3$ clusters. Then $\Sigma_b^j$ is $2$ times the variance of the draws from the cluster in which $\theta^j_b$ belongs. This accommodates multimodality in the objective function. The inner loop, over $b=1,\dots,B$, is run in parallel which speeds up the computation significantly. In the application, the final $n \times \kappa_n^2 = \|\bar{g}_n(\hat\theta_n)\|_{W_n}^2 + 2\log\log(n) = 10.34$ is attained from the initial $n \times \kappa_{1n}^2 = 5500$ after $J=45$ iterations.

The output of Algorithm \ref{algo:pmc} is used both to compute $B_{n,\infty}$ and later for test inversion by picking $\hat j$ such that $\hat j = \inf \{ j =1,\dots,J, \kappa_{jn}^2 \geq \chi^2_{d_g - \hat d_n}(1-\alpha)\}$ and running one more iterations with $\kappa_{\hat j + 1n}^2 = \chi^2_{d_g - \hat d_n}(1-\alpha)$. This yields the 5000 draws shown in Figure \ref{fig:CS_gamma_psi}.

\section{Sample Code to Implement to Procedure} \label{sec:code}
The following provides some sample R code to perform the steps outlined in Section \ref{sec:outline} for the Monte Carlo example in Appendix \ref{apx:MC_additional_simu}.
{\footnotesize
\begin{lstlisting}[basicstyle=\linespread{0.5}]
  require(randtoolbox) # Used to generate the integration grid
  library(pracma)      # Used to compute matrix square root

  library(CVXR)   # CVX for R
  library(Rmosek) # To use the MOSEK solver in CVX

  set.seed(123)
  
  n = 1e3 # Sample size
  B = 1e4 # Number of draws
  
  # Robust and standard critical values:
  critical_R = qchisq(0.95,2) 
  critical_S = qchisq(0.95,1)

  # **************************************************************
  #             Simulate Data, Define Moment Conditions
  # **************************************************************

  c  = 1         # c determines identification strength
  b1 = c/sqrt(n) # theta1 = c/sqrt(n)
  b2 = 5         # theta2 is fixed

  # Simulate data: x1, x2, e, and y = b1*x1 + b1*b2*x2 + e 
  x1 = rnorm(n)
  x2 = rnorm(n)
  e  = rnorm(n)
  y  = b1*x1 + b1*b2*x2 + e

  moments <- function(b,y,x1,x2) { 
      # computes the sample moments and the variance of the moments
      e_hat = y - b[1]*x1 - b[1]*b[2]*x2 # residuals
      mom   = cbind(e_hat,e_hat)*cbind(x1,x2)

      mom_m = apply(mom,2,mean) # g_bar
      V     = var(mom)          # V_hat

      return( list( mom = mom_m, V = V ) )
  }

  objective <- function(b,y,x1,x2) {
      # computes the GMM objective function
      mm = moments(b,y,x1,x2)
      return( t(mm$mom)%*%solve(mm$V,mm$mom) )
  }

  # **************************************************************
  #             Compute the quasi-Jacobian Matrix
  # **************************************************************

  # Set the integration grid: 
  s = sobol(B,2,scrambling=1) 
  p = cbind(rep(b1,B),rep(b2,B)) + 2*(s-1/2)

  objs = rep(NA,B)              # Store GMM objective values
  moms = matrix(NA,B,2)         # Store sample moments mom
  Vs   = array(NA,dim=c(2,2,B)) # Store variances V

  for (b in 1:B) { # Evaluate the moments on the grid
    mm   = moments(p[b,],y,x1,x2)
    objs[b] = t(mm$mom)%*%solve(mm$V,mm$mom)
    moms[b,] = mm$mom
    Vs[,,b] = mm$V
  }

  # Select draws on the level set
  ind = which(objs - min(objs) <= 2*log(log(n))/n) 
  grid_sub = p[ind,]
  moms_sub = moms[ind,]
  Vs_sub   = Vs[,,ind]

  X = cbind(1,grid_sub) # regressors: intercept and theta_b

  # write the optimization problem for CVX
  beta = Variable(dim(X)[2],dim(moms_sub)[2]) # matrix of coefficients (A,B)

  objc <- Minimize(norm( moms_sub - X %*% beta,"I")) # l-infinity loss
  prob <- Problem(objc)                   # compile the problem 
  result <- solve(prob,solver="ECOS_BB")  # compute the solution 
  coef = result$getValue(beta)           # extract solution

  Bn = t(coef[2:3,])                     # quasi-Jacobian matrix

  # Now compute the normalization matrix for the left-hand-side
  V = matrix(0,2,2)     # Compute V_bar the average variance matrix 
  for (b in 1:length(ind)) {
    V = V + Vs_sub[,,b]/length(ind)
  }

  # Now compute the normalization matrix for the right-hand-side
  mu <- Variable(1,2) # vector of means
  one = matrix(1,length(ind),1)
  VV = Variable(2,2) # matrix of variances

  objc <- Minimize( - log_det(VV) + 0.5*norm( (grid_sub%*%VV - kronecker(one,mu))^2,"I"))                        #  setup the minimization problem in CVX
  prob <- Problem(objc)                 # compile
  result2 <- solve(prob,solver="MOSEK") # solve using MOSEK solver

  phi = result2$getValue(VV) # extract solution
  # Note that phi = Sigma^(-1/2), the problem was reparameterized

  # **************************************************************
  #              Identification Category Selection
  # **************************************************************

  v = c(1,0)           # vector which spans theta1
  M = diag(2)-v%*%t(v) # Projection matrix onto the span on theta2

  # Normalized quasi-Jacobian matrix
  # sqrtm computes the matrix square root and Binv its inverse
  Bnorm = ( sqrtm(V)$Binv )%*%( Bn%*%M )%*%phi

  # singular values in decreasing order
  sing   = svd(Bnorm)$d 
  cutoff = sqrt(2*log(n)/n) # cutoff lambda_n for ICS

  print('Singular values without projecting out theta1:')
  print( round(svd(( sqrtm(V)$Binv )%*%( Bn )%*%phi)$d,3) )
  print('Singular values after projecting out theta1:')
  print(round(sing,3))
  print('Cutoff:')
  print(cutoff)

  # Set critical value depending on the singular value and cutoff
  cr = 1*(sing[1]>cutoff)*critical_S + 1*(sing[1]<cutoff)*critical_R

  if (sing[1]>cutoff) {
    print('Nuisance parameter is semi-strongly identified')
  } else {
    print('Nuisance parameter is weakly identified')
  }
  
  # **************************************************************
  #                      Subvector Inference
  # **************************************************************

  # Test H0: b1 = b10 at the 5% significance level
  b10 = 0

  obj <- function(b2,b10,y,x1,x2) {
    return( objective(c(b10,b2),y,x1,x2) ) 
  }

  # Anderson-Rubin test statistic
  AR = n*optimize(obj,c(-20,20),b10=b10,y=y,x1=x1,x2=x2)$objective

  if (AR > cr) {
    print('Reject H0')
  } else {
    print('Cannot reject H0')
  }

  # Compute a 95% confidence set:
  ind = which(objs <= cr)
  print('Confidence Interval for theta1:')
  print(c(min(p[ind,1]),max(p[ind,1])))

  print('True value:')
  print(b1)

  \end{lstlisting}
}

\section{Additional Results for Section \ref{sec:MonteCarlo}} \label{apx:conds_NLS}
\subsection{Verification of the Main Assumptions} \label{apx:NLS_verif_assumptions}
We now verify the main assumptions for the NLS example in Appendix \ref{apx:MC_additional_simu}:
\[ y_i = \theta_1 x_{1i} + \theta_{1}\theta_2 x_{i2} + u_i, \]
where $(x_{1i},x_{2i},u_i)\sim \mathcal{N}(0,I)$ iid. The optimization space is $\Theta = \Theta_1\times \Theta_2 = [\underline{\theta}_1,\overline{\theta}_1]\times [\underline{\theta}_2,\overline{\theta}_2]$, where $-\infty < \underline{\theta}_{1,2} < 0 < \overline{\theta}_{1,2} < \infty$. We can then set $\overline{\Theta} = \overline{\Theta}_1\times \overline{\Theta}_2 = [\underline{\theta}_1 + \varepsilon,\overline{\theta}_1 - \varepsilon]\times [\underline{\theta}_2 + \varepsilon,\overline{\theta}_2 - \varepsilon]$ for any $0<\varepsilon< \min_{j=1,2}(|\underline{\theta}_j|,|\overline{\theta}_j|)$. The parameter space is then $\Gamma = \{ \gamma = (\theta,\omega) \in \overline{\Theta}\times \Omega \}$, where $\Omega$ indexes the distribution $F$ of $(x_{1i},x_{2i},u_i)$ which here is very simple since $\Omega = \{\Phi\}$, the normal distribution above. More general choices of distribution spaces one could consider could take the form: $\Omega = \{ F, \mathbb{E}_F(x_{1i},x_{2i},u_i)=(\mu_1,\mu_2,0), \|(\mu_1,\mu_2)\| \leq c, \mathbb{E}_F((x_{1i}-\mu_1,x_{2i}-\mu_2,u_i)(x_{1i}-\mu_1,x_{2i}-\mu_2,u_i)^\prime) = \Sigma, 0<c \leq \lambda_{\min}(\Sigma) \leq \lambda_{\max}(\Sigma) \leq C < \infty, \mathbb{E}_F(\|(x_{1i},x_{2i},u_i)\|^4) \leq C \}$. See \citet{Andrews2012} for more examples. Assumption \ref{ass:moments} i., ii. hold for this choice of $\overline{\Theta},\Theta$, and $\Gamma.$

The sample moments are $\bar{g}_n(\theta) = \frac{1}{n} \sum_{i=1}^n  (y_i x_{1i} - \theta_1,y_ix_{2i} - \theta_1\theta_2)^\prime$ and their population counterpart is $g(\theta,\gamma_0) = (\theta_{10}-\theta_1,\theta_{10}\theta_{20}-\theta_1 \theta_2)^\prime.$ They can be re-written as:
\[ g(\theta,\gamma_0) = -\left( \begin{array}{cc} 1 & 0\\ -\theta_2 & \theta_{10} \end{array} \right) \left( \begin{array}{c}  \theta_1-\theta_{10}\\ \theta_2-\theta_{20}  \end{array}\right). \]
The lower triangular matrix has two eigenvalues: $1$ and $\theta_{10}$. Hence, we have the following inequality: $\|g(\theta,\gamma_0)\| \geq \min(1,|\theta_{10}|) \times \|\theta-\theta_0\|$. This implies that Assumption \ref{ass:identification} i. holds with $\delta(\gamma_0) = \min(1,|\theta_{10}|)$ which is continuous in $\gamma = (\theta,\omega)$, and $h(\varepsilon)=\varepsilon$. To verify Assumption \ref{ass:identification} ii., take $\theta = (\theta_{10},\theta_2)$, then $\|g(\theta,\gamma_0)\| = |\theta_{10}| \times \|\theta-\theta_0\|$. We have $|\theta_{10}| = \min(1,|\theta_{10}|) \times \frac{|\theta_{10}|}{\min(1,|\theta_{10}|)} \leq \max(1,|\overline{\theta}_{10}|,|\underline{\theta}_{10}|) \times \min(1,|\theta_{10}|)$.

Assumption \ref{ass:ss} i. holds for any $\theta_{1n} \neq 0$. Condition ii. holds if $\sqrt{n}|\theta_{1n}|\to\infty$. Condition iii. is a stochastic equicontinuity condition which can be verified by Lipschitz continuity and conditions on the parameter space and the distribution of the covariates and the errors. Condition iv holds because the quadratic term vanishes at the same rate as the first-order term in the Taylor expansion ($g(\theta,\gamma_n)$ is a polynomial of order $2$ which becomes flat wrt $\theta_2$ when $\theta_{1n}\to 0$). Condition v. can be verified numerically.

For Assumption \ref{ass:weak} i., note that $\|g(\theta,\gamma_n)\|^2 = \|\theta_1-\theta_{1n}\|^2 + \|\theta_{1}\theta_2 - \theta_{1n}\theta_{2n}\|^2 \geq \|\theta_1-\theta_{1n}\|^2$. Here we can use $\tilde \delta(\gamma_n)=1$, $\tilde h(\varepsilon) = \varepsilon$. For $\sqrt{n}\|g(\theta_{1n},\theta_2,\gamma_n)\| = \sqrt{n}|\theta_{1n}| \times \|\theta_2-\theta_{20}\| \leq \sqrt{n}|\theta_{1n}| \times 2 \max(|\overline{\theta}_{2}|,|\underline{\theta}_{2}|) \to (\lim_{n\to\infty} \sqrt{n}|\theta_{1n}|) 2 \max(|\overline{\theta}_{2}|,|\underline{\theta}_{2}|) < \infty$ for weak sequences. Hence, Assumption 5 i. and ii. hold with $\mathcal{B}_{2}^0 = \Theta_2$.

\subsection{Additional Simulation Results} \label{apx:MC_additional_simu}
\paragraph{Consumption Capital Asset Pricing Model (CAPM)} 

\begin{figure}[H] \caption{CAPM - distribution of estimates $\hat\theta_n = (\hat\delta_n,\hat\gamma_n)$} \label{fig:CAPMest} \centering
    \setlength\tabcolsep{2.5pt}
  \renewcommand{\arraystretch}{0.9} 
    {\small \begin{tabular}{cc} 
      $\bf{n=100}$ & $\bf{n=250}$ \\
      \includegraphics[scale=0.45]{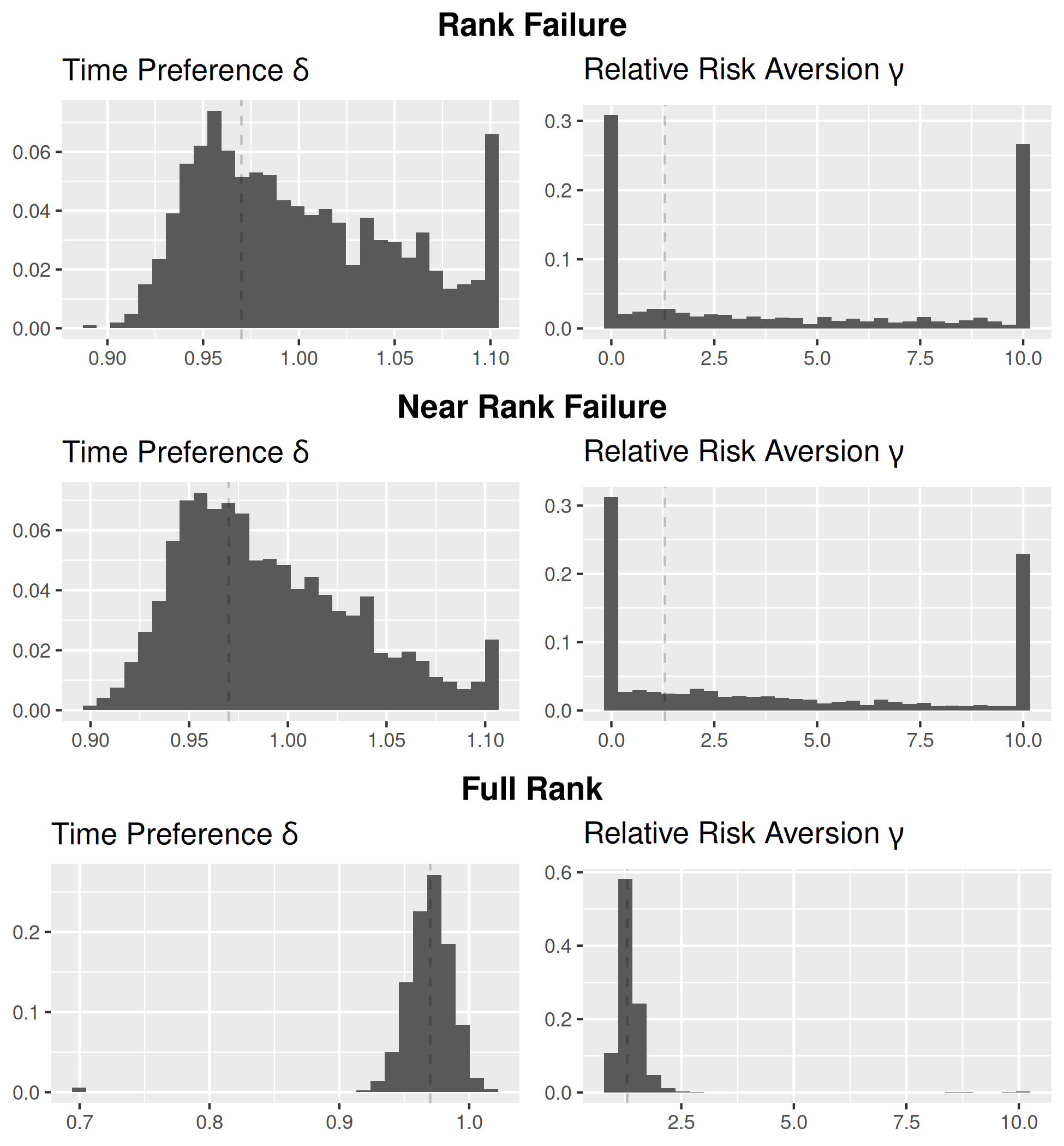} & \includegraphics[scale=0.45]{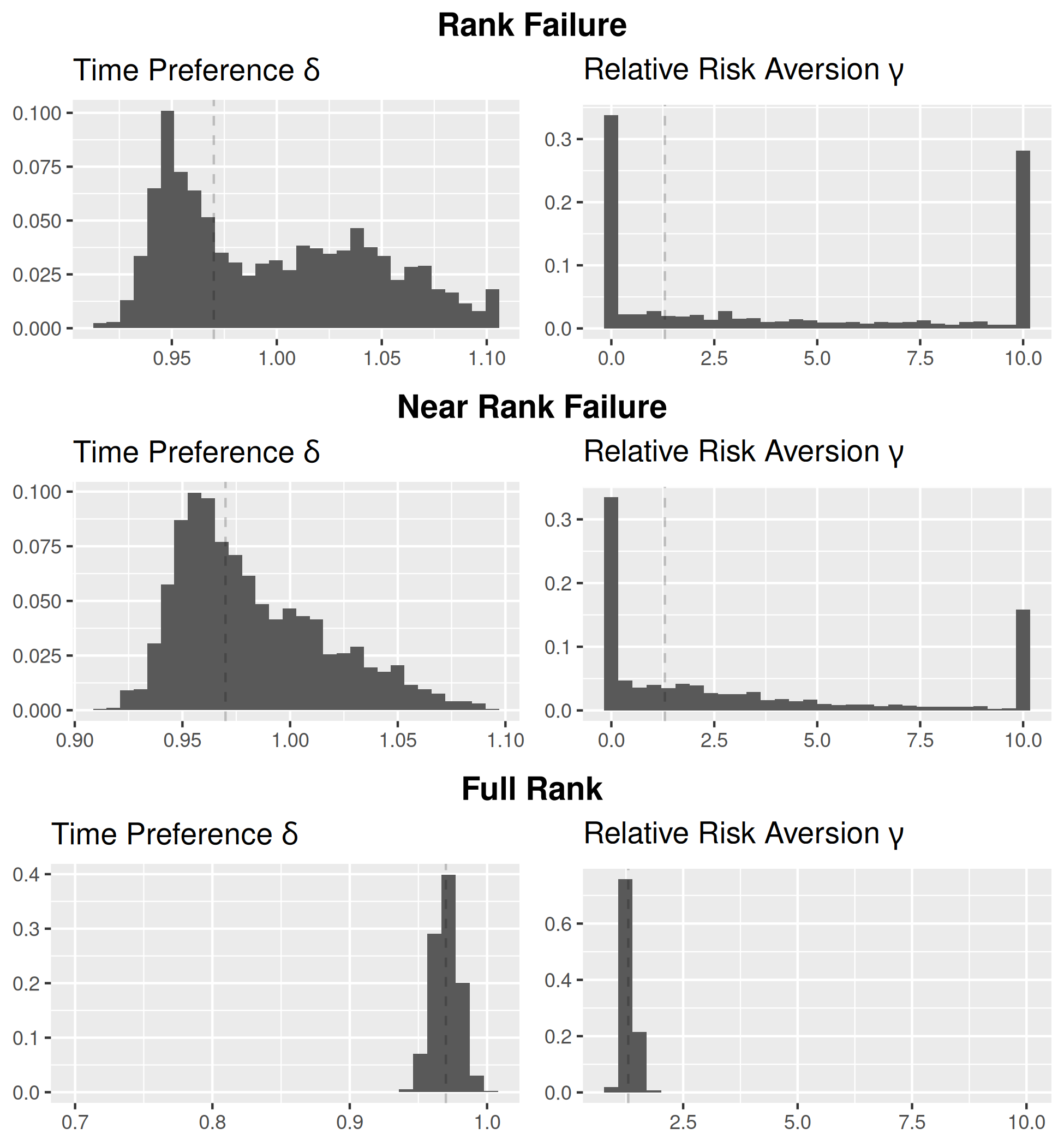}\\ 
      $\bf{n=500}$ & $\bf{n=1000}$ \\
    \includegraphics[scale=0.45]{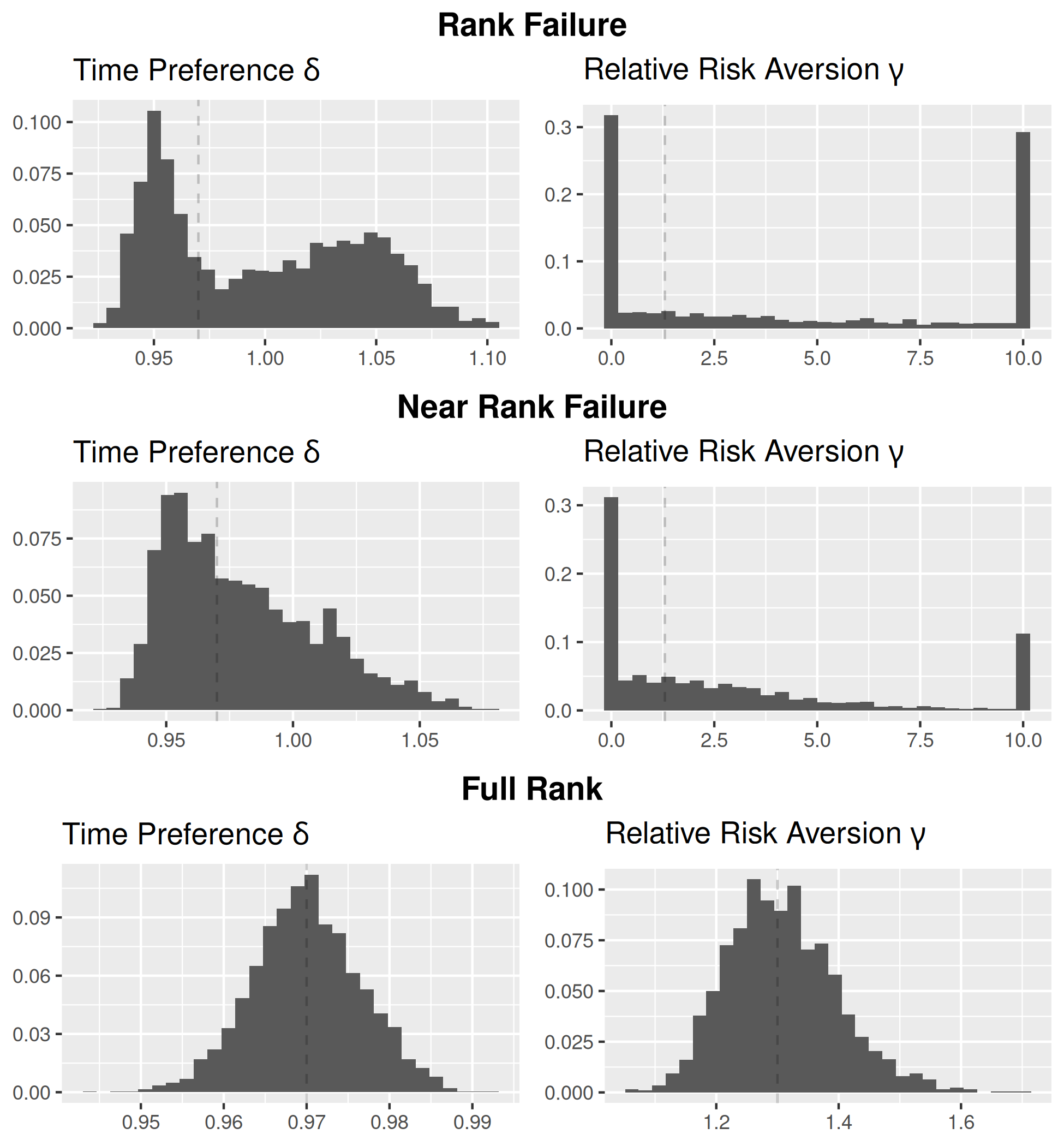} & \includegraphics[scale=0.45]{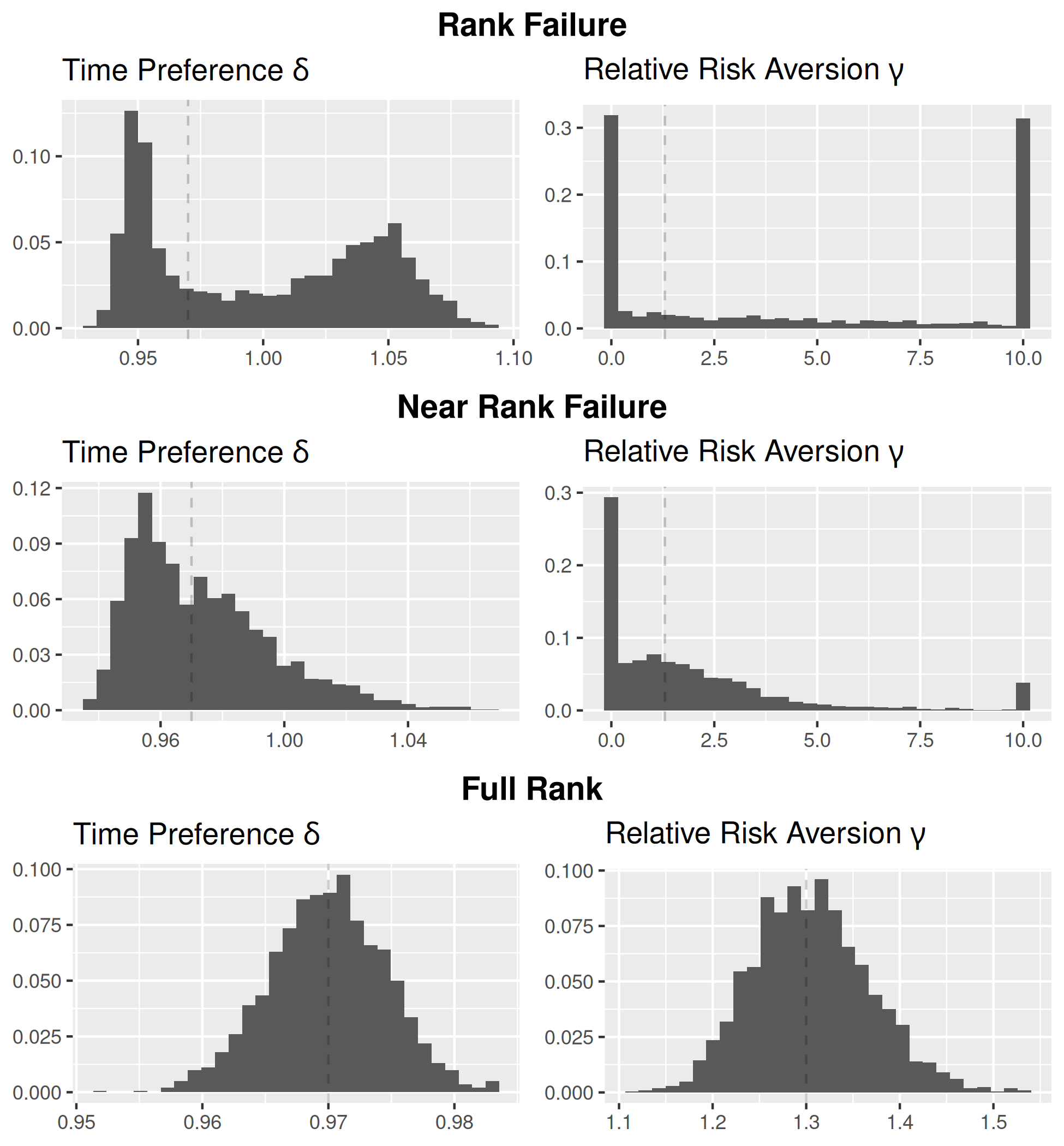}
    \end{tabular}}
    \notes{\textbf{Note:} true value $(\delta_0,\gamma_0) = (0.97,1.3)$: dashed vertical lines. 2000 Monte Carlo replications. Estimates computed for continuously-updated GMM with $W_n = \hat{V}_n(\theta)^{-1}$ where $\hat{V}_n$ is a HAC estimate of $\text{var}[\sqrt{n}\overline{g}_n(\theta)]$. }
  \end{figure}

Figure \ref{fig:CAPMest} shows the sampling distribution of the CAPM estimates $\hat\theta_n = (\hat\delta_n,\hat\gamma_n)$. 

  \begin{table}[H] \caption{CAPM (larger $\kappa_n$) -- size of 95\% CIs for $\delta$ and $\gamma$, frequency for detecting identification failure}
    \label{tab:CAPM_2x}
    \centering \setlength\tabcolsep{2.5pt}
    \renewcommand{\arraystretch}{0.9} 
    {\small
    \begin{tabular}{c|c|gccc|g||gccc|g||gccc|g}
      \hline \hline
     & & \multicolumn{5}{c||}{Rank Failure} & \multicolumn{5}{c||}{Near Rank Failure} & \multicolumn{5}{c}{Full Rank}\\ \hline \rowcolor{white}
     n & & AR$_1$ & AR$_2$ & AR$_3$ & $t_n$ & $<\underline{\lambda}_n$ & AR$_1$ & AR$_2$ & AR$_3$ & $t_n$ & $<\underline{\lambda}_n$ & AR$_1$ & AR$_2$ & AR$_3$ & $t_n$ & $<\underline{\lambda}_n$ \\ 
      \hline
      \multirow{2}{*}{100} & $\delta$ & 0.01 & 0.01 & 0.03 & 0.02 & 1.00 & 0.01 & 0.01 & 0.03 & 0.02 & 1.00 & 0.02 & 0.02 & 0.05 & 0.07 & 0.97  \\ 
      & $\gamma$ & 0.05 & 0.02 & 0.05 & 0.00 & 0.00 & 0.05 & 0.02 & 0.05 & 0.00 & 0.00 & 0.04 & 0.01 & 0.04 & 0.06 & 0.00 \\ \hline
      \multirow{2}{*}{250} & $\delta$ & 0.02 & 0.02 & 0.05 & 0.09 & 1.00 & 0.02 & 0.02 & 0.03 & 0.07 & 1.00 & 0.03 & 0.02 & 0.04 & 0.05 & 0.35 \\ 
      & $\gamma$ & 0.05 & 0.02 & 0.05 & 0.00 & 0.00 & 0.05 & 0.02 & 0.05 & 0.00 & 0.00 & 0.04 & 0.02 & 0.04 & 0.06 & 0.00 \\ \hline
      \multirow{2}{*}{500} & $\delta$ & 0.02 & 0.02 & 0.04 & 0.17 & 1.00 & 0.01 & 0.01 & 0.04 & 0.08 & 1.00 & 0.05 & 0.02 & 0.05 & 0.05 & 0.00 \\ 
      & $\gamma$ & 0.04 & 0.02 & 0.04 & 0.00 & 0.00 & 0.04 & 0.02 & 0.04 & 0.00 & 0.00 & 0.06 & 0.02 & 0.06 & 0.05 & 0.00 \\ \hline
      \multirow{2}{*}{1000} & $\delta$ & 0.02 & 0.02 & 0.05 & 0.22 & 1.00 & 0.02 & 0.02 & 0.04 & 0.05 & 1.00 & 0.05 & 0.02 & 0.05 & 0.05 & 0.00 \\ 
      & $\gamma$ & 0.05 & 0.02 & 0.05 & 0.00 & 0.00 & 0.04 & 0.02 & 0.04 & 0.00 & 0.00 & 0.05 & 0.02 & 0.05 & 0.05 & 0.00 \\ 
       \hline \hline
    \end{tabular}}
    \notes{ 
    \textbf{Note:} 2000 Monte Carlo replications. AR$_1$, AR$_2$, AR$_3$: projection inference using AR statistic and $\chi^2$ critical values with $3-\hat{d}_n$, $2$, and $1$ degrees of freedom; $\hat{d}_n \in \{0,1\}$. $t_n$: t-test with standard normal critical values. $<\underline{\lambda}_n$: frequency (in \%) of singular values below cutoff $\underline{\lambda}_n$ after projecting out the parameter of interest. Rows for $\delta$ show results for $H_0:\delta=\delta_0$. Rows for $\gamma$ show results for $H_0:\gamma=\gamma_0$. $\kappa_n = 2\sqrt{2\log(\log[n])/n}$, $\underline{\lambda}_n = \sqrt{2\log(n)/n}$, $W_n = \hat{V}_n(\theta)^{-1}$ where $\hat{V}_n$ is a HAC estimate of $\text{var}[\sqrt{n}\overline{g}_n(\theta)]$.  }
    \end{table}

    \begin{figure} \caption{CAPM (larger $\kappa_n$) - distribution of largest and smallest singular values} \label{fig:CAPM_JI_sing} \centering
      \setlength\tabcolsep{2.5pt}
    \renewcommand{\arraystretch}{0.9} 
      \includegraphics[scale=0.45]{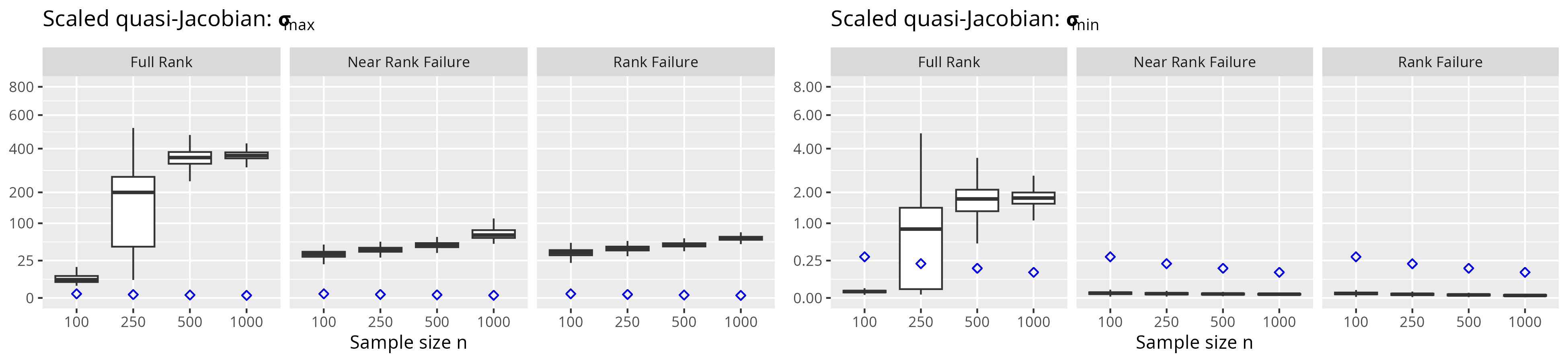}
      \notes{\textbf{Note:} True value $(\delta_0,\gamma_0) = (0.97,1.3)$. 2000 Monte Carlo replications. $W_n = \hat{V}_n(\theta)^{-1}$ where $\hat{V}_n$ is a HAC estimate of $\text{var}[\sqrt{n}\overline{g}_n(\theta)]$. $\kappa_n = 2\sqrt{2\log(\log[n])/n}$, $\underline{\lambda}_n = \sqrt{2\log(n)/n} =0.30, 0.21, 0.16, 0.12$ for $n=100,250,500,1000$. $\sigma_{\max},\sigma_{\min}$: largest and smallest singular values. Median values of $\sigma_{\min}$ for $n=100,250,500,1000$: $3\cdot 10^{-3},2\cdot 10^{-3},2\cdot 10^{-3},1\cdot 10^{-3}$ (RF), $4\cdot 10^{-3}$, $3\cdot 10^{-3}$, $3\cdot 10^{-3}$, $2\cdot 10^{-3}$ (NRF), and $7 \cdot 10^{-3}$, $0.85$, $1.75$, $1.79$ (FR).}
    \end{figure}

    \begin{figure} \caption{CAPM (larger $\kappa_n$) - power comparison} \label{fig:CAPM2x_pow} \centering
      \setlength\tabcolsep{2.5pt}
    \renewcommand{\arraystretch}{0.9} 
      {\small \begin{tabular}{cc} 
        $\bf{n=100}$ & $\bf{n=250}$ \\
        \includegraphics[scale=0.45]{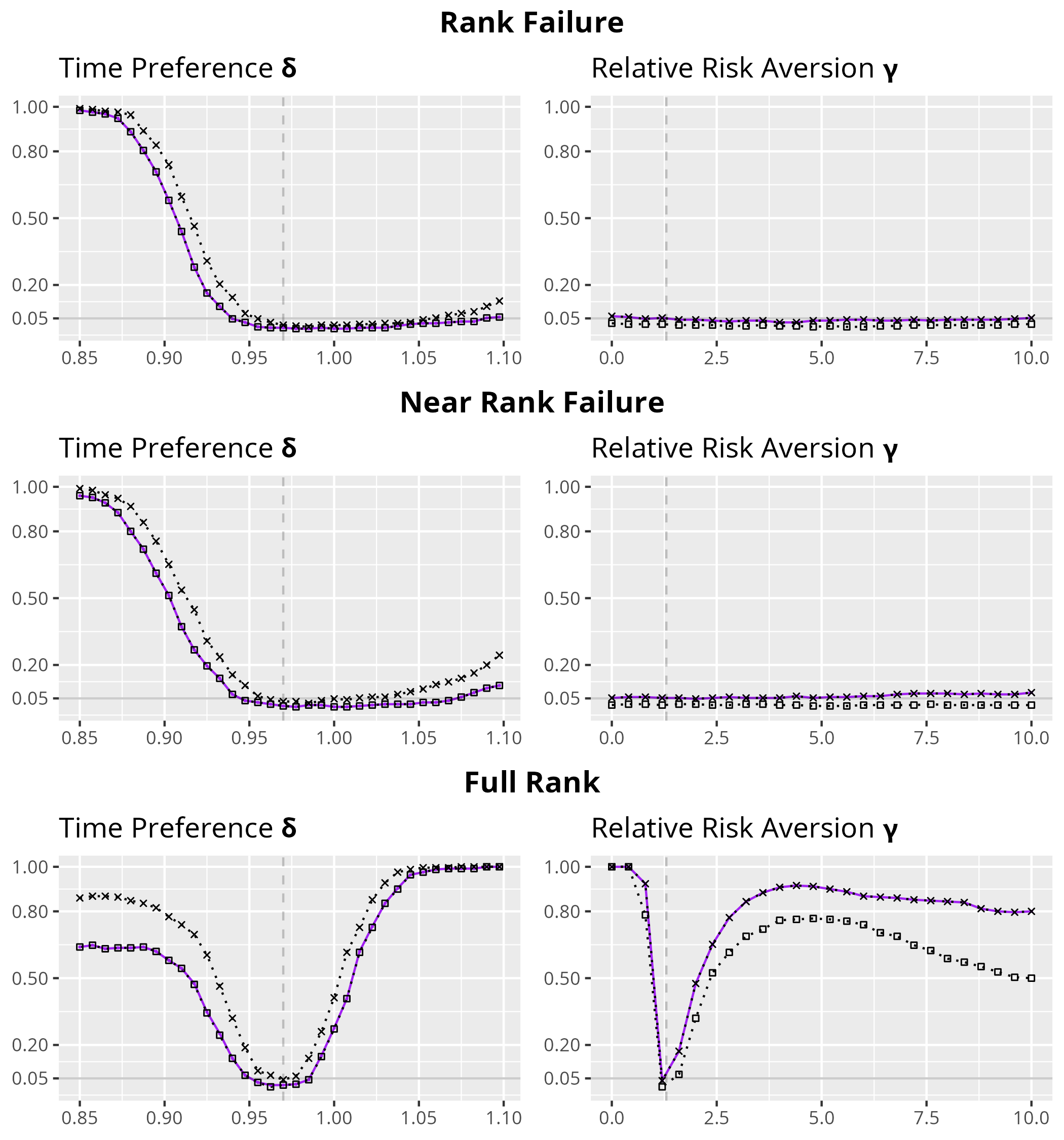} & \includegraphics[scale=0.45]{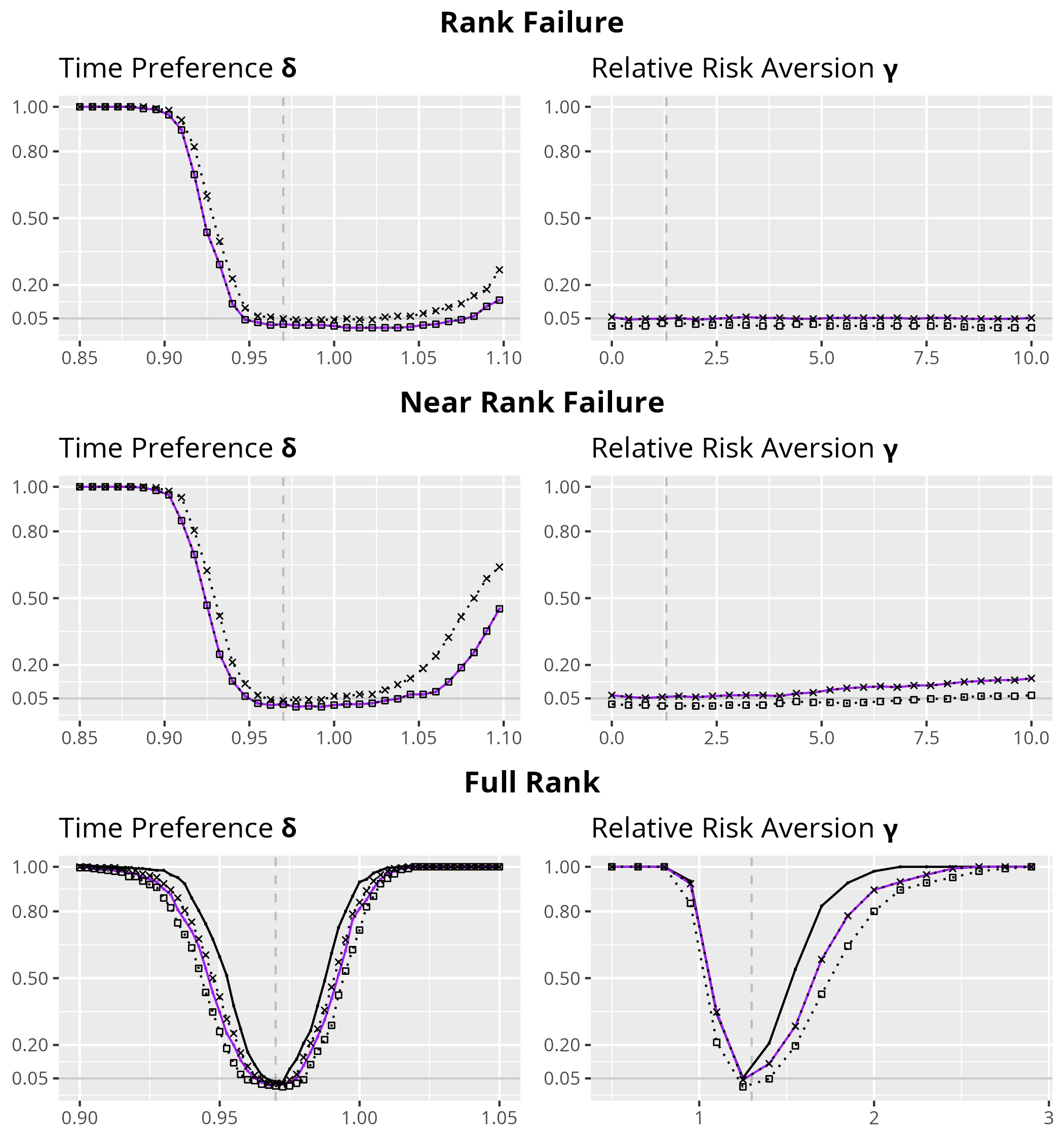}\\ 
        $\bf{n=500}$ & $\bf{n=1000}$ \\
      \includegraphics[scale=0.45]{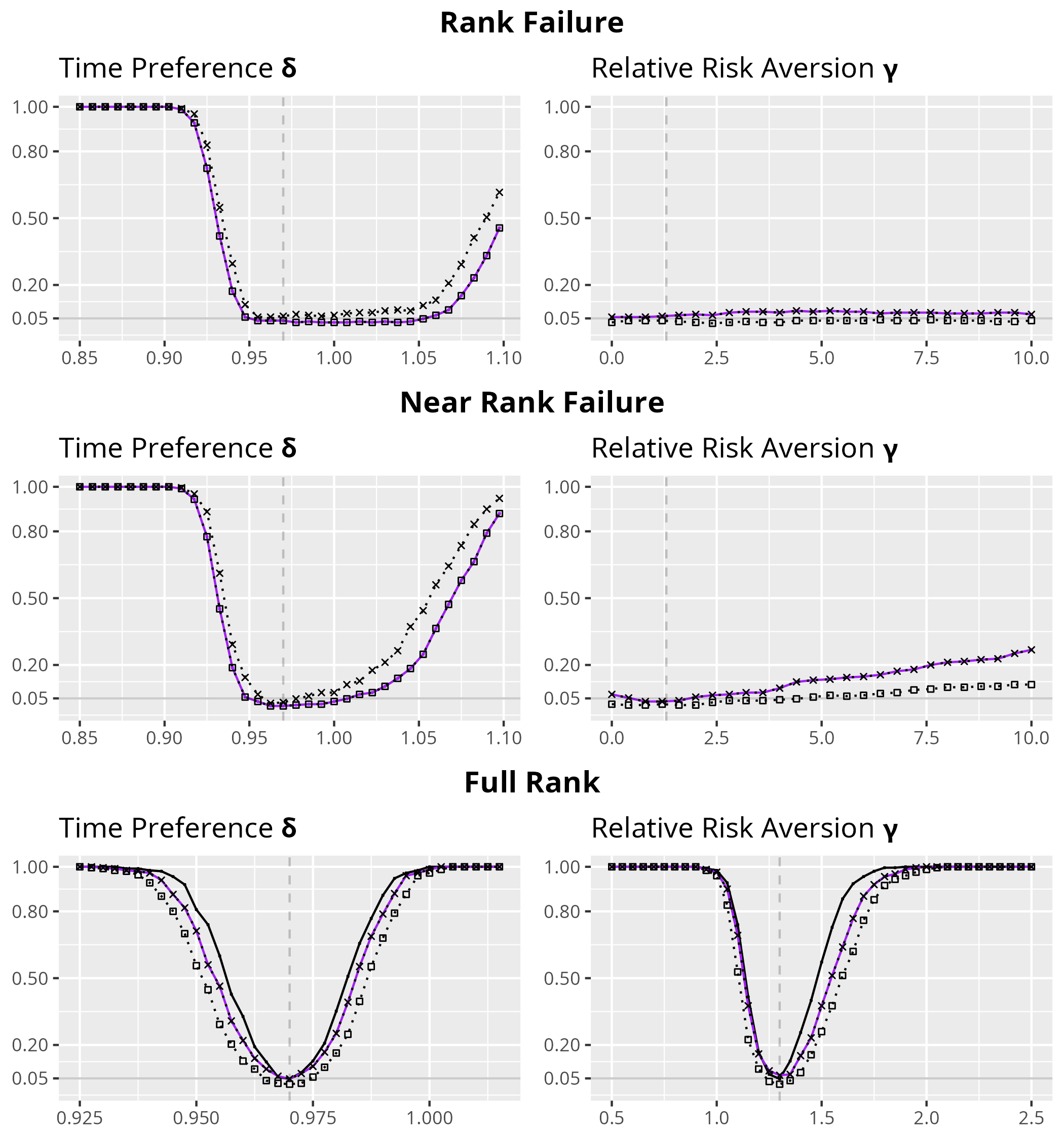} & \includegraphics[scale=0.45]{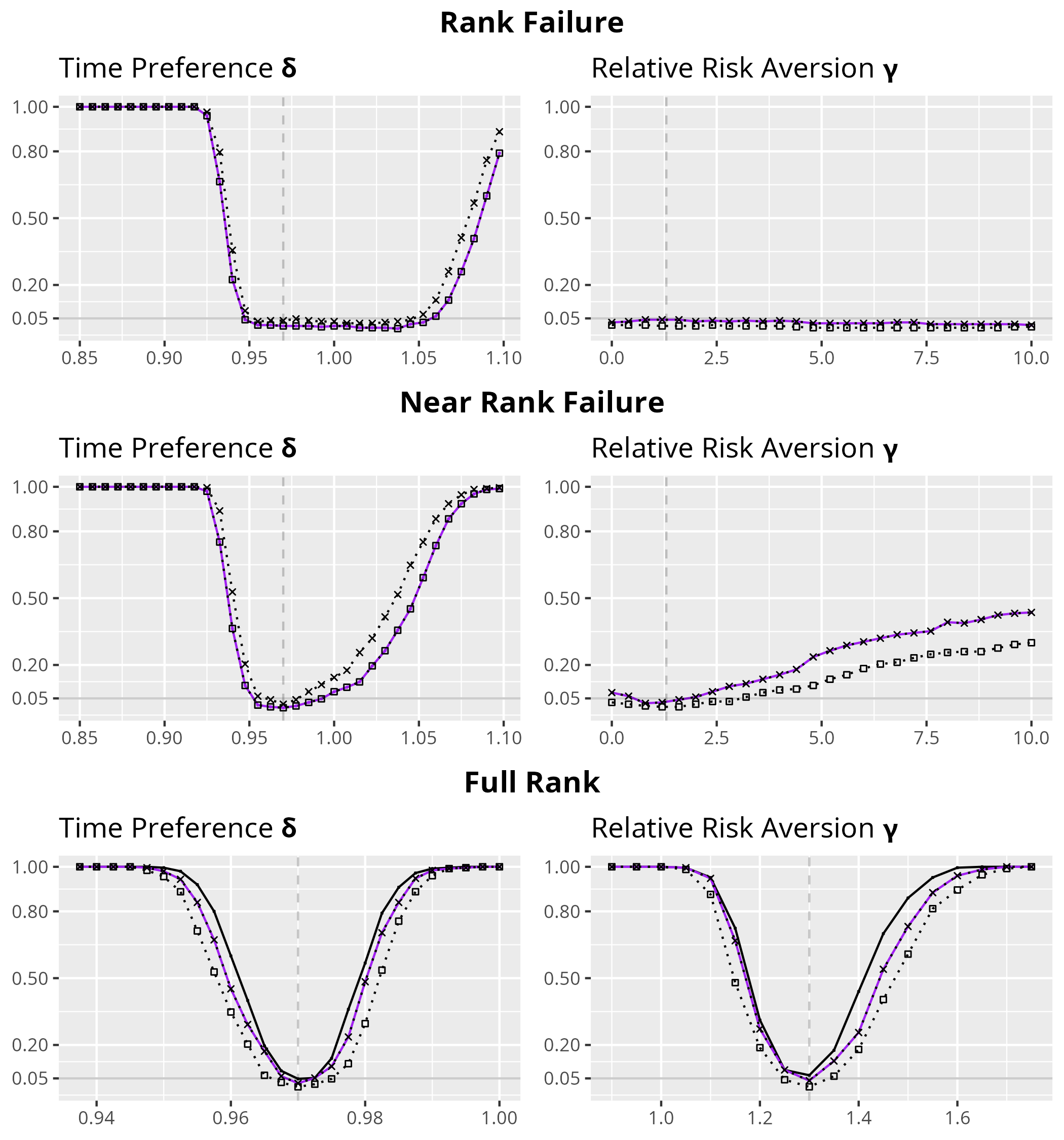}\\
      \includegraphics[scale=0.65]{CAPMpowLegend.png}
      \end{tabular}}
      \notes{\textbf{Note:} Nominal size = $5\%$. True value $(\delta_0,\gamma_0) = (0.97,1.3)$: dashed vertical lines. 250 Monte Carlo replications. Estimates computed for continuously-updated GMM with $W_n = \hat{V}_n(\theta)^{-1}$ where $\hat{V}_n$ is a HAC estimate of $\text{var}[\sqrt{n}\overline{g}_n(\theta)]$. AR$_1$, AR$_2$, AR$_3$: projection inference using AR statistic and $\chi^2$ critical values with $3-\hat{d}_n$, $3$, and $2$ degrees of freedom; $\hat{d}_n \in \{0,1\}$. $t_n$: t-test with standard normal critical values.}
    \end{figure}

Table \ref{t2xCAPM_JI} and Figure \ref{fig:CAPM_JI_est} replicate the results in the main text for a just-identified specification where $Z_t = (1,R_t)^\prime$.
\begin{table}[H] \caption{CAPM (Just-Identified) -- size of 95\% CIs for $\delta$ and $\gamma$, frequency for detecting identification failure}
  \label{tab:CAPM_JI}
  \centering \setlength\tabcolsep{2.5pt}
  \renewcommand{\arraystretch}{0.9} 
  {\small
  \begin{tabular}{c|c|gccc|g||gccc|g||gccc|g}
    \hline \hline
   & & \multicolumn{5}{c||}{Rank Failure} & \multicolumn{5}{c||}{Near Rank Failure} & \multicolumn{5}{c}{Full Rank}\\ \hline \rowcolor{white}
   n & & AR$_1$ & AR$_2$ & AR$_3$ & $t_n$ & $<\underline{\lambda}_n$ & AR$_1$ & AR$_2$ & AR$_3$ & $t_n$ & $<\underline{\lambda}_n$ & AR$_1$ & AR$_2$ & AR$_3$ & $t_n$ & $<\underline{\lambda}_n$ \\ 
    \hline
    \multirow{2}{*}{100} & $\delta$ & 0.01 & 0.01 & 0.04 & 0.01 & 1.00 & 0.01 & 0.01 & 0.04 & 0.01 & 1.00 & 0.04 & 0.01 & 0.05 & 0.05 & 0.38  \\ 
    & $\gamma$ & 0.05 & 0.02 & 0.05 & 0.00 & 0.00 & 0.05 & 0.01 & 0.05 & 0.00 & 0.00 & 0.05 & 0.01 & 0.05 & 0.06 & 0.00 \\ \hline
    \multirow{2}{*}{250} & $\delta$ & 0.02 & 0.02 & 0.04 & 0.03 & 1.00 & 0.01 & 0.01 & 0.03 & 0.05 & 1.00 & 0.05 & 0.02 & 0.05 & 0.04 & 0.04 \\ 
    & $\gamma$ & 0.05 & 0.02 & 0.05 & 0.00 & 0.00 & 0.04 & 0.01 & 0.04 & 0.00 & 0.00 & 0.05 & 0.01 & 0.05 & 0.05 & 0.00 \\ \hline
    \multirow{2}{*}{500} & $\delta$ & 0.02 & 0.02 & 0.05 & 0.08 & 1.00 & 0.01 & 0.01 & 0.04 & 0.11 & 1.00 & 0.05 & 0.01 & 0.05 & 0.04 & 0.00 \\ 
    & $\gamma$ & 0.05 & 0.02 & 0.05 & 0.00 & 0.00 & 0.05 & 0.01 & 0.05 & 0.00 & 0.00 & 0.06 & 0.02 & 0.06 & 0.05 & 0.00 \\ \hline
    \multirow{2}{*}{1000} & $\delta$ & 0.01 & 0.01 & 0.05 & 0.09 & 1.00 & 0.01 & 0.01 & 0.04 & 0.14 & 1.00 & 0.06 & 0.01 & 0.06 & 0.05 & 0.00 \\ 
    & $\gamma$ & 0.05 & 0.01 & 0.05 & 0.00 & 0.00 & 0.04 & 0.01 & 0.04 & 0.00 & 0.00 & 0.05 & 0.01 & 0.05 & 0.05 & 0.00 \\ 
     \hline \hline
  \end{tabular}}
  \notes{ 
  \textbf{Note:} 2000 Monte Carlo replications. AR$_1$, AR$_2$, AR$_3$: projection inference using AR statistic and $\chi^2$ critical values with $2-\hat{d}_n$, $2$, and $1$ degrees of freedom; $\hat{d}_n \in \{0,1\}$. $t_n$: t-test with standard normal critical values. $<\underline{\lambda}_n$: frequency (in \%) of singular values below cutoff $\underline{\lambda}_n$ after projecting out the parameter of interest. Rows for $\delta$ show results for $H_0:\delta=\delta_0$. Rows for $\gamma$ show results for $H_0:\gamma=\gamma_0$. $\kappa_n = \sqrt{2\log(\log[n])/n}$, $\underline{\lambda}_n = \sqrt{2\log(n)/n}$, $W_n = \hat{V}_n(\theta)^{-1}$ where $\hat{V}_n$ is a HAC estimate of $\text{var}[\sqrt{n}\overline{g}_n(\theta)]$.  }
  \end{table}

  \begin{figure} \caption{CAPM (Just-Identified) - distribution of largest and smallest singular values} \label{fig:CAPM_JI_sing} \centering
    \setlength\tabcolsep{2.5pt}
  \renewcommand{\arraystretch}{0.9} 
    \includegraphics[scale=0.45]{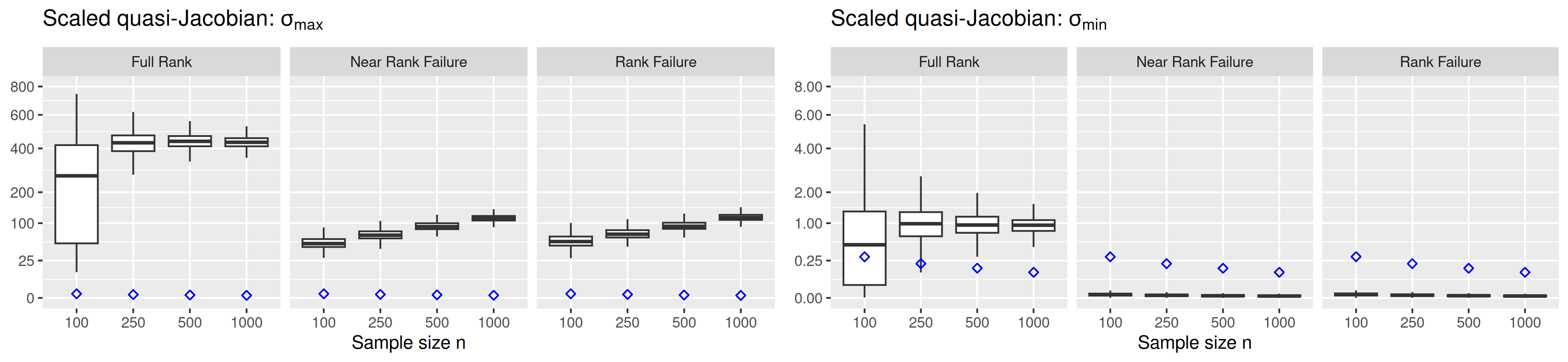}
    \notes{\textbf{Note:} True value $(\delta_0,\gamma_0) = (0.97,1.3)$. 2000 Monte Carlo replications. $W_n = \hat{V}_n(\theta)^{-1}$ where $\hat{V}_n$ is a HAC estimate of $\text{var}[\sqrt{n}\overline{g}_n(\theta)]$. $\kappa_n = \sqrt{2\log(\log[n])/n}$, $\underline{\lambda}_n = \sqrt{2\log(n)/n} =0.30, 0.21, 0.16, 0.12$ for $n=100,250,500,1000$. $\sigma_{\max},\sigma_{\min}$: largest and smallest singular values. Median values of $\sigma_{\min}$ for $n=100,250,500,1000$: $2\cdot 10^{-3},1\cdot 10^{-3},9\cdot 10^{-4},6\cdot 10^{-4}$ (RF), $2\cdot 10^{-3}$, $1\cdot 10^{-3}$, $8\cdot 10^{-4}$, $6\cdot 10^{-4}$ (NRF), and $0.51$, $0.98$, $0.95$, $0.95$ (FR).}
  \end{figure}

  \begin{figure} \caption{CAPM (Just-Identified) - power comparison} \label{fig:CAPM_JI_pow} \centering
    \setlength\tabcolsep{2.5pt}
  \renewcommand{\arraystretch}{0.9} 
    {\small \begin{tabular}{cc} 
      $\bf{n=100}$ & $\bf{n=250}$ \\
      \includegraphics[scale=0.45]{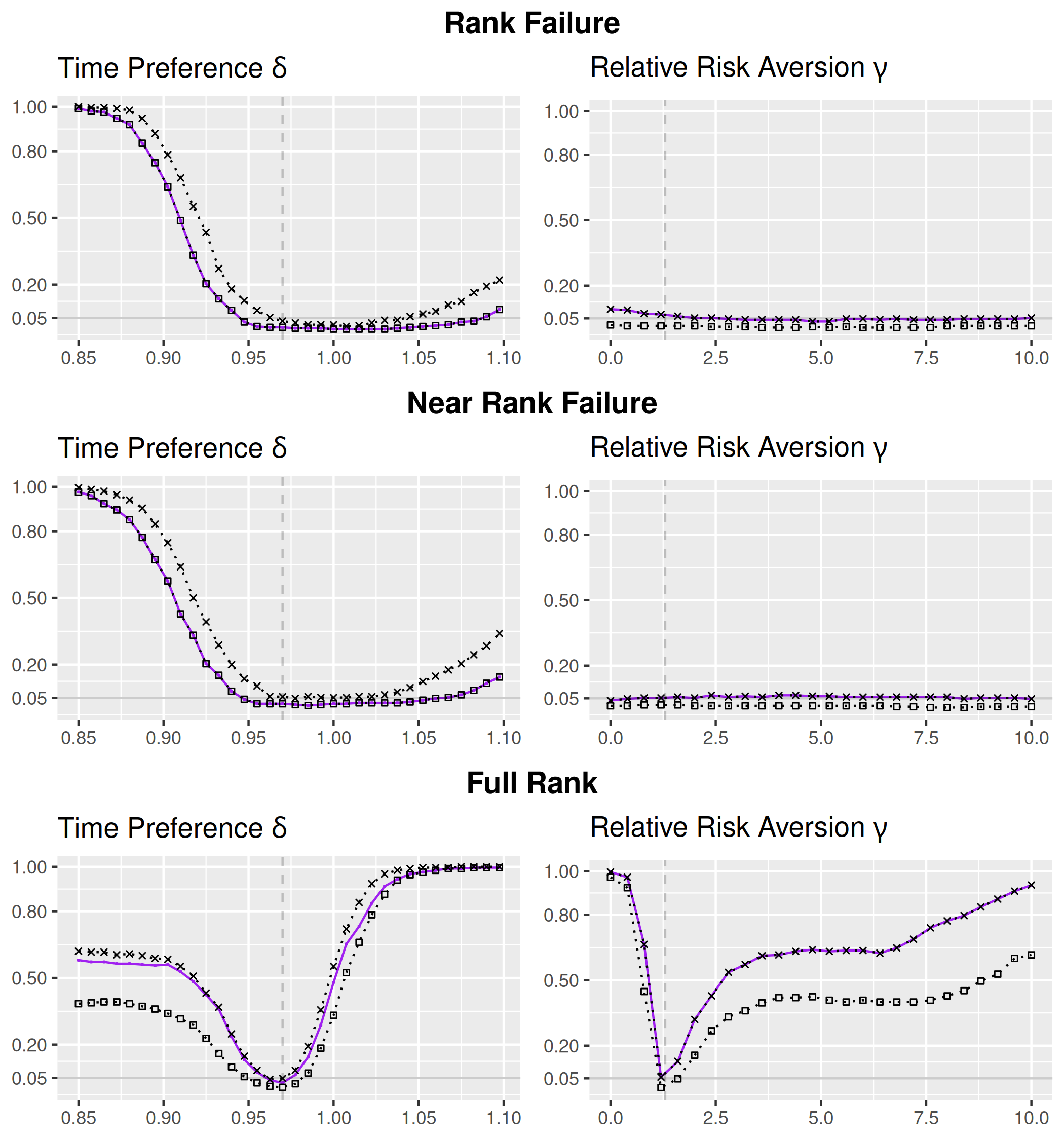} & \includegraphics[scale=0.45]{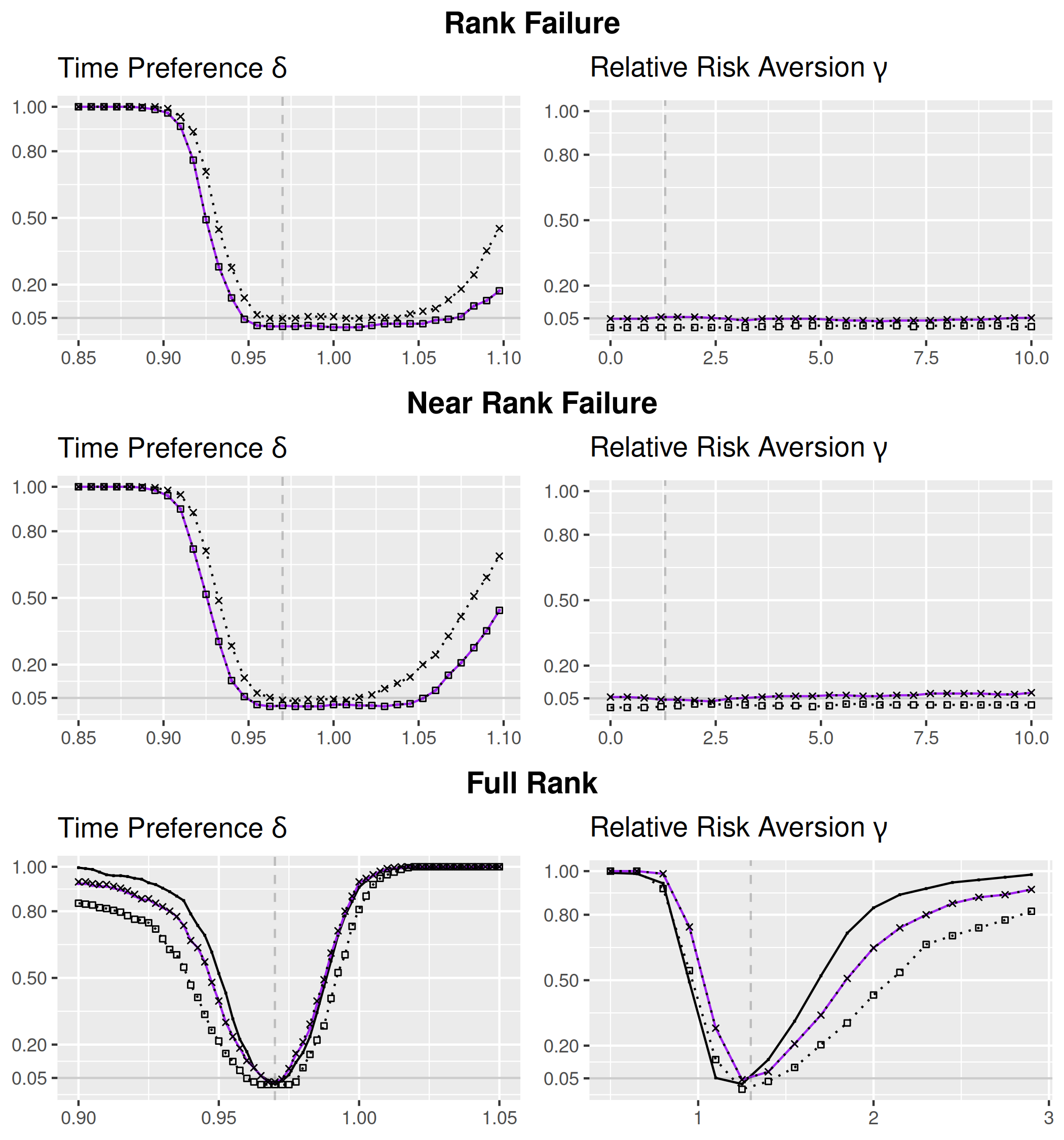}\\ 
      $\bf{n=500}$ & $\bf{n=1000}$ \\
    \includegraphics[scale=0.45]{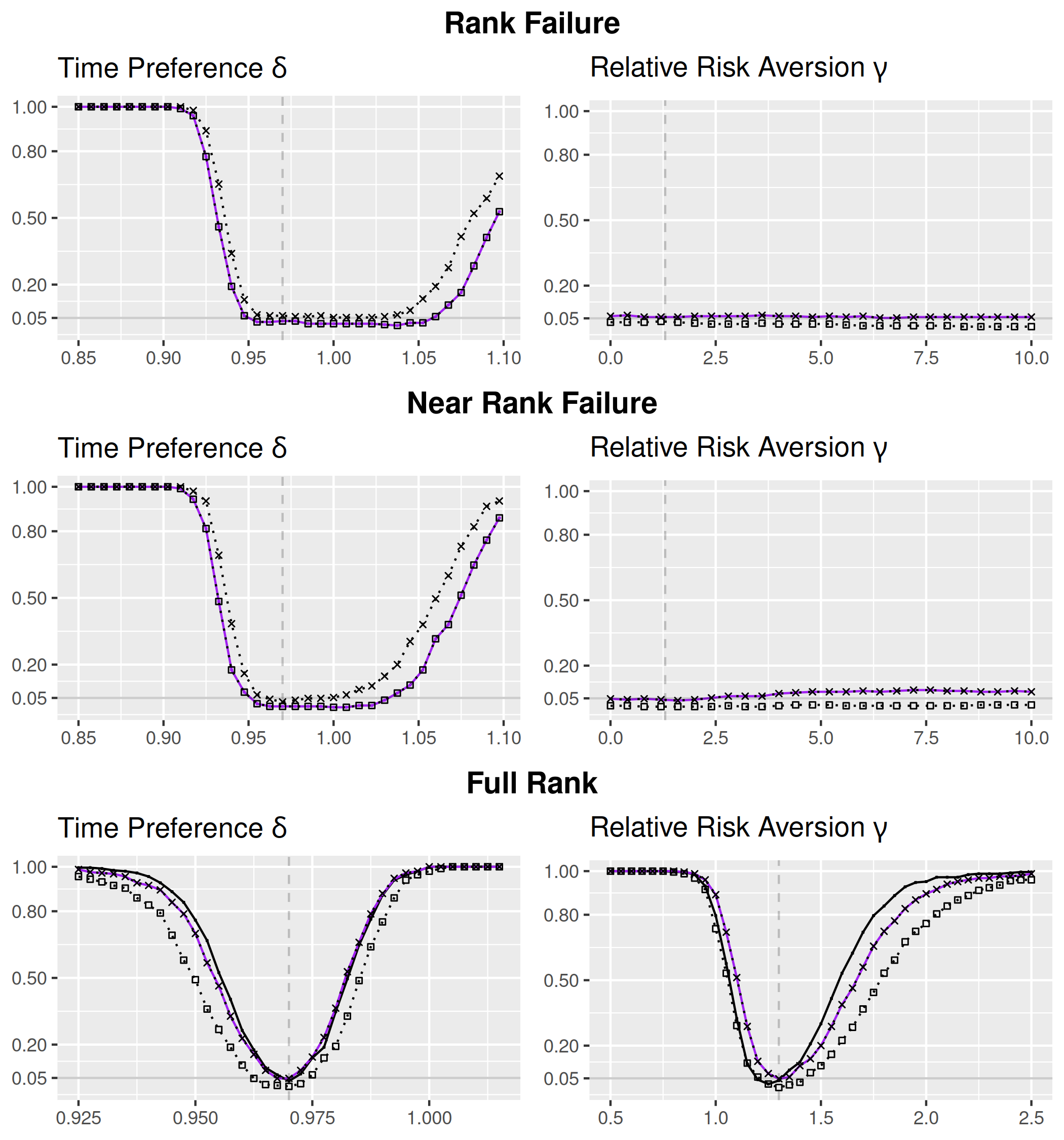} & \includegraphics[scale=0.45]{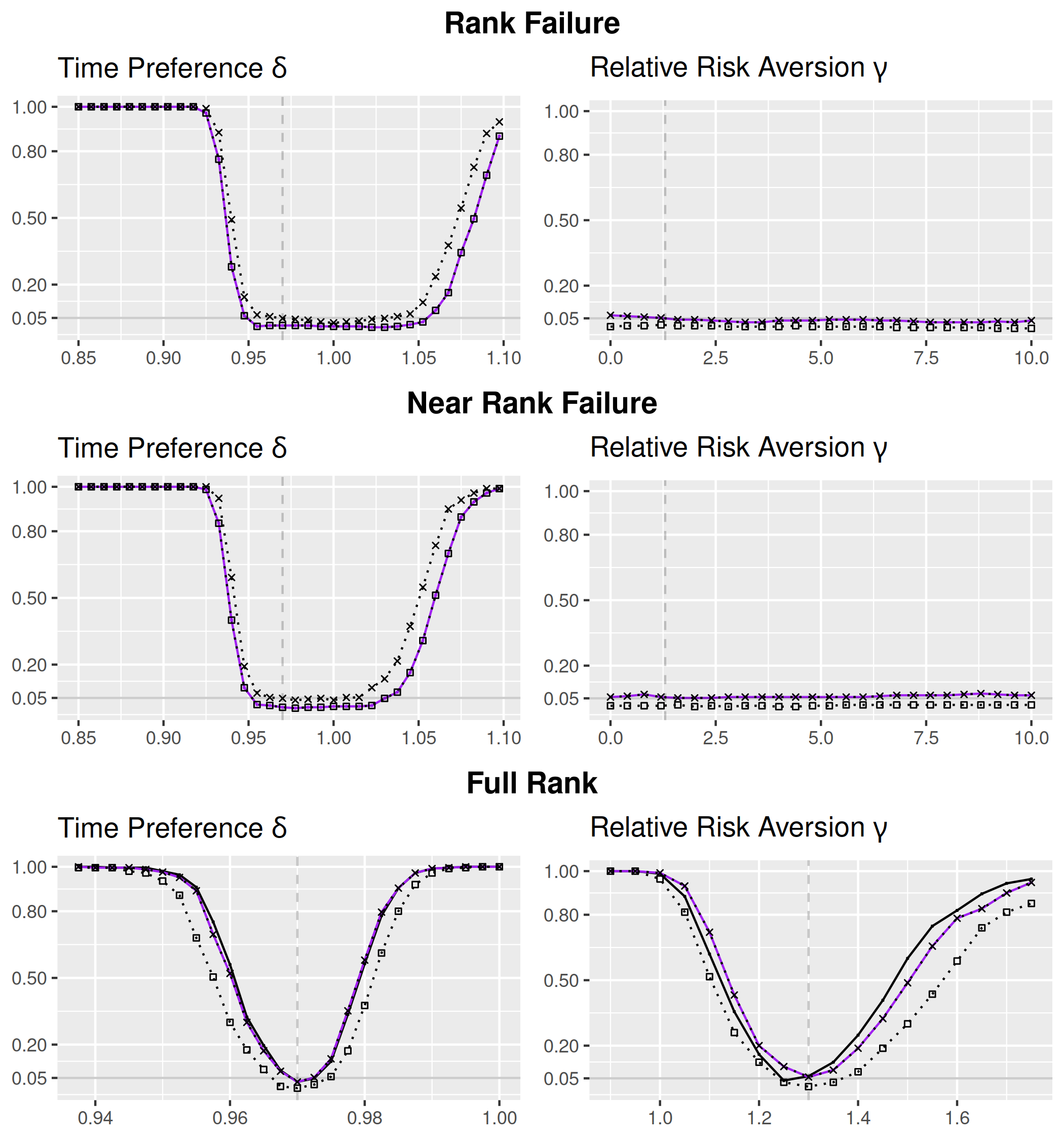}\\
    \includegraphics[scale=0.65]{CAPMpowLegend.png}
    \end{tabular}}
    \notes{\textbf{Note:} Nominal size = $5\%$. True value $(\delta_0,\gamma_0) = (0.97,1.3)$: dashed vertical lines. 250 Monte Carlo replications. Estimates computed for continuously-updated GMM with $W_n = \hat{V}_n(\theta)^{-1}$ where $\hat{V}_n$ is a HAC estimate of $\text{var}[\sqrt{n}\overline{g}_n(\theta)]$. AR$_1$, AR$_2$, AR$_3$: projection inference using AR statistic and $\chi^2$ critical values with $2-\hat{d}_n$, $2$, and $1$ degrees of freedom; $\hat{d}_n \in \{0,1\}$. $t_n$: t-test with standard normal critical values.}
  \end{figure}

  \begin{figure}[H] \caption{CAPM (Just-Identified) - distribution of estimates $\hat\theta_n = (\hat\delta_n,\hat\gamma_n)$} \label{fig:CAPM_JI_est} \centering
    \setlength\tabcolsep{2.5pt}
  \renewcommand{\arraystretch}{0.9} 
    {\small \begin{tabular}{cc} 
      $\bf{n=100}$ & $\bf{n=250}$ \\
      \includegraphics[scale=0.45]{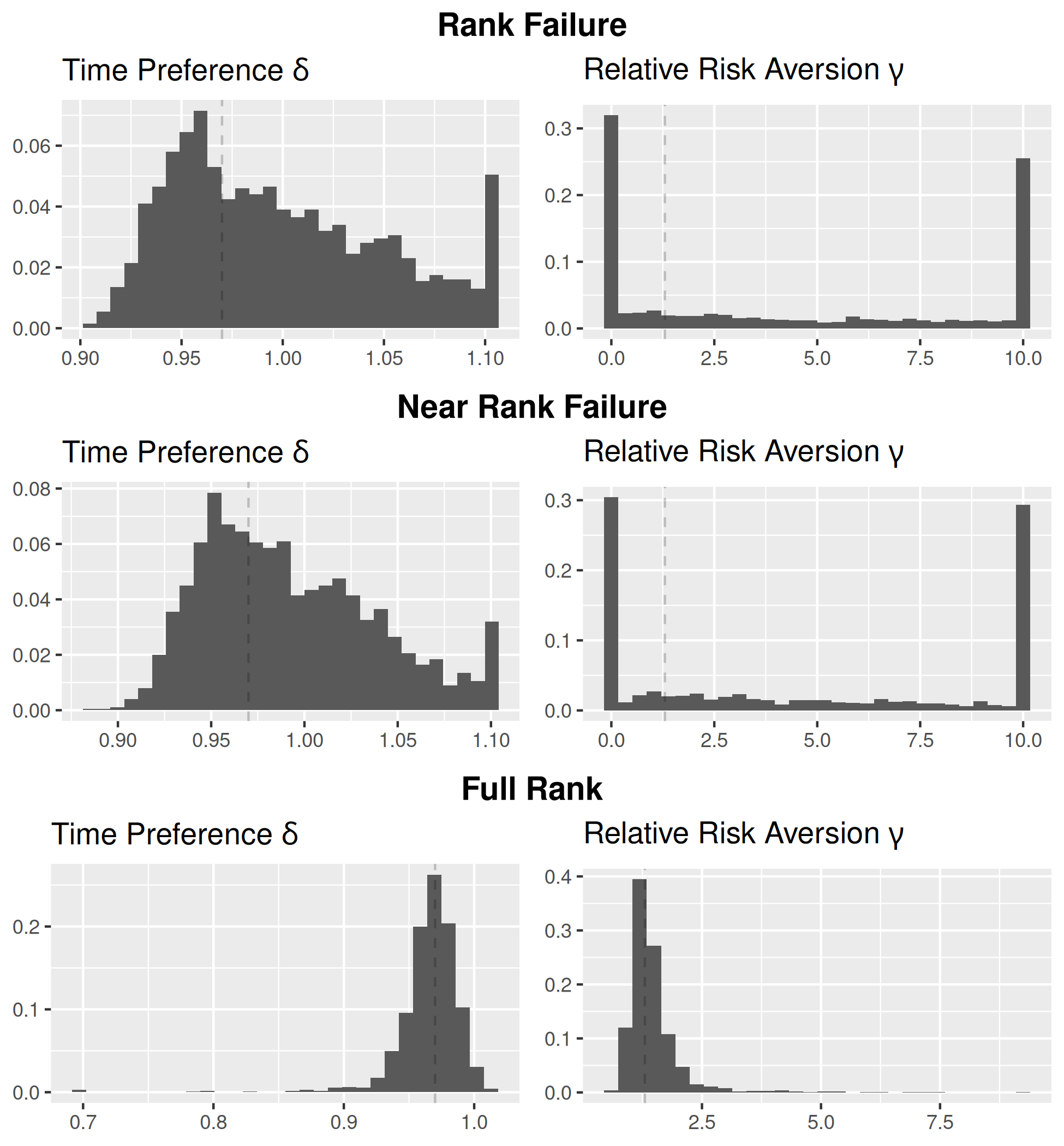} & \includegraphics[scale=0.45]{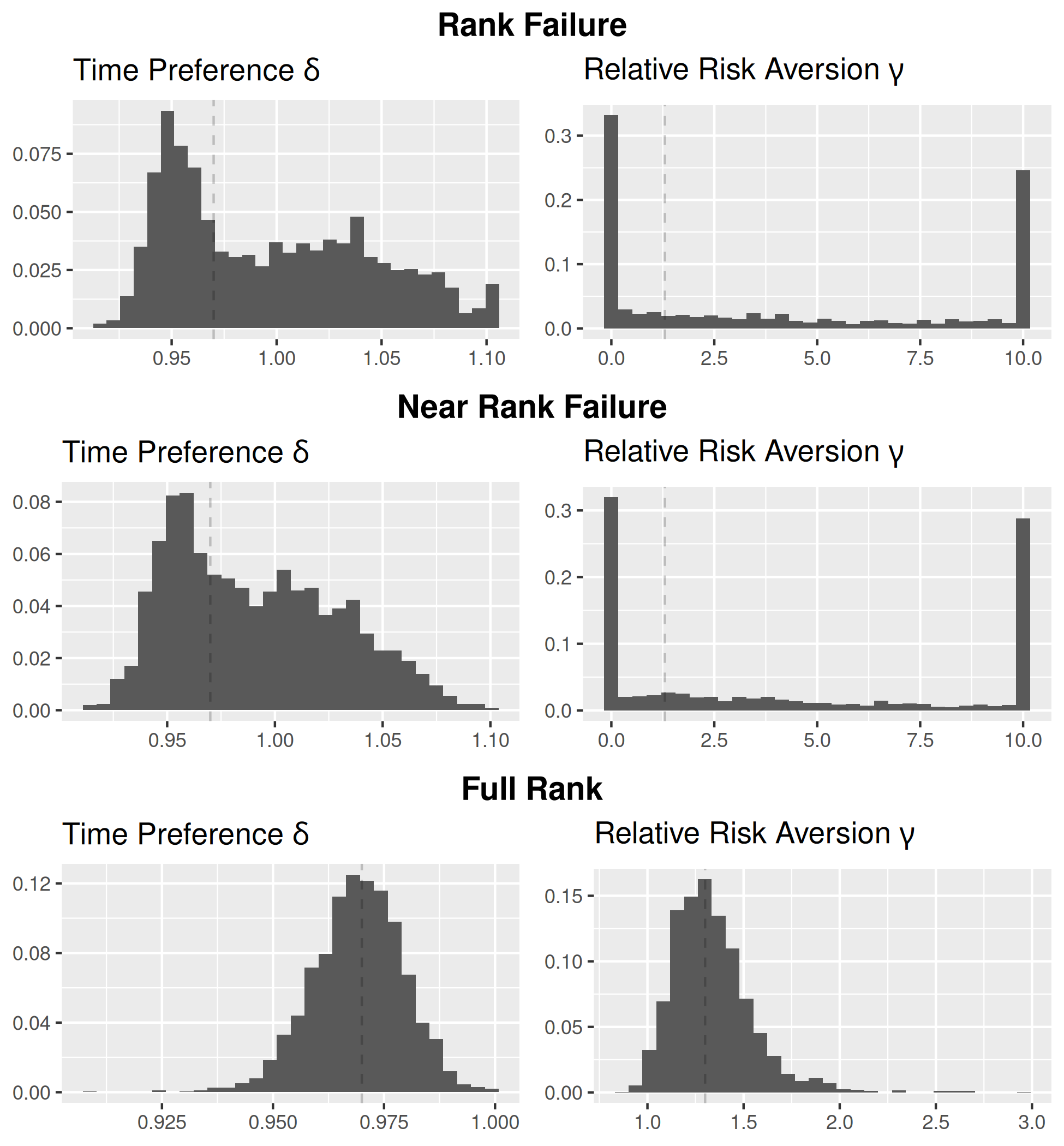}\\ 
      $\bf{n=500}$ & $\bf{n=1000}$ \\
    \includegraphics[scale=0.45]{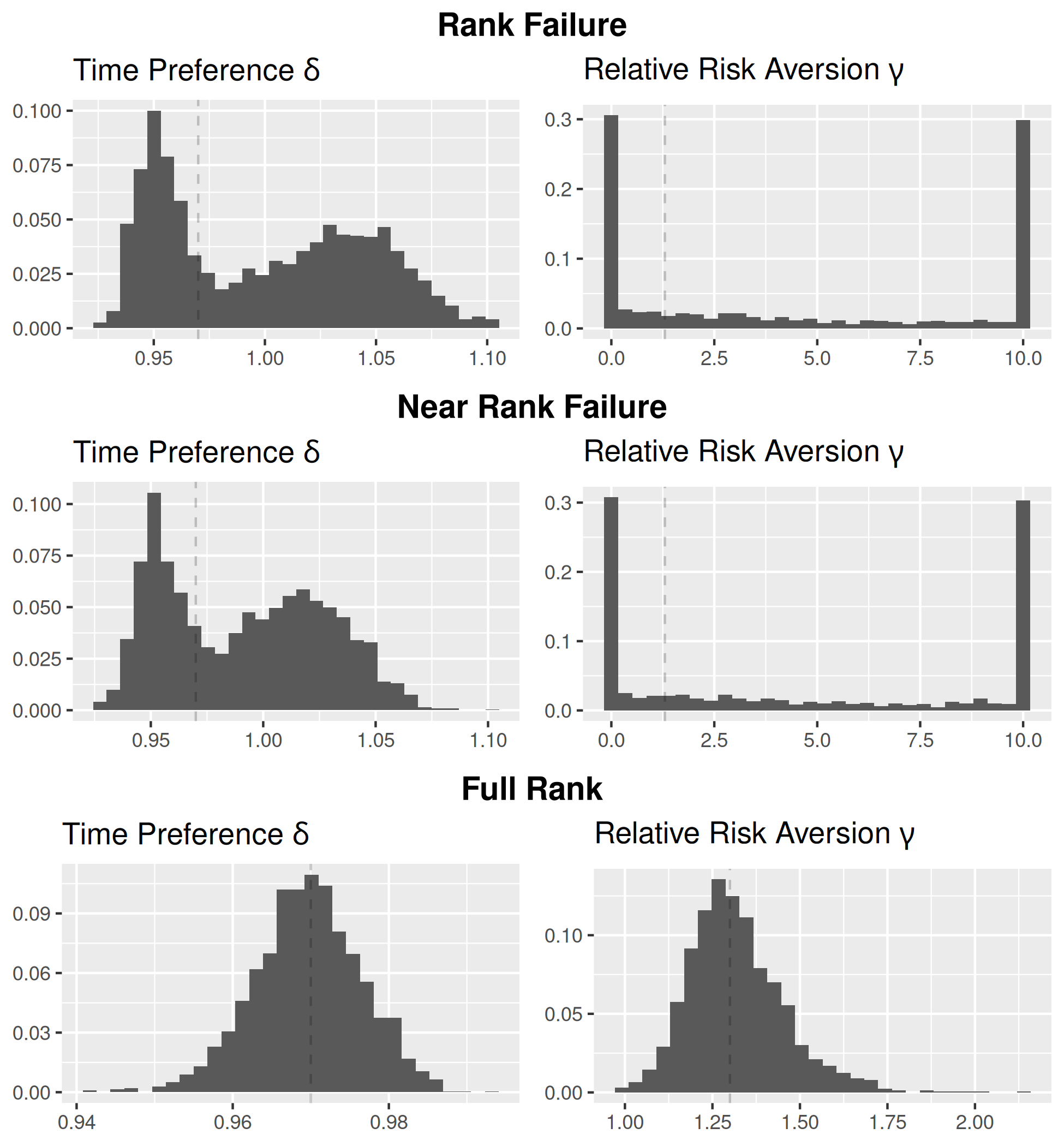} & \includegraphics[scale=0.45]{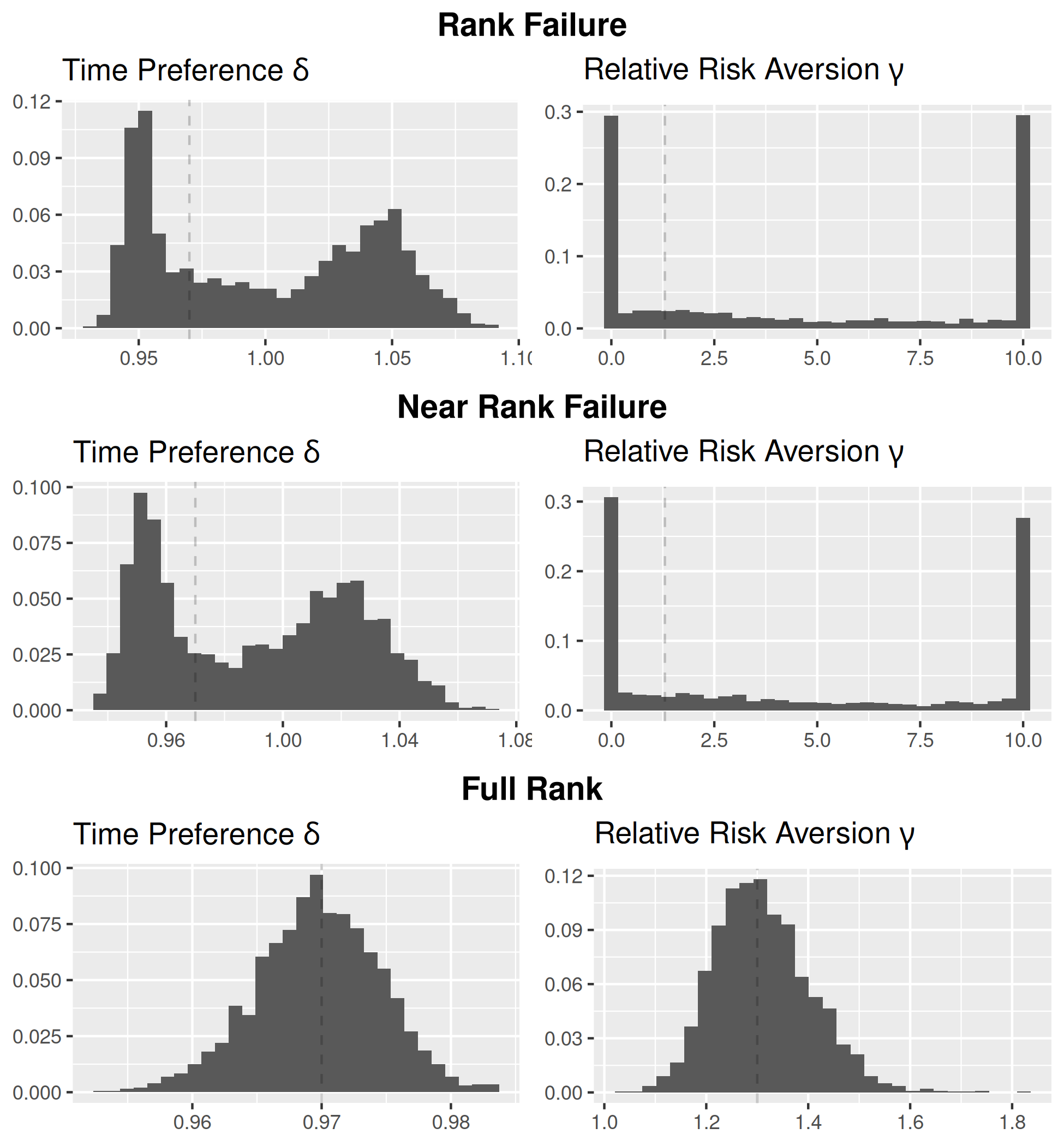}
    \end{tabular}}
    \notes{\textbf{Note:} true value $(\delta_0,\gamma_0) = (0.97,1.3)$: dashed vertical lines. 2000 Monte Carlo replications. Estimates computed for continuously-updated GMM with $W_n = \hat{V}_n(\theta)^{-1}$ where $\hat{V}_n$ is a HAC estimate of $\text{var}[\sqrt{n}\overline{g}_n(\theta)]$.}
  \end{figure}

\paragraph{Non-Linear Regression Model.}
To illustrate the finite-sample properties of the quasi-Jacobian matrix and the test procedure, consider the following nonlinear regression model:
\begin{align} y_i = \theta_1 x_{1i} + \theta_1 \theta_2 x_{2i} + u_i, \label{eq:DGP_MC} \end{align}
where $(x_{1i},x_{2i},u_i) \sim \mathcal{N}(0,I)$ iid. The sample moment conditions are $\bar{g}_n(\theta) = \frac{1}{n}\sum_{i=1}^n (y_i x_{1i} - \theta_1, y_i x_{2i} - \theta_1\theta_2)^\prime$ with population counterpart $g(\theta,\gamma_0) = (\theta_{10} - \theta_1, \theta_{10}\theta_{20} - \theta_1\theta_2)^\prime$. 
For $\theta_{10} = 0$, $\theta_2$ is unidentified and for $\theta_{1n} = c n^{-1/2}$, $c > 0$, $\theta_2$ is weakly identified, even if $\theta_1 = \theta_{1n}$ is known and fixed. The reparameterization $\beta = M\theta$ here is $\beta_1 = \theta_1$, $\beta_2 = \beta_{22} = \theta_2$, $M=I$, and $\mathcal{B}_2^0 = \mathcal{B}_{22}^0 = [\underline{\theta_2},\overline{\theta}_2]$ where $[\underline{\theta}_1,\overline{\theta}_1] \times [\underline{\theta}_2,\overline{\theta}_2] = \Theta_1 \times \Theta_2=\Theta$. The assumptions used for the main results are verified for this model in Appendix \ref{apx:NLS_verif_assumptions}. 

In this simple example, the source of the identification failure is known so that the type I test procedure in \citet[AC12]{Andrews2012} will be used as a benchmark. Let $\text{ICS}_n = |\hat\theta_{1n}| / \hat{\sigma}_{\hat \theta_{1n}}$, where $\hat\theta_{n}=(\hat\theta_{1n},\hat\theta_{2n})^\prime$ is the sample minimizer of $\|\bar{g}_n(\theta)\|$ and $\hat{\sigma}_{\hat \theta_{1n}}^2$ estimates the asymptotic variance of $\hat \theta_{1n}$ using the sandwich formula. The test statistic is $\text{QLR}_n(\theta_1) = \text{AR}_n(\theta_1)$ since the model is just-identified. Let $\underline{\lambda}_n$ be as in Section \ref{sec:test}. When $\text{ICS}_n > \underline{\lambda}_n$, the test rejects $H_0$ if $\text{AR}_n(\theta_1) > \chi^2_1(1-\alpha)$. When $\text{ICS}_n \leq \underline{\lambda}_n$, the test rejects $H_0$ if $\text{AR}_n(\theta_1) > c_{LF,1-\alpha}$ where $c_{LF,1-\alpha}$ is the least-favorable $1-\alpha$ quantile of $\text{AR}_n(\theta_1)$ over $(\theta_2,\gamma) \in \Theta_2 \times \Gamma$. Note that under $H_0:\theta_{10}=0$, $\text{AR}_n(\theta_{10}) \overset{d}{\to} \chi^2_2$, regardless of $\theta_{20}$. Hence, the projection-based critical value in Section \ref{sec:test} is the least-favorable critical value, $c_{LF,1-\alpha} = \chi^2_2(1-\alpha)$.\footnote{A null-imposed least-favorable critical value can also be computed by simulating the distribution of $\text{AR}_n(\theta_{10})$ for each $H_0: \theta_{1}=\theta_{10}$ and all possible $\theta_{20}$. This will not be used here to keep computation manageable.} To summarize, this implementation of the \citet{Andrews2012} procedure relies on the same test statistic and critical values as in Section \ref{sec:test}; the only difference is the choice of ICS statistic. 

\begin{figure}[h] \centering \caption{Model (\ref{eq:DGP_MC}): finite Sample properties of the test and ICS procedures} \label{fig:NLS_rej}
  \includegraphics[scale=0.9]{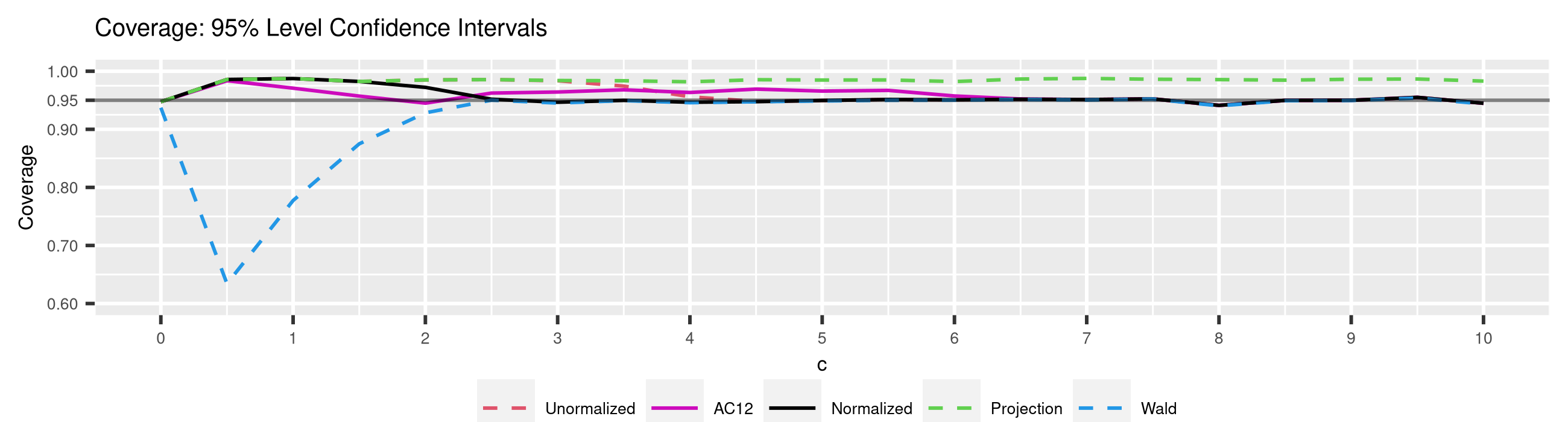}\\
  \includegraphics[scale=0.9]{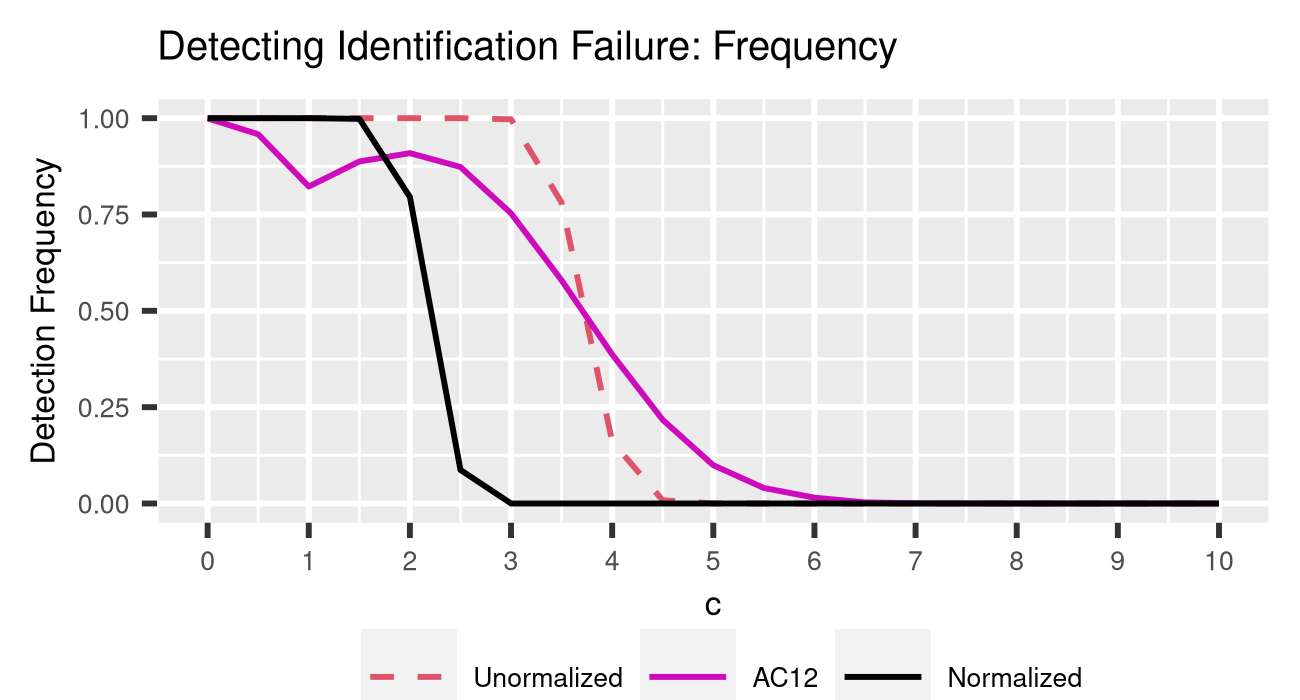}\includegraphics[scale=0.9]{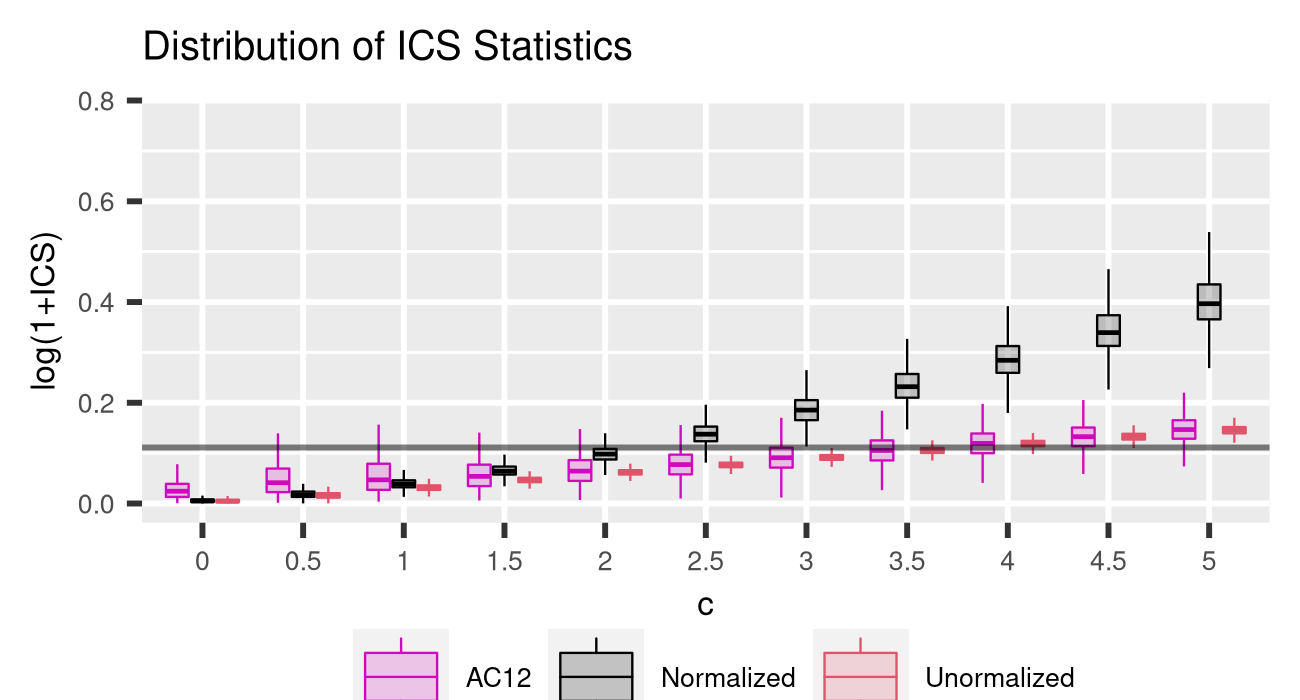}
\end{figure}

Figure \ref{fig:NLS_rej} reports the finite sample properties of several tests and ICS procedures. The top panel shows coverage for $H_0: \theta_1 = \theta_{1n} = cn^{-1/2}$, $c \in [0,10]$, $n=1000$, using a Wald statistic, full projection inference, AC12, and the test procedure from Section \ref{sec:test} using the normalized and unnormalized quasi-Jacobian $B_{n,\infty}$. The Wald test suffers from severe size distortion for $c \in [0,2]$ but is accurate for larger values of $c$. Full projection inference is robust regardless of $c$ but conservative for $c>0$. AC12 and the present procedures have coverage above the 95\% nominal level, the unnormalized procedure is more conservative, AC12 is non-monotonic. To better understand these patterns, the bottom two panels provide further information on the ICS procedures. The left panel shows how often $\text{ICS}_n \leq \underline{\lambda}_n$. The normalized statistic sees a large decline around $c=2$ when size distortion is less severe. AC12 is non-monotonic around $c=1$ where the Wald statistic, on which it is based, has large size distortion. The unnormalized statistic declines sharply but later than the normalized one. To further understand these differences, the right panel plots the distribution of $\log(1+\text{ICS}_n)$. The solid horizontal line indicates the cutoff $\log(1+\underline{\lambda}_n)$. The normalized statistic diverges quickly with $c$, as identification becomes stronger. This matches the above discussion on the role of post-multiplying the quasi-Jacobian by $\Sigma_n^{-1/2}$. AC12 is more dispersed, resulting in more variable outcomes for the ICS procedure as seen in the slow decline in the left panel. AC12 increases with $c$ at a similar rate as the unnormalized statistic. Finite sample power properties of these test procedures are reported in Appendix \ref{apx:MC_additional_simu} as well as results using a larger $\kappa_n$.

Figure \ref{fig:NLS_pow} below presents the finite-sample power properties of the test procedures used in Section \ref{sec:MonteCarlo}. It shows rejection rates against local alternatives $H_0: \theta_1 = \theta_{1n} + a n^{-1/2}$ where the true $\theta_{1n} = c n^{-1/2}$. The nuisance parameter $\theta_2$ is unidentified for $c=0$ and weakly identified for $c \simeq 0$. Each panel summarizes the finite-sample power properties for a specific level of identification strength $c$. 

\begin{figure}[H] \centering \caption{Finite Sample Properties Power of Test Procedures} \label{fig:NLS_pow}
  \includegraphics[scale=1]{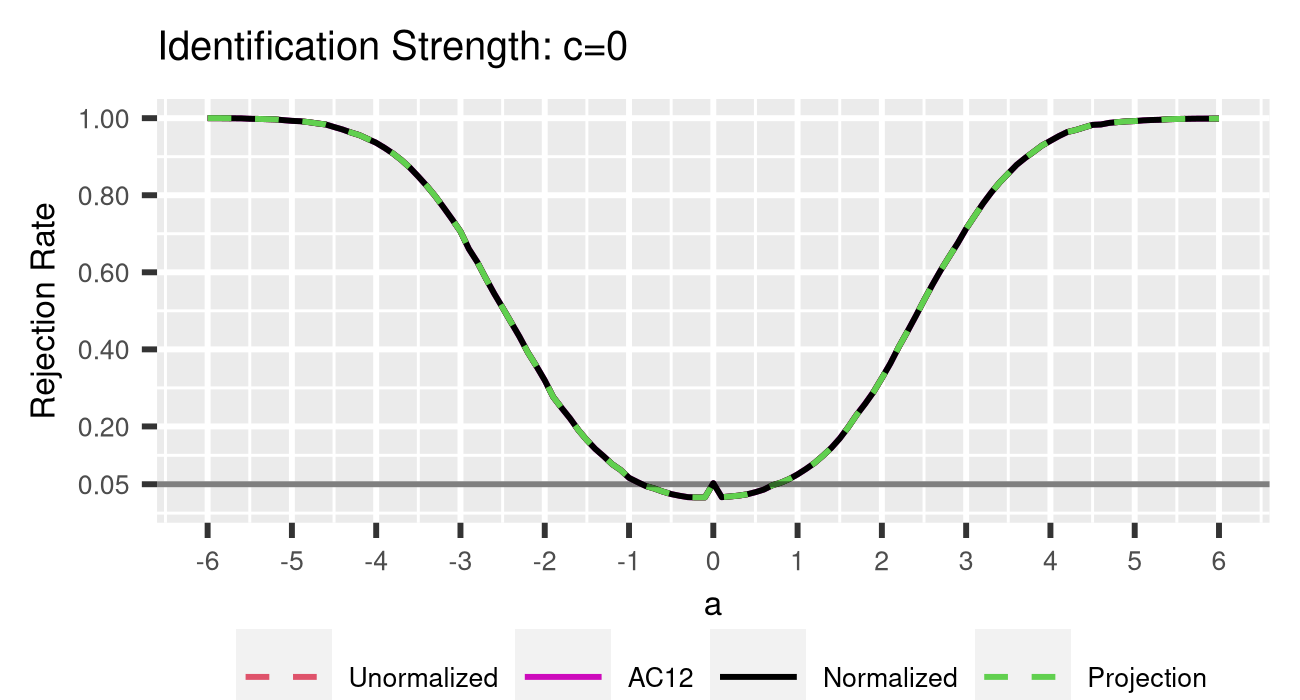}\includegraphics[scale=1]{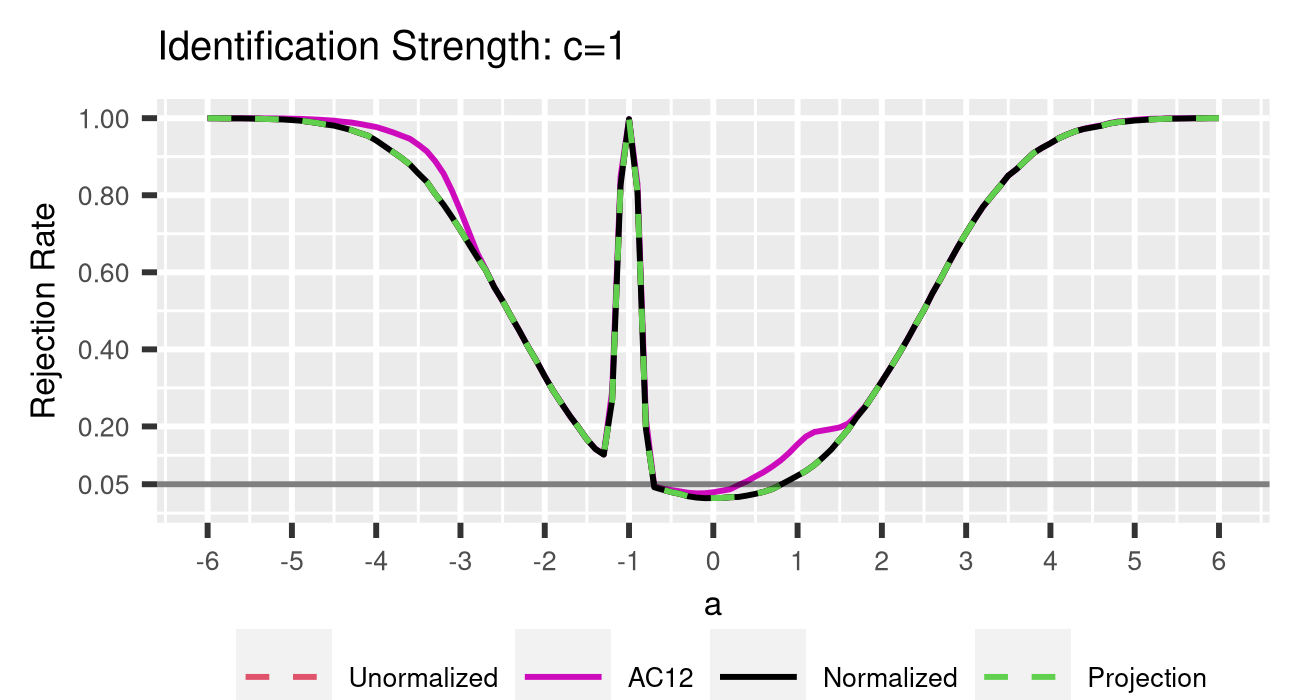}\\
  \includegraphics[scale=1]{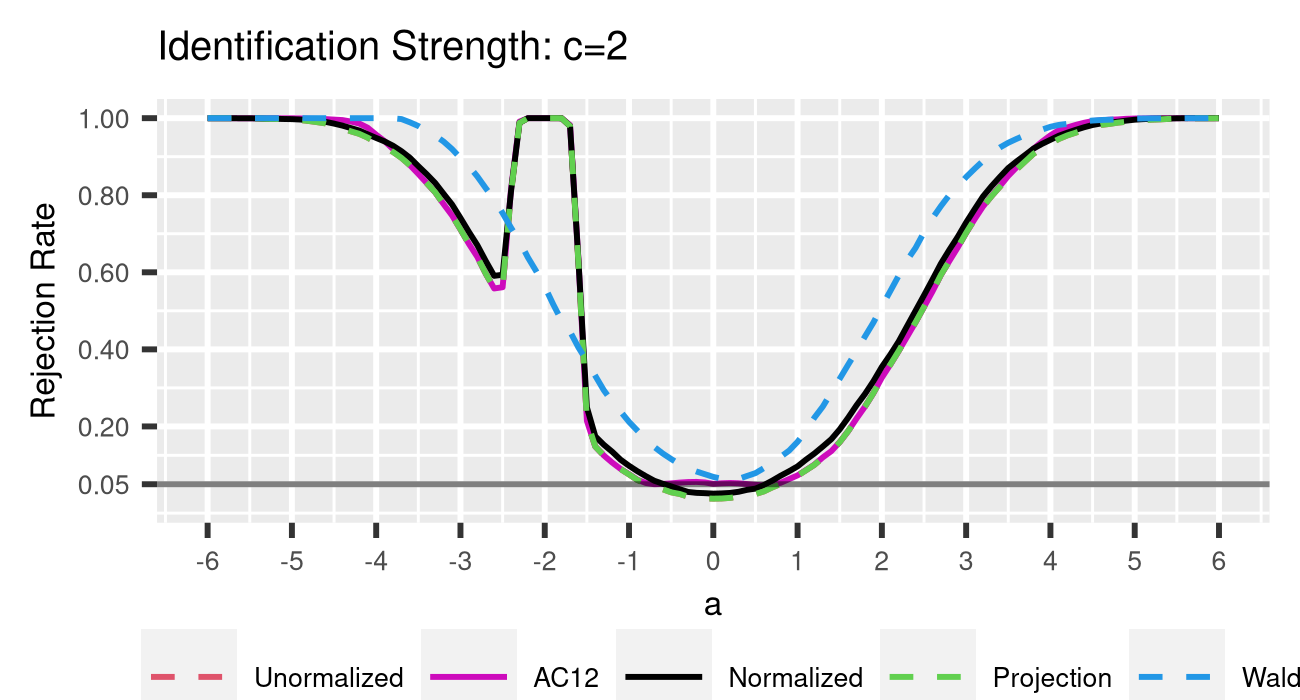}\includegraphics[scale=1]{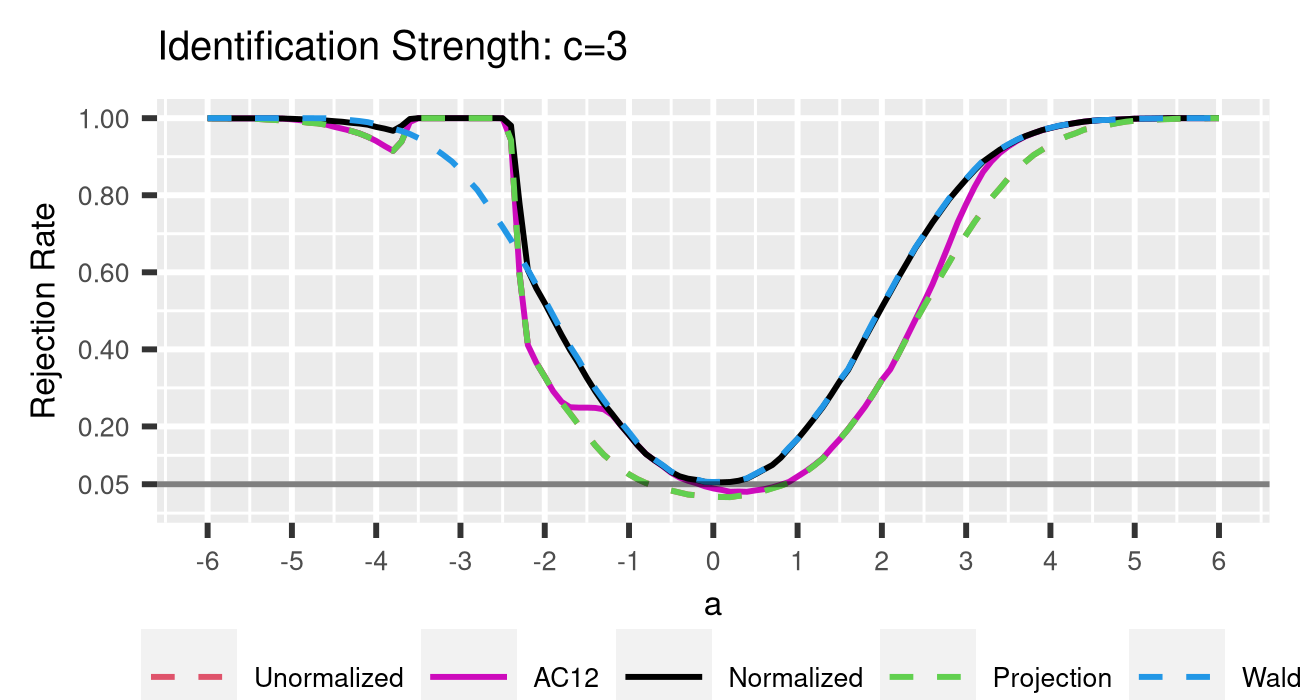}\\
  \includegraphics[scale=1]{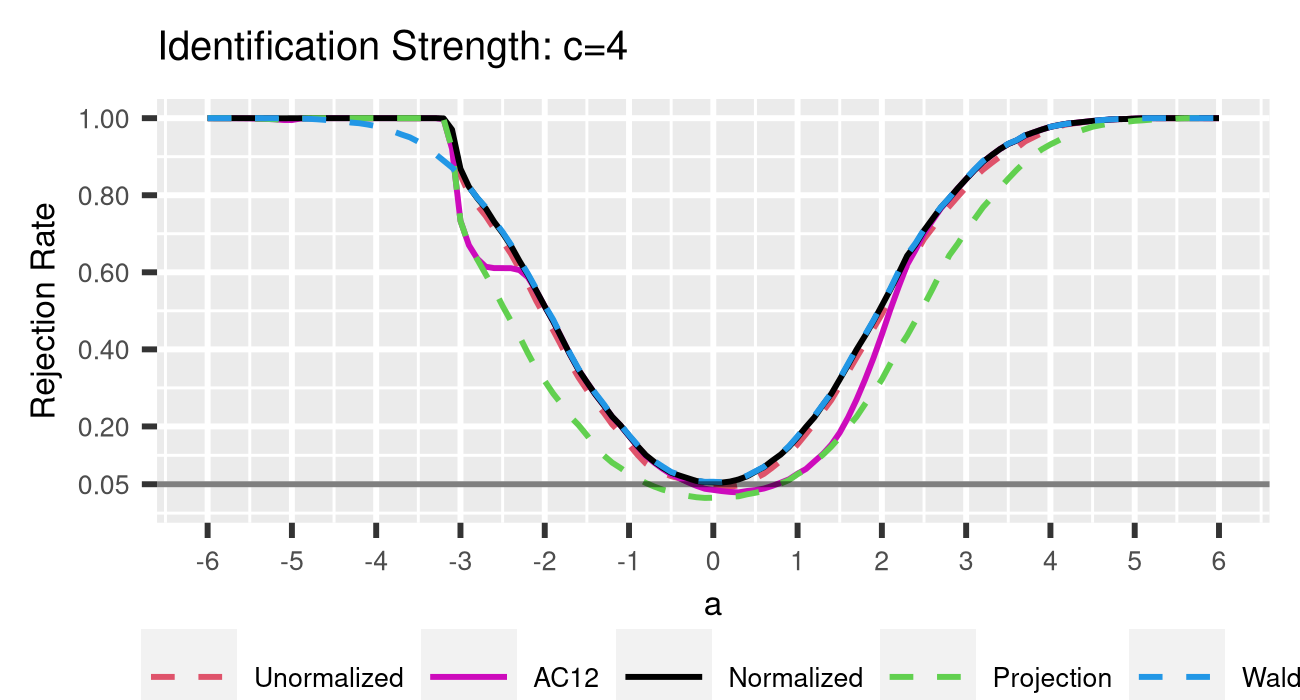}\includegraphics[scale=1]{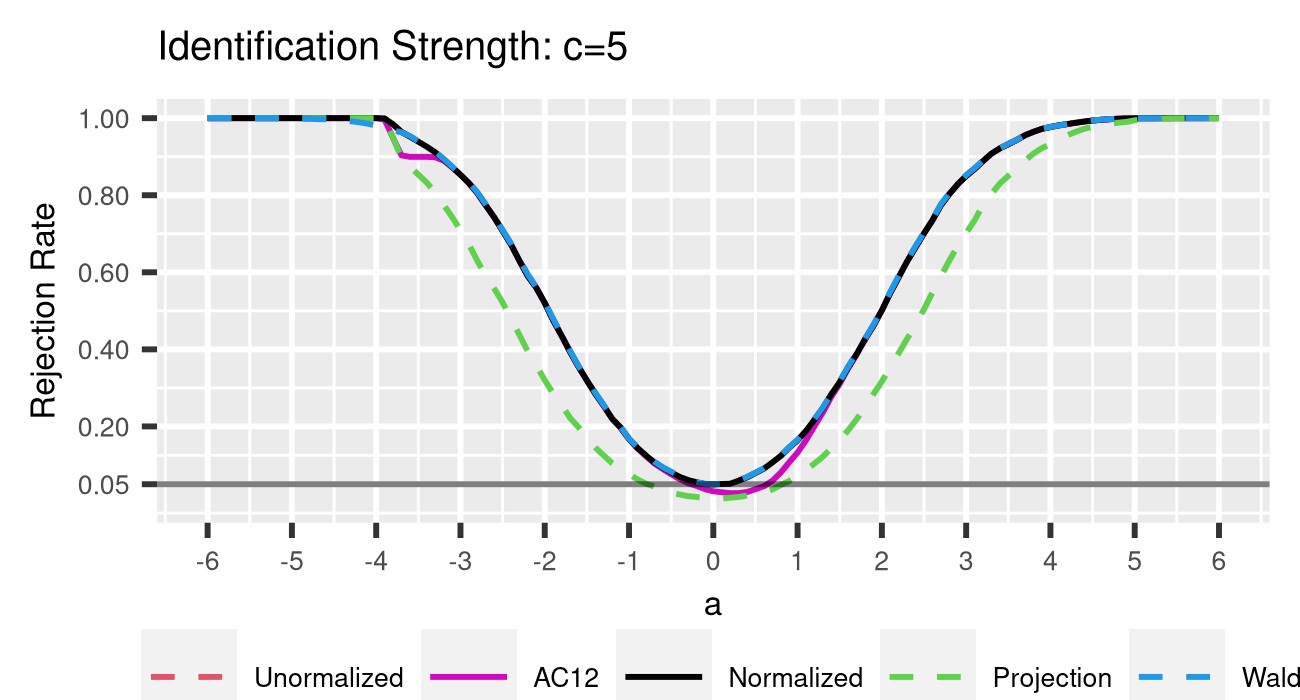}
\end{figure}

For $c=0$ the Projection, AC12, and (un)normalized procedures have identical properties. For $c=1$, AC12 does not detect identification all the time (see Figure \ref{fig:NLS_rej}, bottom left panel) which leads to small critical value and higher rejection rates than the other methods. For $c \in [2,3]$, the normalized test procedure relies on $\chi^2_1$ critical values and has comparable power to the Wald test except for $a+c \simeq 0$. Recall that for just-identified models, the test procedure in Section \ref{sec:test} is equivalent to a standard QLR test when $\hat{d}_n=d_{\theta_2}$ which can be more powerful than the Wald test in finite samples. The normalized ICS procedure is thus more powerful since it almost always picks $\hat{d}_n=d_{\theta_2}$ when $c \geq 2$ (see Figure \ref{fig:NLS_rej}, bottom left panel). The Wald test is not reported for $c < 2$ where it suffers from important size distortion. AC12 has lower power for $c \in [2,4]$ and similar power properties for $c \geq 5$. The unnormalized procedure is comparable to AC12 for $c \in [2,3]$ and is more powerful for $c = 3$.

\begin{figure}[h] \centering \caption{Finite Sample Properties of Test and ICS Procedures} \label{fig:NLS2_rej}
  \includegraphics[scale=0.9]{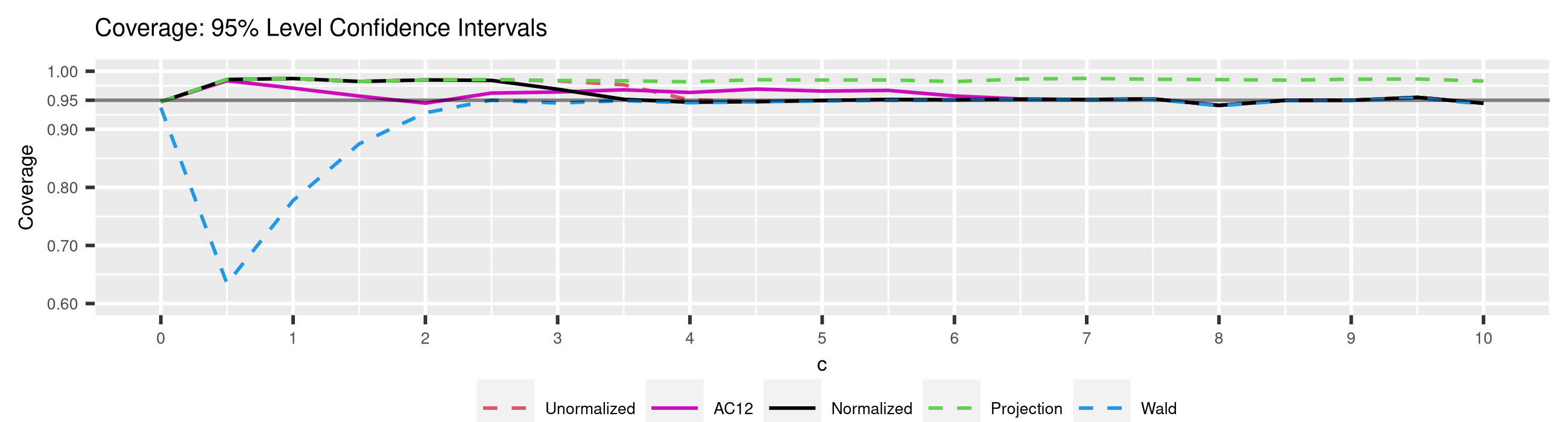}\\
  \includegraphics[scale=0.9]{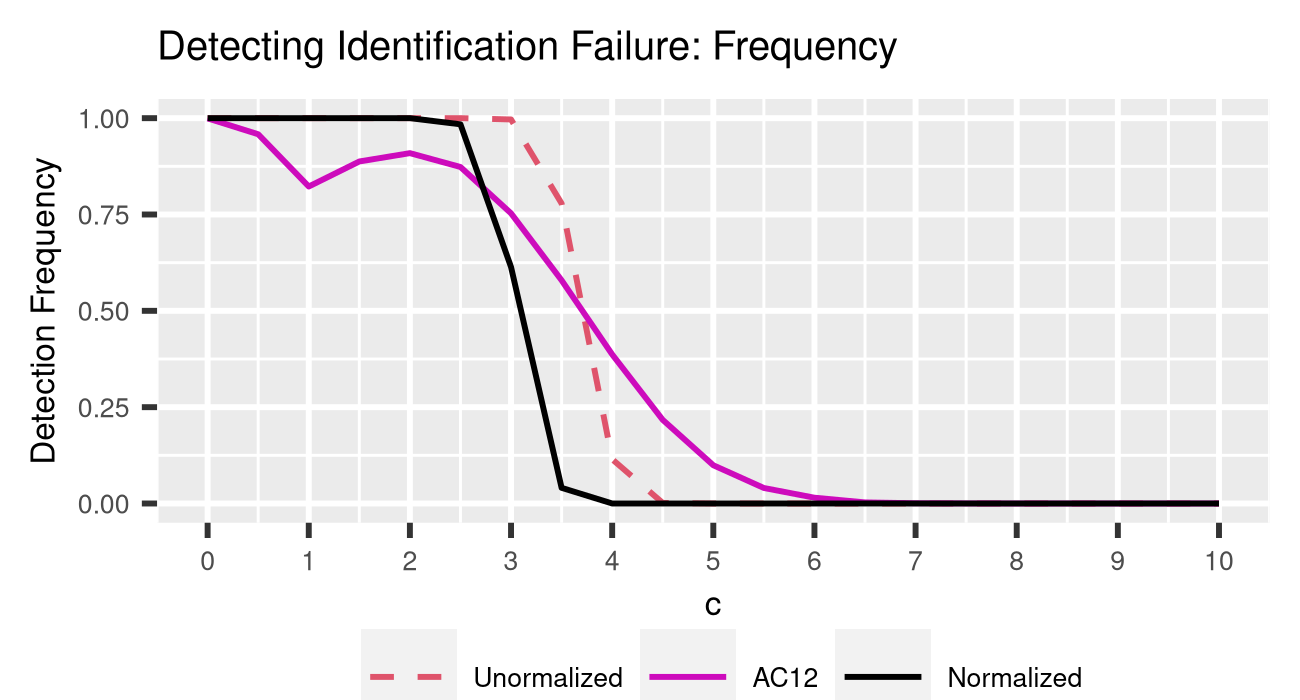}\includegraphics[scale=0.9]{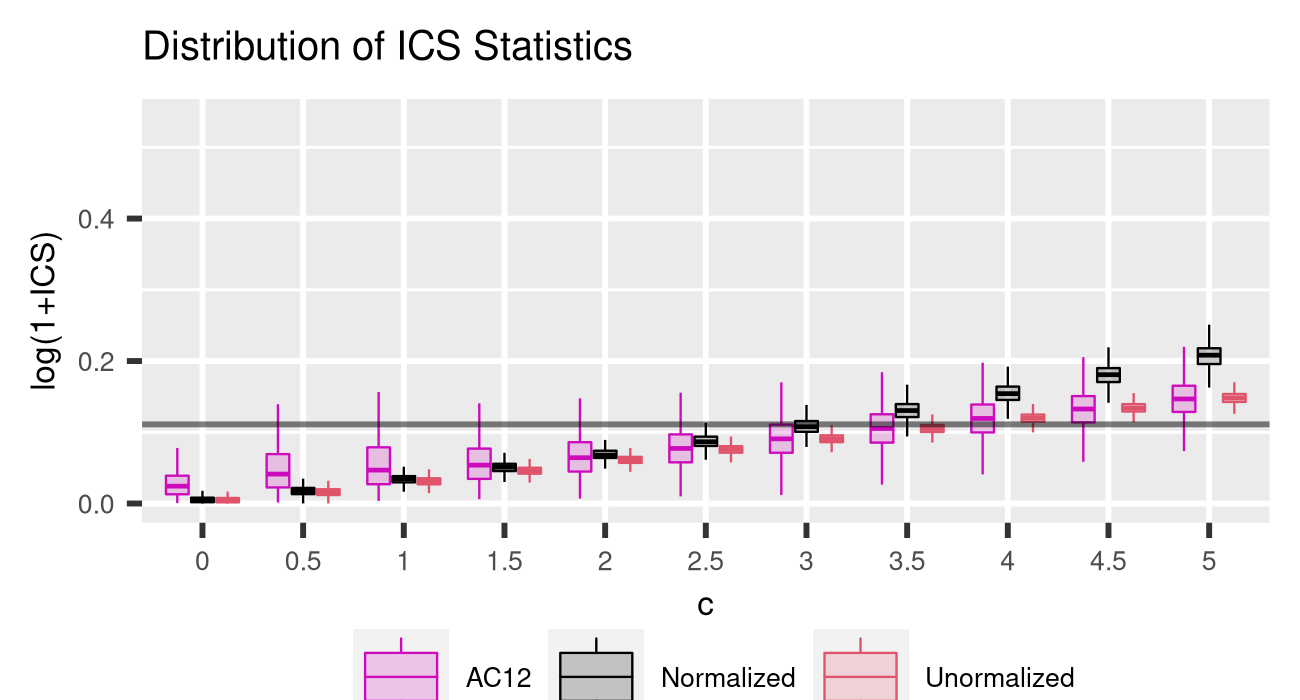}
  \notes{\textbf{Note:}  this choice of $\kappa_n$ corresponds to a 99.99981\% AR confidence set for the full parameter vector $\theta$.}
\end{figure}

\subsection{Additional Empirical Results} \label{apx:extra_empirical}
Confidence sets for $\gamma$ and $\psi^{-1}$ with a $\chi^2_6$ critical value: $[5.28,25]$ and $[0.01,0.87]$, respectively. Using a $\chi^2_6$ critical value amounts to using $\underline{\lambda}_n \in [0.25,387)$ in the baseline results (Table \ref{tab:LRR_eigen}) and $\underline{\lambda}_n \in [0.51,130)$ with the larger value for $\kappa_n$ (Table \ref{tab:extra_LRR_eigen} below).

\begin{table}[ht] \centering
  \caption{Long-Run Risks: singular values of Jacobian and quasi-Jacobian, larger $\kappa_n$} \label{tab:extra_LRR_eigen}
  \begingroup
  \setlength\tabcolsep{4.5pt}
    \renewcommand{\arraystretch}{0.935} 
  {\footnotesize
  \begin{tabular}{l|cccccccccccc}
    \hline \hline
   & $\lambda_{1}$ & $\lambda_{2}$ & $\lambda_{3}$ & $\lambda_{4}$ & $\lambda_{5}$ & $\lambda_{6}$ & $\lambda_{7}$ & $\lambda_{8}$ & $\lambda_{9}$ & $\lambda_{10}$ & $\lambda_{11}$ & $\lambda_{12}$ \\ 
    \hline
  $\bar{V}_n^{-1/2} \partial_\theta \bar{g}_n(\hat\theta_n) \Sigma_n^{-1/2}$ & 
  $1 . 10^6$ & $6 . 10^5 $ & $2 . 10^5$ & $3. 10^4$ & $1 . 10^4$ & 145 & 20 & 0.61 & 0.20 & 0.02 & $<10^{-2}$ & $<10^{-2}$\\  
  $\bar{V}_n^{-1/2} B_{n,\infty} \Sigma_n^{-1/2}$ & 
  $5 . 10^6$ & $1 . 10^6$ & $4 . 10^5$ & $7 . 10^4$ & $1 . 10^3$ & 169 & 0.46 & 0.33 & 0.12 & 0.01 & $<10^{-2}$  & $<10^{-2}$ \\ 
  $\bar{V}_n^{-1/2} B_{n,\infty} P_{\theta_1}^\perp \Sigma_n^{-1/2} P_{\theta_1}^\perp$ & 
  $5 . 10^6$ & $1. 10^6$ & $4. 10^5$ & $7. 10^4$ & $1 . 10^3$ & 169 & 0.43 & 0.32 & 0.02 & $<10^{-2}$ & 0.00 & 0.00 \\ 
     \hline \hline
  \end{tabular} }
  \endgroup
  \notes{\textbf{Note:}  $B_{n,\infty}, \Sigma_n$ computed using $B=1000$ draws, $\kappa_n = 0.43$, $n_s = n \times (1+1/S)$. This choice of $\kappa_n$ corresponds to a 99.99981\% AR confidence set for the full parameter vector $\theta$.}
\end{table}

\begin{figure}[H] \centering \caption{Joint 95\% Confidence Set for $(\gamma,\psi^{-1})$ with $\chi^2_{5}$ critical value} \label{fig:CS_gamma_psi_6}
  \includegraphics[scale=0.9]{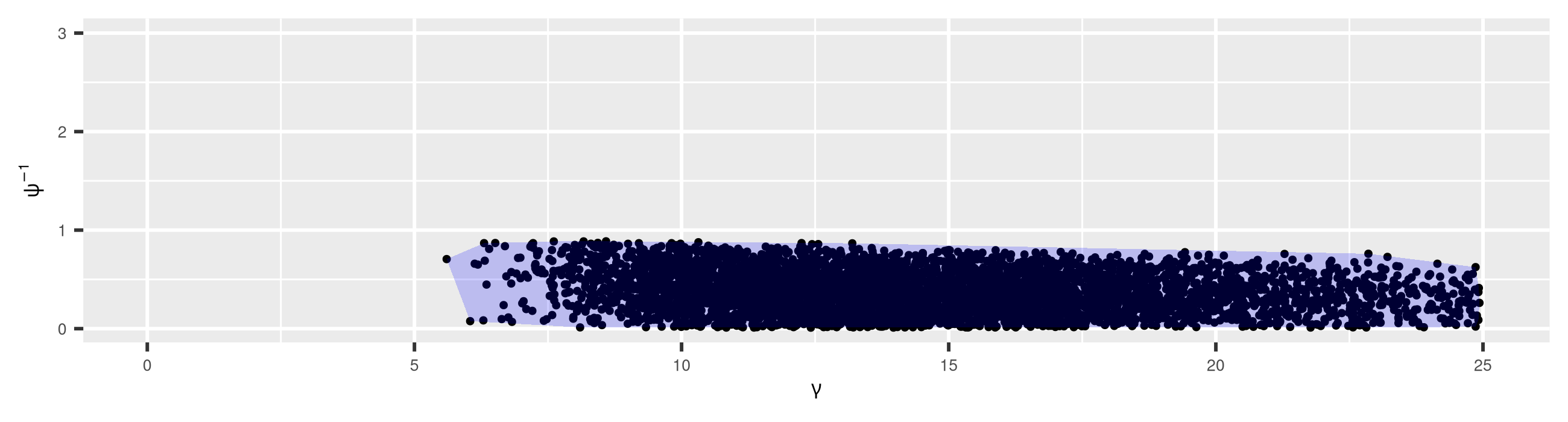}
\end{figure}

\section{Asymptotic Properties of the quasi-Jacobian under Higher-Order Identification} \label{apx:ho}
The following provides pointwise asymptotic results for the quasi-Jacobian matrix when the model is globally but not locally identified.
\begin{assumption}[Higher-Order Identification] \label{ass:ho}
Let $(\theta_0,\gamma_0) \in \overline{\Theta} \times \Gamma$ be such that for some $\varepsilon > 0$ the moments satisfy:
\[ \inf_{\|\theta-\theta_0\| \geq \varepsilon} \|g(\theta,\gamma_0)\| \geq \underline{\delta},  \]
where $\underline{\delta} > 0$. For some $r \geq 2$, there exists orthogonal projection matrices $P_1,\dots,P_r$ and constants $C_1 \geq 0,\dots,C_{r-1} \geq 0, C_r > 0$ where $\sum_{j} C_j P_j$ has rank $d_\theta$ and $C_j C_\ell P_j P_\ell =0$ for any $1 \leq j < \ell \leq r$. These constants and projection matrices are such that for some $\overline{C}>0$ and any $\|\theta-\theta_0\| \leq \varepsilon$:
\[ \overline{C}[\sum_{j=1}^r C_j\|P_j(\theta-\theta_0)\|^j] \geq \|g(\theta,\gamma_0)\| \geq  \sum_{j=1}^r C_j\|P_j(\theta-\theta_0)\|^j. \]
\end{assumption}
Assumption \ref{ass:ho} implies that the model is globally identified but local identification fails so that around $\theta=\theta_0$, the moment function is not linear but approximately polynomial of order $r \geq 2$. If $C_j >0$ then $\|g(\theta,\gamma_0)\|$ is approximately a polynomial of order $j$ in the directions spanned by $P_j$. This contrasts with locally identified models where $g(\theta,\gamma_0)  \approx \partial_\theta g(\theta_0,\gamma_0)(\theta-\theta_0)$ which is locally linear when $\partial_\theta g(\theta_0,\gamma_0)$ is full rank and the non-linear remainder terms are negligible. Under this type of local identification failure, the parameters are consistently estimable but $\hat\theta_n$ has non-standard limiting distribution. Full vector inference using the \citet{Anderson1949} statistic remains valid. As in weakly identified models, concentrating out locally identified nuisance parameters leads to more powerful and asymptotically valid inferences.
\begin{theorem} \label{th:ho}
Suppose Assumption \ref{ass:moments} ii-iii, \ref{ass:kernel_bandwidth}, and \ref{ass:ho} hold for $\gamma=\gamma_0$, then:
\[ \lambda_{\min}(B_{n,\infty}^\prime B_{n,\infty}) = O_p(\kappa_n^{2[1-1/r]}). \]
For any $v_j$ such that $P_j v_j = v_j$ and $C_j >0$: $\|B_{n,\infty}v_j\| = O_p(\kappa_n^{1-1/j})$.
\end{theorem}

\paragraph{Proof of Theorem \ref{th:ho} for $B_{n,\infty}$:}
  Pick $h \in \mathbb{R}$ and $v_j \in \text{span}(P_j)$ with $\|v_j\|=1$ for some $j \in \{2,\dots,r\}$ with $C_j \neq 0$. Let $\theta_{jn} = \theta_0 + \kappa_n^{1/j} h v_j$, by Assumption \ref{ass:ho} we have:
  \[ \|\bar{g}_n(\theta_{jn})/\kappa_n\|_W \leq \overline{\lambda}_WC_j|h|^j + n^{-1/2}\kappa_n^{-1}\overline{\lambda}_W \sup_{\theta \in \Theta}\sqrt{n}\|\bar{g}_n(\theta)-g(\theta,\gamma_0)\| \leq 3/4, \]
  wpa 1 for all $|h| \leq 1/2 [\overline{\lambda}_WC_j]^{-1/j}$. This implies that $\hat{K}_n(\theta_{jn}) \geq \underline{K} = \in_{x \in [0,3/4]}K(x) >0$, wpa 1 uniformly in $|h| \leq 1/2 [\overline{\lambda}_WC_j]^{-1/j}$. Using similar arguments as in the proof of Theorem \ref{th:weak_sup}, we have: $\|\bar{g}_n(\theta) - A_{n,\infty} - B_{n,\infty}\theta\|\hat{K}_n(\theta) \leq \overline{K} \underline{\lambda}_W^{-1}\kappa_n + o(\kappa_n)$, for all $\theta \in \Theta$. Using the triangular inequality we have for any $h_1 \neq h_2$ such that $|h_{1,2}| \leq 1/2 [\overline{\lambda}_WC_j]^{-1/j}$:
  \[ \|B_{n,\infty}\kappa_n^{1/j}[h_1-h_2]v_j\| \leq \underline{K}^{-1}\overline{K} \underline{\lambda}_W^{-1}\kappa_n + o(\kappa_n),\]
wpa 1. Since $j>1$, this implies that:
\[ \|B_{n,\infty}v_j\| \leq |h_1-h_2|^{-1}\underline{K}^{-1}\overline{K} \underline{\lambda}_W^{-1}\kappa_n^{1-1/j} + o(\kappa_n^{1-1/j}), \]
wpa 1 for each $j$ such that $C_j \neq 0$. In particular, we have for $j=r$ that: $\|B_{n,\infty}v_r\| \leq O_p(\kappa_n^{1-1/r})$ so that $\lambda_{\min}(B_{n,\infty}^\prime B_{n,\infty}) \leq v_r^\prime B_{n,\infty}^\prime B_{n,\infty} v_r \leq O_p(\kappa_n^{2[1-1/r]})$.
\qed

\end{appendices}
\end{document}